\documentclass[showpacs,preprint]{revtex4}
\usepackage{graphicx}
\usepackage{subfigure}
\usepackage[section]{placeins}
\usepackage{morefloats}
\usepackage{amssymb}
\usepackage{amsmath}
\usepackage{bm}
\usepackage{latexsym}
\usepackage{hyperref}

\newcommand{\Tr}{{\text Tr}}
\begin{document}

\title{Suppression of decoherence by bath ordering}

\author{Jun Jing \footnote{Email: jingjun@sjtu.edu.cn}, H.R. Ma}
\affiliation{Institute of Theoretical Physics,
Shanghai Jiao Tong University\\
800 DongChuan Road, MinHang, Shanghai 200240, China}
\date{\today}

\begin{abstract}

The dynamics of two coupled spins-1/2 coupled to a spin-bath is
studied as an extended model of the Tessieri-Wilkie Hamiltonian
\cite{TWmodel}. The pair of spins served as an open subsystem were
prepared in one of the Bell states and the bath consisted of some
spins-1/2 is in a thermal equilibrium state from the very beginning.
It is found that with the increasing the coupling strength of the
bath spins, the bath forms a resonant antiferromagnetic order. The
polarization correlation between the two spins of the subsystem and
the concurrence are recovered in some extent to the isolated
subsystem. This suppression of the subsystem decoherence may be used
to control the quantum devices in practical applications.

\end{abstract}

\maketitle

\section{Introduction}

Quantum decoherence is a common and inevitable phenomenon in an open
quantum system due to its interaction with the surrounding bath or
environment. An initial superposition state of the system,
$\rho_s(t=0)=|\psi_0\rangle\langle\psi_0|$, has to decay into a
classical mixture of states, $\rho_s=\sum_ip_i\rho_i$,
$\sum_ip_i=1$, after a decoherence time $\tau_d$. During the same
period of time, some of the information in the initial state of the
system might irreversibly lose into the bath \cite{Zurek, Joos,
Zurek1981, Zurek2003, Leggett1987}. A real-world system, for
instance a quantum device or qubit, cannot be completely isolated
from the environment. In the last two decades, there are great
interests in the search for realizations of {\it quantum
computation} as well as {\it quantum communications}, such
technologies rely on the possibilities that quantum devices can be
manufactured with negligible decoherence in the period of operation.
Since decoherence is intrinsic to open quantum systems, the problem
is transformed to a problem on how to reduce or eliminate the
decoherence of an open quantum system. Many works have been devoted
into the research of the influences caused by the subsystem-bath
coupling and the decoherence dynamics of open quantum systems
\cite{breuer, nielson}, for the possible realizations of quantum
communication and quantum computation. \\

A very important part of any theoretical research of decoherence is
the modeling of the bath or environment. There had been mainly two
important types of bath schemes: (i) the bath consisted of harmonic
oscillators, e.g., the Caldeira-Leggett model \cite{Leggett}; (ii)
the bath made up of spins-1/2, e.g., model used in Ref.
\cite{Prokoev}. For both types of bath model, there are typically
two kinds of approximations, Markovian \cite{Gardiner} or
non-Markovian \cite{Shresta}, used in the studies. The focus of most
of the researches were on the eliminating the destructive effects of
the environment to the system. However, Tessieri and Wilkie proposed
a new idea to the reducing of decoherence \cite{TWmodel} of the open
subsystem, which is a spin-1/2 coupled to a low-temperature bath of
spins. They introduced the coupling terms between bath spins into
their spin-bath Hamiltonian, which is a simplification of the
vibronic Hamiltonian of the impurity and crystal \cite{Estreicher}
using spin-1/2 modes. One of the most important results they found
is that the strong antiferromagnetic intra-bath interactions could
be utilized to make the dynamics of the central spin almost
autonomous from the bath around. And Dawson et al. \cite{Milburn}
also found the decoherence effect of the bath can be suppressed by
the increasing intra-bath coupling in Tessieri-Wilkie model. \\

Inspired by their works and the papers of Xiang et al. \cite{Xiang}
and Liu et al. \cite{Liu}, the aim of this paper is to study the
dynamics of correlation characteristics between the \emph{two}
coupled center spins, which constitute our subsystem. The bath in
our study is the same model used by the Ref. \cite{TWmodel,
Milburn}, which is prepared in a thermal equilibrium state at time
$t=0$. The dynamics of the subsystem and the bath then calculated,
phenomena such as decoherence oscillation were observed, which can
be quantified by the amplitudes of polarization components
$\langle\sigma_x\rangle$, $\langle\sigma_y\rangle$ and
$\langle\sigma_z\rangle$. We studied the evolution of the spatial
correlation between polarizations of the two subsystem spins along
three directions $\vec{x}$, $\vec{y}$ and $\vec{z}$. we also
discussed the entanglement between the two system spins \cite{Shan},
which is an essential ingredient in the quantum communication
\cite{Ekert, Bennett} and has no classical counterpart. The
concurrence of the subsystem was chosen to be a measure of the
entanglement between the two spins within the subsystem
\cite{Wootters1, Wootters2}. It will be demonstrated that in our
extended model, both the polarization correlation and entanglement
of the subsystem can be restored to a great extent to the isolated
case by the increasing pairwise couplings between the bath spins.
And we also try to clarify the physics behind this revival function
of intra-bath coupling. The rest of this paper is organized as
follows. In section \ref{Hamiltonian}, we introduce the Hamiltonian
for our two-center-spins-spin-bath model; In section
\ref{calculation}, we explain the computation procedures about the
evolution of the correlation and of the concurrence for the
subsystem; Detailed results and discussions are in section
\ref{discussion}; The conclusion of our study is given in section
\ref{conclusion}.

\section{The extended spin-spin-bath model}\label{Hamiltonian}

The subsystem we studied here consists of two spins
antiferromagneticly coupled in the $x$ direction, and aligned in $x$
and $z$ direction respectively by applied fields. The bath consists
of $N-2$ spins, every pair of the spins are also coupled
antiferromagneticly in the $x$ direction, and aligned in both $x$
and $z$ direction. The subsystem interacts with the bath by the
antiferromagnetic coupling in the $x$ direction. This model is an
extension to the Tessieri-Wilkie model \cite{TWmodel}, which comes
from the simplification of an experimentally realizable model
system. The original model is based on the system that an
interstitial He atom in an otherwise pure silicon or diamond
cluster, the He atom causing small lattice distortion in the cluster
\cite{Davies}, thus inducing vibronic coupling between the
electronic states and phonon states of the cluster. This mechanism
is further simplified by considering only the ground state and the
first excited state of the interstitial He atom, representing the
states with a spin-1/2 Pauli matrix, and representing the phonon
states as spin-1/2 modes, which resulting in the Tessieri-Wilkie
model. This model system is regarded as a potential realization for
quantum computing device. Our extension is that instead of one
center spin, we have two center spins coupled together by an
antiferromagnetic coupling. In this extension we merely regard the
Tessieri-Wilkie model as a model system and did not go into details
of the physical realization. So that we will only focus our
attention to the extended model itself in the following
discussions.\\

The Hamiltonian of our system can be written as:
\begin{eqnarray}\label{whole}
H &=& H_S+H_B+H_{SB}, \\ \label{Hs} H_S &=&
\frac{\omega_s}{2}\sigma_z^{(1)}+\beta\sigma_x^{(1)}+\frac{\omega_s}{2}\sigma_z^{(2)}+
\beta\sigma_x^{(2)}+\lambda_{ss}\sigma_x^{(1)}\sigma_x^{(2)},\\
H_B &=&
\sum_{i=3}^{N}\frac{\omega_b}{2}\sigma_z^{(i)}+\beta\sum_{i=3}^N\sigma_x^{(i)}+
\lambda_{bb}\sum_{i=3}^{N-1}\sum_{j=i+1}^{N}\sigma_x^{(i)}\sigma_x^{(j)},\\
\label{H_SB} H_{SB} &=&
\lambda_{sb}\sum_{i=3}^{N}(\sigma_x^{(1)}\sigma_x^{(i)}+\sigma_x^{(2)}\sigma_x^{(i)}).
\end{eqnarray}
Where $H_S$ is the Hamiltonian of the subsystem with two interacting
spins (labeled by 1,2) coupled in $x$ direction with coupling
strength $\lambda_{ss}$. $\beta$, $\omega_s/2$ and $\omega_b$ may be regarded
as applied fields acting in $x$ and $z$ directions to the sub-system and
bath spins. The $\sigma_x$ and $\sigma_z$ are the Pauli
matrices:
\begin{equation}
\sigma_x=\left(\begin{array}{cc}
      0 & 1 \\
      1 & 0
    \end{array}\right), \quad
\sigma_z=\left(\begin{array}{cc}
      1 & 0\\
      0 & -1
    \end{array}\right).
\end{equation}
The bath consists of $N-2$ spins labeled by $3$ to $N$, coupled each
other in the $x$ direction with coupling strength $\lambda_{bb}$,
denoted as $H_B$. The coupling between subsystem and bath is given
by $H_{SB}$, which is also the coupling of each subsystem spin with
every bath spin in $x$ direction, with coupling strength
$\lambda_{sb}$. All the interactions are antiferromagnetic, that is
\begin{equation}
\lambda_{ss}\geq0,\quad \lambda_{bb}\geq0,\quad \lambda_{sb}\geq0.
\end{equation}
In the following calculation, we use the system of units where the
Boltzmann constant $k_B=1$ and Plank constant $\hbar=1$. The other
parameters in the Hamiltonian take the following values in the
subsequent computations:
\begin{eqnarray*}
\omega_s&=&0.8,\quad
\beta = 0.1, \quad
\omega_b = 1.0.
\end{eqnarray*}

\section{Calculation procedures}\label{calculation}

In this section, we describe in detail the algorithms for the
calculation of the dynamics of the subsystem modeled by the
Hamiltonian (\ref{whole}). The initial state of the subsystem is
chosen to be one of the four Bell states:
\begin{eqnarray}\label{Bell1}
|\psi_S(0)\rangle^1&=&1/\sqrt{2}(|11\rangle+|00\rangle),
\\\label{Bell2}
|\psi_S(0)\rangle^2&=&1/\sqrt{2}(|10\rangle+|01\rangle),
\\\label{Bell3}
|\psi_S(0)\rangle^3&=&1/\sqrt{2}(|11\rangle-|00\rangle),
\\\label{Bell4}
|\psi_S(0)\rangle^4&=&1/\sqrt{2}(|10\rangle-|01\rangle).
\end{eqnarray}
where $|0\rangle$ and $|1\rangle$ refer to the spin ``down'' and
spin ``up'' in the $z$ direction, respectively. The reason that we
use these states as initial states is that among all the quantum
states for a pair of coupled spin-1/2, Bell bases have the largest
degree of entanglement and many other interesting characteristics
\cite{Bell}. However, it should be noted that the last Bell state
given in equation (\ref{Bell4}) is a state free from the bath, the
action of the interaction part of the Hamiltonian, $H_{SB}$, on the
state gives zero. So that the state will evolve with time just as an
isolated system. Further more, the state is also an eigenstate of
the subsystem Hamiltonian $H_s$:
\begin{equation}
H_s(1/\sqrt{2}(|10\rangle-|01\rangle))=-1/\sqrt{2}\lambda_{ss}(|10\rangle-|01\rangle).
\end{equation}
The time evolution of the state is simply an exponential factor
$e^{i\omega t}$, with $\omega = -\lambda_{ss}$. So that the physical
quantities will not vary with time in this state and we will not
consider it in the following calculations. The initial state of the
bath is taken to be the thermal equilibrium state:
$\rho_B(0)=Z^{-1}\exp(-H_B/T)$, where $Z$ is the partition function
of the environment $Z=\Tr(\exp(-H_B/k_BT))$. The density matrix
$\rho(t)$ of the whole system can formally be represented as:
\begin{eqnarray}
\rho(t)&=&\exp(-iHt)\rho(0)\exp(iHt)\\ \label{rhoB}
\rho(0)&=&\rho_S(0)\otimes\rho_B(0)\\ \label{rho_s}
\rho_S(0)&=&|\psi_S(0)\rangle\langle\psi_S(0)|
\end{eqnarray}
In order to find the density matrix $\rho(t)$, we follow the method
suggested by Tessieri et al \cite{TWmodel}. The thermal bath state
$\rho_B(0)$ can be expanded with the eigenstates of the environment
Hamiltonian:
\begin{eqnarray}\label{rho_B}
\rho_B(0)&=&\sum_{n=1}^{2^m}|\phi_n^{(B)}\rangle\omega_n\langle\phi_n^{(B)}|,\\
\label{weight}
\omega_n&=&\frac{e^{-E_n/T}}{Z},\\
Z&=&\sum_{n=1}^{2^m}e^{-E_n/T}.
\end{eqnarray}
Here $|\phi_n^{(B)}\rangle$, $n=1, 2, 3, \cdots, 2^{m}$, $m=N-2$,
are the eigenstates of the environment Hamiltonian $H_B$, and $E_n$
the corresponding eigen energies. With this expansion, the density
matrix $\rho(t)$ can be written as:
\begin{equation}\label{equ:2m}
\rho(t)=\sum_{n=1}^{2^m}\omega_n|\Psi_n(t)\rangle\langle\Psi_n(t)|.
\end{equation}
Where
\begin{equation}
|\Psi_n(t)\rangle =\exp(-iHt)|\Psi_n(0)\rangle =U(t)|\Psi_n(0)\rangle.
\end{equation}
The initial state is
\[
|\Psi_n(0)\rangle =|\psi_S(0)\rangle |\phi_n^{(B)}\rangle.
\]
The evolution operator $U(t)$ can be evaluated by the efficient
algorithm of polynomial schemes \cite{Dobrovitski1, Hu, Jing}. The
method used in this calculation is the Laguerre polynomial expansion
method we proposed in Ref. \cite{Jing}, which is pretty well suited
to this problem and can give accurate result in a comparatively
smaller computation load. More precisely, the evolution operator
$U(t)$ is expanded in terms of the Laguerre polynomial of the Hamiltonian as:
\begin{eqnarray*}
U(t) &=& e^{-iHt} \\ &=& \left(\frac{1}{1+it}\right)^{\alpha+1}
\sum^{\infty}_{k=0}\left(\frac{it}{1+it}\right)^kL^{\alpha}_k(H)
\end{eqnarray*}
where $\alpha$ distinguishes different types of Laguerre polynomials
\cite{Arfken}, $k$ is the order of the Laguerre polynomial. In real
calculations the expansion has to be cut at some value of
$k_{\text{max}}$, which was taken to be $24$ in this study. With the
largest order of the expansion fixed, the time step $t$ is
restricted to some value in order to get accurate results of the
evolution operator. For longer times the evolution can be achieved
by more steps. The action of the Laguerre polynomial of Hamiltonian
to the states is calculated by recurrence relations of the Laguerre
polynomial. The efficiency of this polynomial scheme \cite{Jing} is
about $8$ times as that of the Runge-Kutta algorithm used in Ref.
\cite{TWmodel}. When the states $|\Psi_n(t)\rangle$ are obtained,
the density matrix can be obtained by performing a summation in
equation (\ref{equ:2m}).

\begin{table}
\caption{the number of the bath states, $M$, needed in the
calculation for different temperatures and thresholds of
$\omega_{th}$.}
\begin{center}
\begin{tabular}{|c|c|c|c|c|c|c|}
\hline {\sl Temperature $T$} & {\sl $\leq0.04$} & {\sl $0.05$} &
{\sl $0.08$} & {\sl\, $0.10$} & {\sl\, $0.15$} & {\sl\, $0.20$}\\
\hline {\sl $M(\omega_n>0.00001)$} & {\sl $8$} & {\sl $8$} & {\sl
$28$} & {\sl $70$} & {\sl $70$} & {\sl $70$}\\ \hline {\sl
$M(\omega_n>0.0001)$} & {\sl $1$} & {\sl $8$} & {\sl $8$} & {\sl
$28$} & {\sl $28$} & {\sl $70$}
\\ \hline {\sl $M(\omega_n>0.001)$}
& {\sl $1$} & {\sl $1$} & {\sl $8$} & {\sl $8$} & {\sl $28$} & {\sl
$70$}
\\ \hline
\end{tabular}
\end{center}
\label{N8L4T01}
\end{table}

When the temperature is not very high, the weights for high energy
states will be very small so that only few lowest states need to be
considered. The maximum states to be included in the summation can
be determined by setting a threshold of weight $\omega_{th}$, and
keep only those states whose weight $\omega_n \ge \omega_{th}$. As
an example, table (\ref{N8L4T01}) gives the number of states to be
used for different temperatures and threshold. In this case the
number of environment spins $m=8$, and couplings among the
environment spins $\lambda_{bb}=4.0$. As can be seen from the table,
the number of states needed is much less than the total number of
states, $2^8=256$. In real calculation the up bound of the summation
in equation (\ref{equ:2m}) will be $M<< 2^m$, chosen by the criteria
specified for the accuracy of results. That is to say, equation
(\ref{equ:2m}) could be changed into the following equation:
\begin{equation}\label{equ:M}
\rho(t)=\sum_{n=1}^M\omega_n|\Psi_n(t)\rangle\langle\Psi_n(t)|.
\end{equation}

After obtaining the density matrix of the whole system, the reduced
density matrix is calculated by a partial trace operation to
$\rho(t)$, which trace out the degrees of freedom of the
environment:
\begin{equation}\label{final}
\rho_S(t)=\Tr_B(\rho(t)).
\end{equation}
For the model of this paper, $\rho_S(t)$ is the density matrix of
the open subsystem consists of two center spins, which can be
expressed as a $4\times4$ matrix in the Hilbert space of the
subsystem spanned by the orthonormal vectors $|00\rangle$,
$|01\rangle$, $|10\rangle$ and $|11\rangle$.\\

After the reduced density matrix is obtained, any physical
quantities of the subsystem can be obtained easily. In the following
we concentrate on two important physical quantities of the subsystem
which reflect the decoherence and entanglement degree of the
subsystem state. These two quantities are the spatial polarization
correlation and the concurrence.

\subsection{Polarization correlation}\label{correlation}

The polarization of either spin in the subsystem is defined as
\begin{equation}\label{polar}
\vec{P}^{(i)}(t)=\Tr(\rho^{(i)}(t)\vec{\sigma}^{(i)}), \quad i=1,2,
\end{equation}
where $\vec{\sigma}=\sigma_x\vec{i}+\sigma_y\vec{j}
+\sigma_z\vec{k}$. To simplify the calculation, we may also trace
out one of the spin degrees of freedom to obtain the $2 \times 2$
sub-reduced density matrix for each spin:
\[
\rho^{(i)}(t)= \Tr_{\bar i}(\rho_S(t))=\left(\begin{array}{cc}
      \rho^{(i)}(t)_{11} & \rho^{(i)}(t)_{10} \\
      \rho^{(i)}(t)_{01} & \rho^{(i)}(t)_{00}
    \end{array}\right)
\]
Here $\Tr_{\bar i}$ means to trace out the other degrees of freedom
of $i$, i.e. trace out $2$ when $i=1$ and trace out $1$ when $i=2$.
Then the three components of $\vec{P}^{(i)}(t)$ are expressed as:
\begin{eqnarray}\label{sigmax}
\langle\sigma_x^{i}\rangle &=& \Tr\left(\rho^{(i)}(t)\sigma_x\right)
=\rho^{(i)}(t)_{10}+\rho^{(i)}(t)_{01},\\
\label{sigmay}\langle\sigma_y^{i}\rangle &=&
\Tr\left(\rho^{(i)}(t)\sigma_y\right)
=i\left(\rho^{(i)}(t)_{10}-\rho^{(i)}(t)_{01}\right),\\
\label{sigmaz} \langle\sigma_z^{i}\rangle &=&
\Tr\left(\rho^{(i)}(t)\sigma_z\right)=\rho^{(i)}(t)_{11}-\rho^{(i)}(t)_{00}.
\end{eqnarray}
The polarization $\vec P^{(i)}(t)$ may be viewed as an indicator of
quantum decoherence. \\

The correlation between the two subsystem spins can be described
by the correlation functions defined bellow:
\begin{eqnarray}\label{corrx}
C_{xx}&=&\langle\sigma_x^{(1)}\sigma_{x}^{(2)}\rangle
  -\langle\sigma_x^{(1)}\rangle\langle\sigma_x^{(2)}\rangle,\\
\label{corry}C_{yy}&=&\langle\sigma_y^{(1)}\sigma_{y}^{(2)}\rangle
  -\langle\sigma_y^{(1)}\rangle\langle\sigma_y^{(2)}\rangle,\\
\label{corrz}C_{zz}&=&\langle\sigma_z^{(1)}\sigma_{z}^{(2)}\rangle
  -\langle\sigma_z^{(1)}\rangle\langle\sigma_z^{(2)}\rangle.
\end{eqnarray}
where $\langle \sigma_\alpha^{(1)}\sigma_\alpha^{(2)} \rangle \equiv
\Tr\left(\rho_S(t) \sigma_\alpha^{(1)}\sigma_\alpha^{(2)} \right)$,
$\alpha=x,y,z$. Since $\sigma_x^2=\sigma_y^2=\sigma_z^2=1$, so that
$\left(\sigma_\alpha^{(1)}+\sigma_\alpha^{(2)} \right)^2= 2 +2
\sigma_\alpha^{(1)}\sigma_\alpha^{(2)}$, the correlation function is
thus also a measurement of the fluctuations of the total spin of the
system.

\subsection{Concurrence}\label{sec:con}

The concurrence of the two spin-1/2 system is an indicator of their
intra entanglement, which is defined as \cite{Wootters1}:
\begin{equation}\label{Concurrence}
C=\max\{\lambda_1-\lambda_2-\lambda_3-\lambda_4,~0\},
\end{equation}
where $\lambda_i$ are the square roots of the eigenvalues of the
product matrix $\rho_S\tilde{\rho}_S$ in decreasing order. Equation
(\ref{Concurrence}) applies to all kinds of states, either mixed or
pure. The matrix $\tilde{\rho}_S$ is constructed as
$(\sigma_y\otimes\sigma_y)\rho_S^*(\sigma_y\otimes\sigma_y)$. If the
bipartite quantum state $\rho_S$ is pure \cite{Shi}, such as the
states in equations (\ref{Bell1})--(\ref{Bell4}). They can be
written as:
\begin{eqnarray*}
\rho_S &=& |\psi\rangle\langle\psi|,\\
|\psi\rangle &=& a|00\rangle+b|01\rangle+c|10\rangle+d|11\rangle,
\end{eqnarray*}
then equation (\ref{Concurrence}) could be simplified to
\begin{equation}\label{Concurrence2}
C(|\psi\rangle)=2|ad-bc|.
\end{equation}

\section{Results and discussions}\label{discussion}

In this section we give the calculated results of polarization
correlations and concurrence with discussions. In all the
calculations given here, the parameters
$\lambda_{bb}=\lambda_{ss}=1.0$ unless otherwise specified. And the
temperature is set as $T=0.1$ much higher than $T=0.02$ in Ref.
\cite{TWmodel} in order to stress the generality of the algorithm
and conclusion. The results for isolated systems $\lambda_{sb}=0.0$
are presented as a standard for comparison, in which there is no
decoherence occurs. On the other hand, the case of
$\lambda_{bb}=0.0$ will also be considered, in which there is a
strong decoherence occurs and the subsystem initial state is beyond
retrieval.

\FloatBarrier
\subsection{Polarization correlation}\label{dis:deco}

\begin{figure}[htbp]
  \centering
  \includegraphics[scale=0.6]{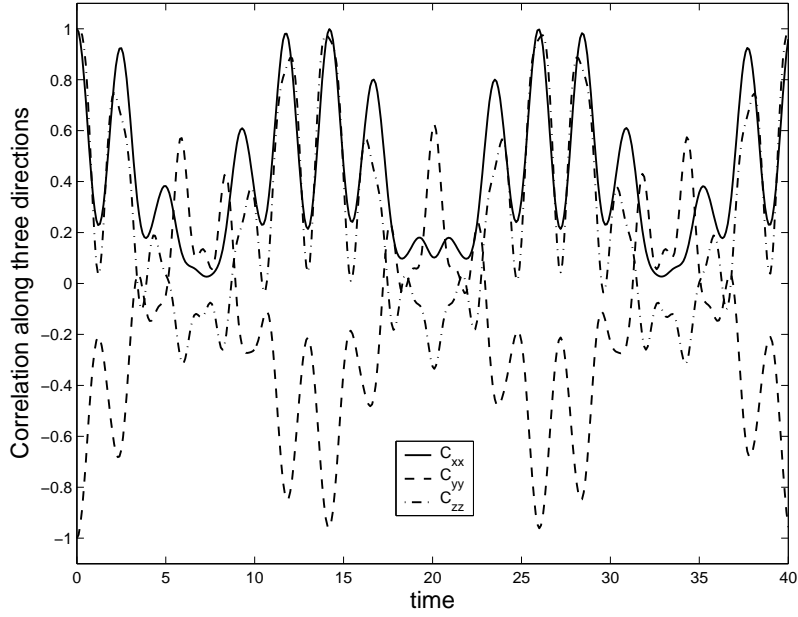}\\
  \caption{Evolution for polarization correlation of the
isolated subsystem,
$C_{\alpha\alpha}=\langle\sigma_{\alpha}^{(1)}\sigma_{\alpha}^{(2)}\rangle
-\langle\sigma_{\alpha}^{(1)}\rangle\langle\sigma_{\alpha}^{(2)}\rangle$,
$\alpha$ is $x$, $y$ or $z$, between the two spins in the subsystem.
The initial state of the  subsystem is
$1/\sqrt{2}(|00\rangle+|11\rangle)$.} \label{fig:Corr1}
\end{figure}

\begin{figure}[htbp]
  \centering
  \includegraphics[scale=0.6]{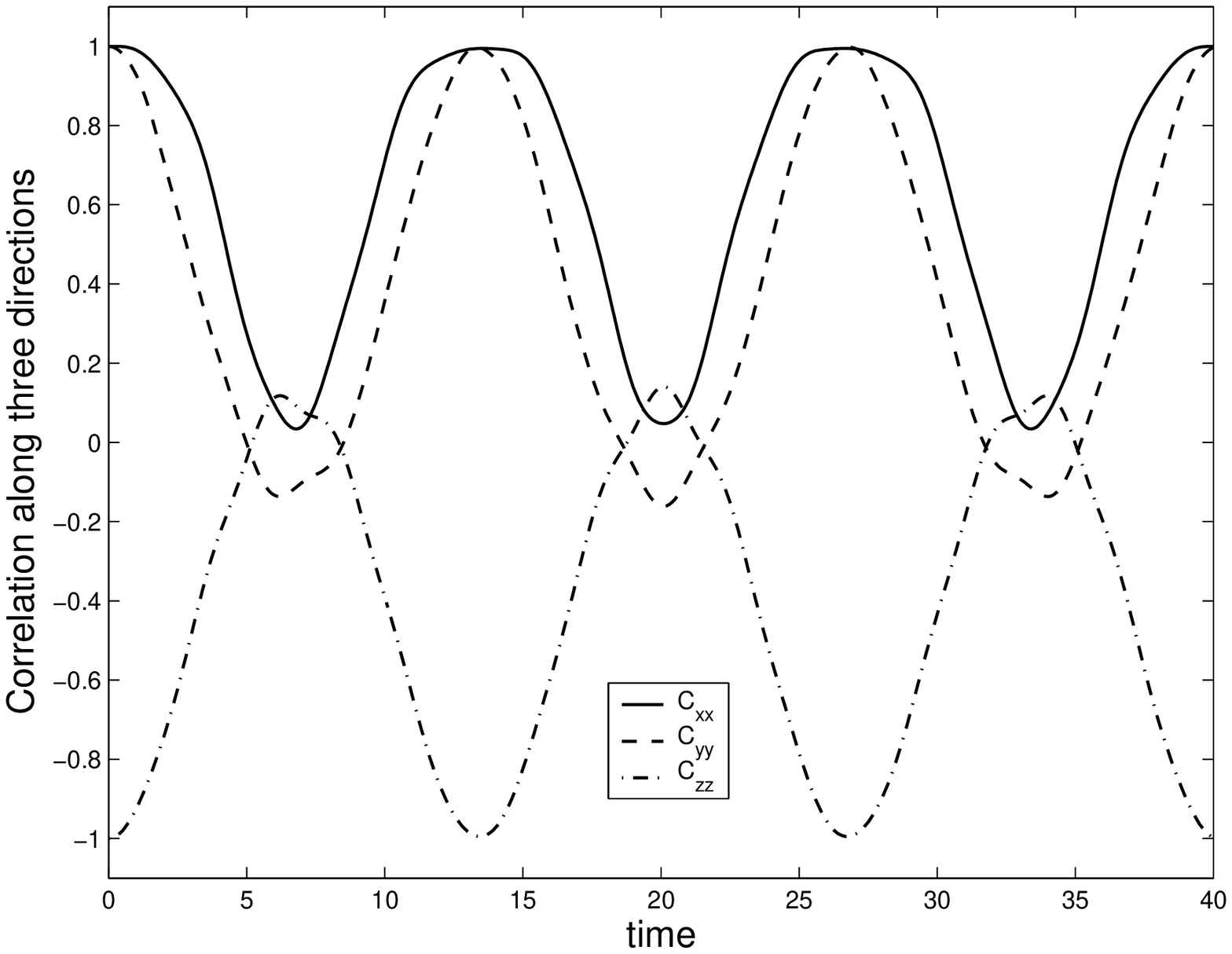}\\
  \caption{Evolution for polarization correlation of the
isolated subsystem,
$C_{\alpha\alpha}=\langle\sigma_{\alpha}^{(1)}\sigma_{\alpha}^{(2)}\rangle
-\langle\sigma_{\alpha}^{(1)}\rangle\langle\sigma_{\alpha}^{(2)}\rangle$,
$\alpha$ is $x$, $y$ or $z$, between the two spins in the subsystem.
The initial state of the  subsystem is
$1/\sqrt{2}(|01\rangle+|10\rangle)$.} \label{fig:Corr2}
\end{figure}

\begin{figure}[htbp]
  \centering
  \includegraphics[scale=0.6]{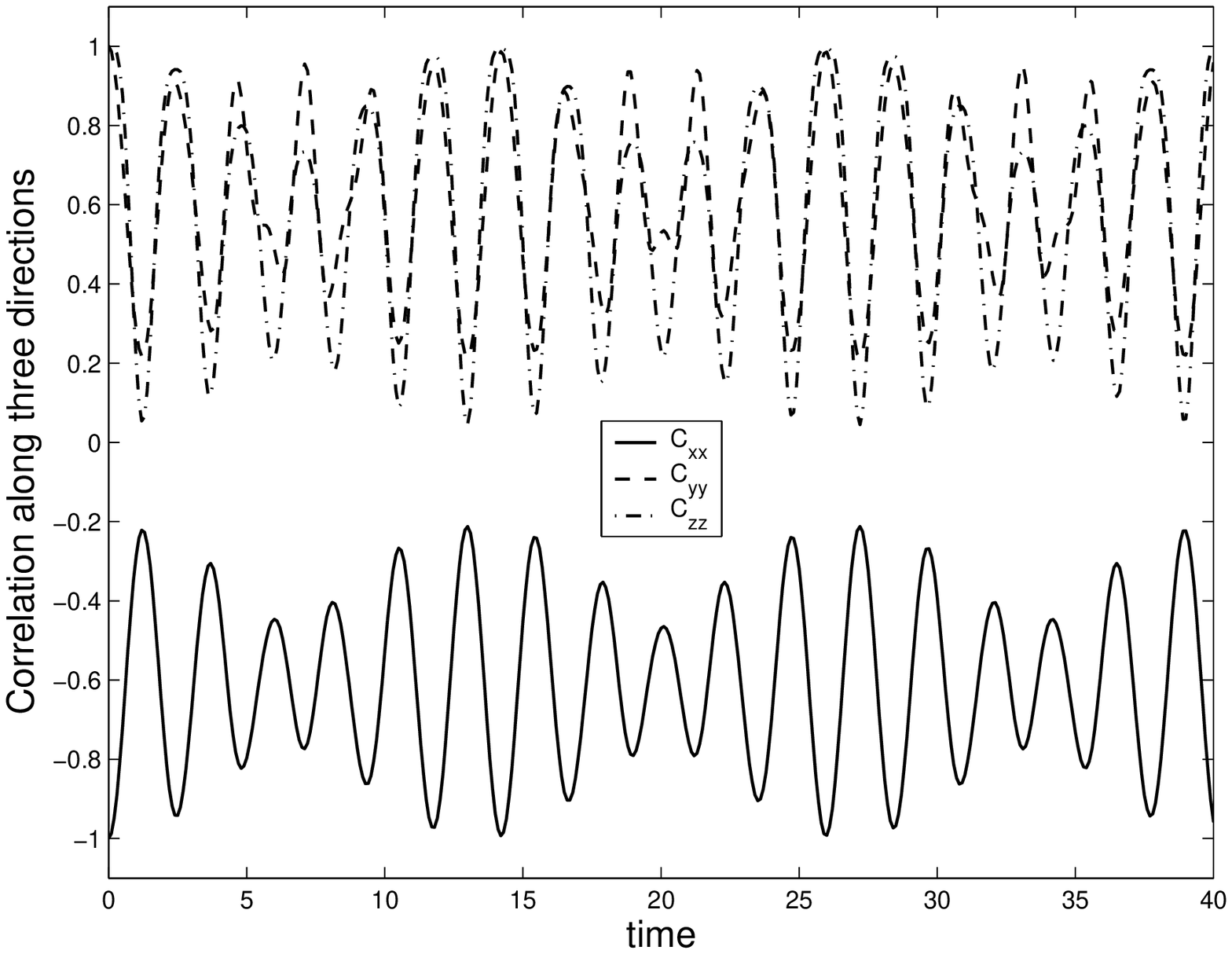}\\
  \caption{Evolution for polarization correlation of the
isolated subsystem,
$C_{\alpha\alpha}=\langle\sigma_{\alpha}^{(1)}\sigma_{\alpha}^{(2)}\rangle
-\langle\sigma_{\alpha}^{(1)}\rangle\langle\sigma_{\alpha}^{(2)}\rangle$,
$\alpha$ is $x$, $y$ or $z$, between the two spins in the subsystem.
The initial state of the  subsystem is
$1/\sqrt{2}(|00\rangle-|11\rangle)$.} \label{fig:Corr3}
\end{figure}

\begin{figure}[htbp]
\centering \subfigure[$\langle\sigma_x^{1}\sigma_x^{2}\rangle
-\langle\sigma_x^{1}\rangle\langle\sigma_x^{2}\rangle$,
$\lambda_{bb}=0.0$] {\label{fig2:N60:xx}
\includegraphics[width=3in]{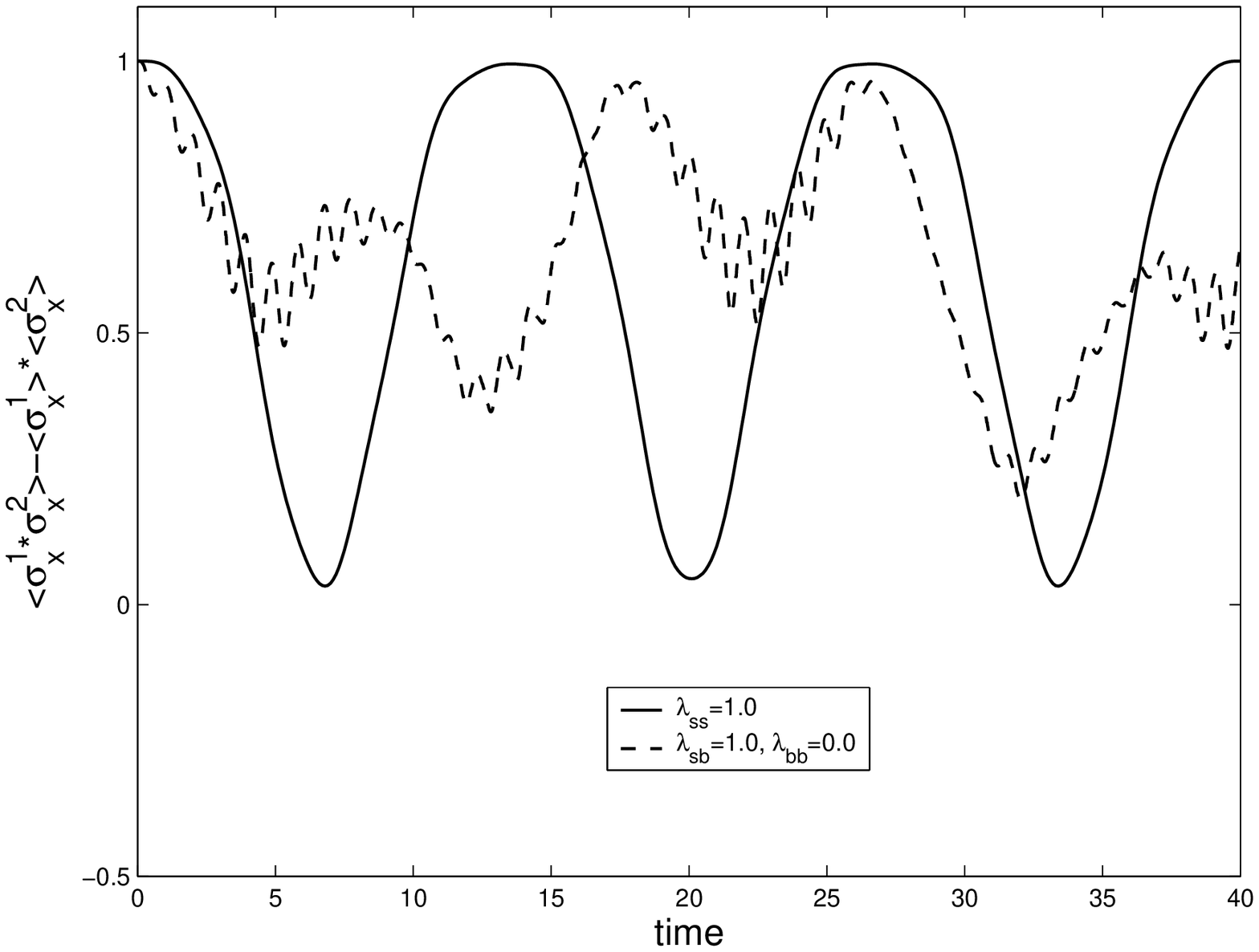}}
\subfigure[$\langle\sigma_y^{1}\sigma_y^{2}\rangle
-\langle\sigma_y^{1}\rangle\langle\sigma_y^{2}\rangle$,
$\lambda_{bb}=0.0$] { \label{fig2:N60:yy}
\includegraphics[width=3in]{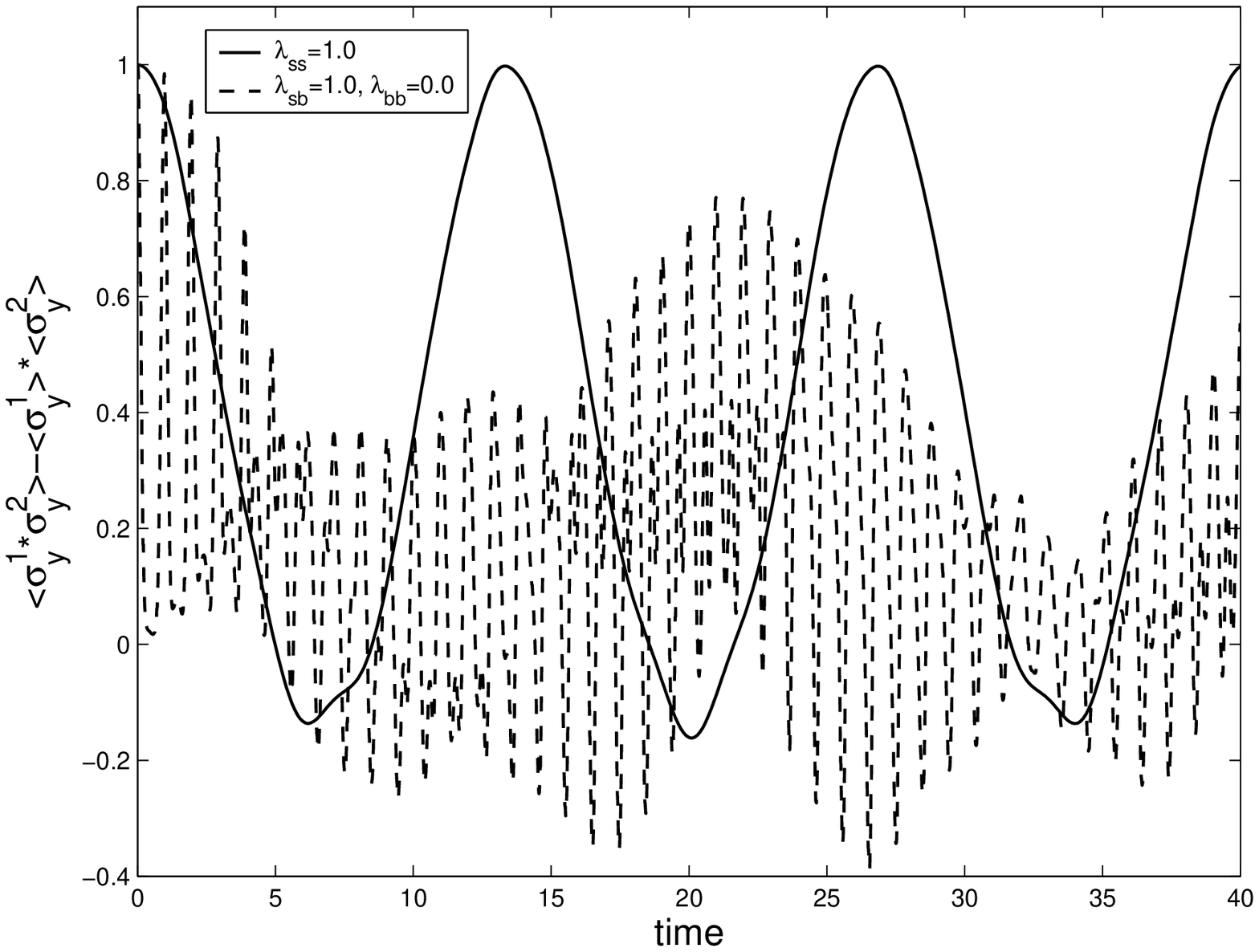}}
\\
\subfigure[$\langle\sigma_z^{1}\sigma_z^{2}\rangle
-\langle\sigma_z^{1}\rangle\langle\sigma_z^{2}\rangle$,
$\lambda_{bb}=0.0$] { \label{fig2:N60:zz}
\includegraphics[width=3in]{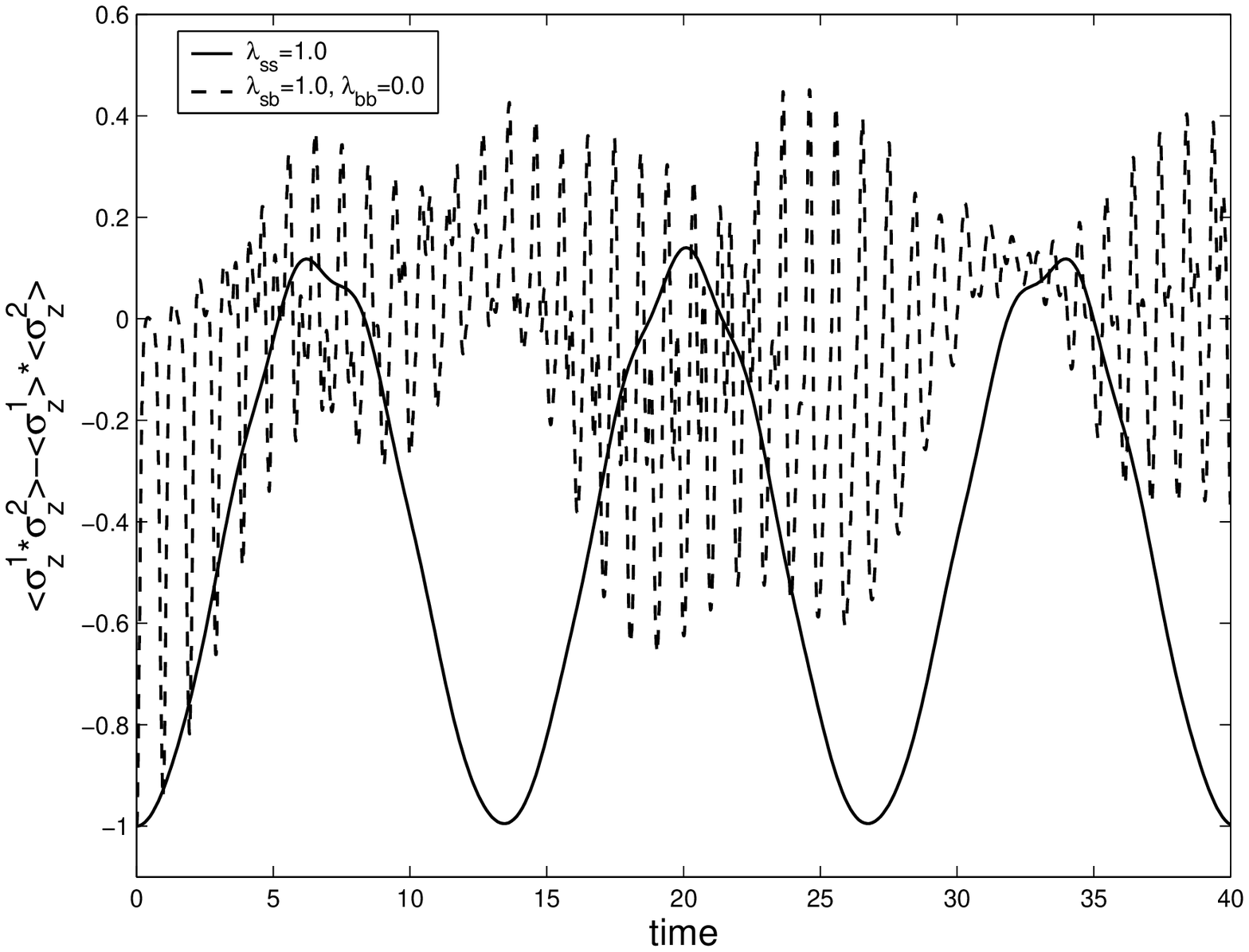}}
\caption{The evolution of polarization correlation along three
directions of the open subsystem. Where there are $6$ spins in the
bath and the initial state of the subsystem is
$1/\sqrt{2}(|10\rangle+|01\rangle)$ and
$\lambda_{ss}=\lambda_{sb}=1.0$, $\lambda_{bb}=0.0$.}
\label{fig2:N60}
\end{figure}

The polarization correlations of the isolated system are plotted in
figures \ref{fig:Corr1}, \ref{fig:Corr2} and \ref{fig:Corr3}, we see
that the evolutions of the polarization correlations are periodic in
time. Then we take into account the case that $\lambda_{bb}=0.0$,
which means there is no coupling among bath spins, in order to
highlight the effect of strong intra-bath coupling. In figure
\ref{fig2:N60}, we show the pure destruction effect on polarization
correlation of subsystem spins imposed by the bath spins ($m=6$) to
the open subsystem, where we neglect the intra-bath coupling
strength. The initial state of the subsystem is
$|\psi_S(0)\rangle^2=1/\sqrt{2}(|01\rangle+|10\rangle)$.\\

\begin{figure}[htbp]
\centering \subfigure[$\lambda_{bb}=2.0$]{ \label{fig:1xx2}
\includegraphics[width=3in]{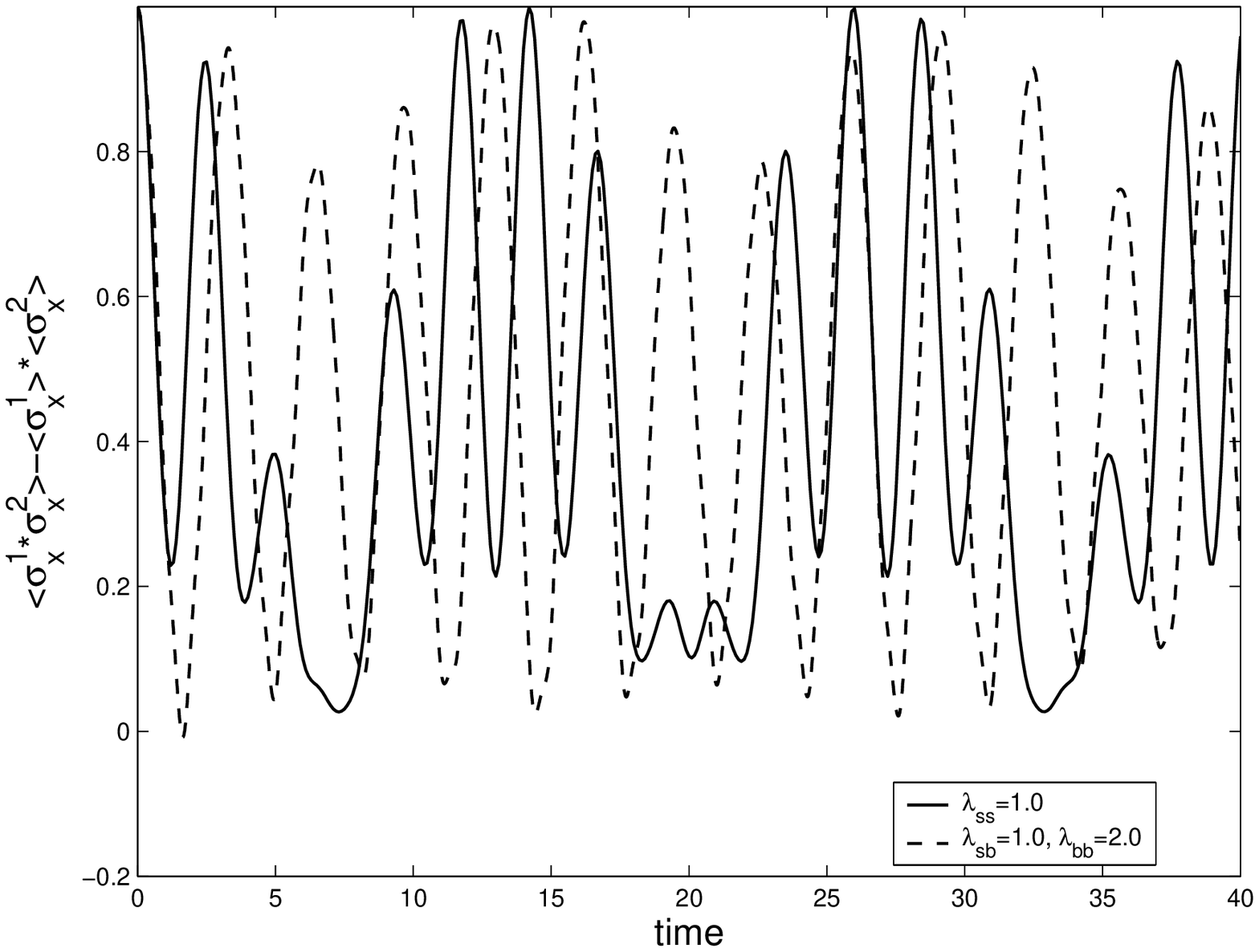}}
\subfigure[$\lambda_{bb}=4.0$]{ \label{fig:1xx4}
\includegraphics[width=3in]{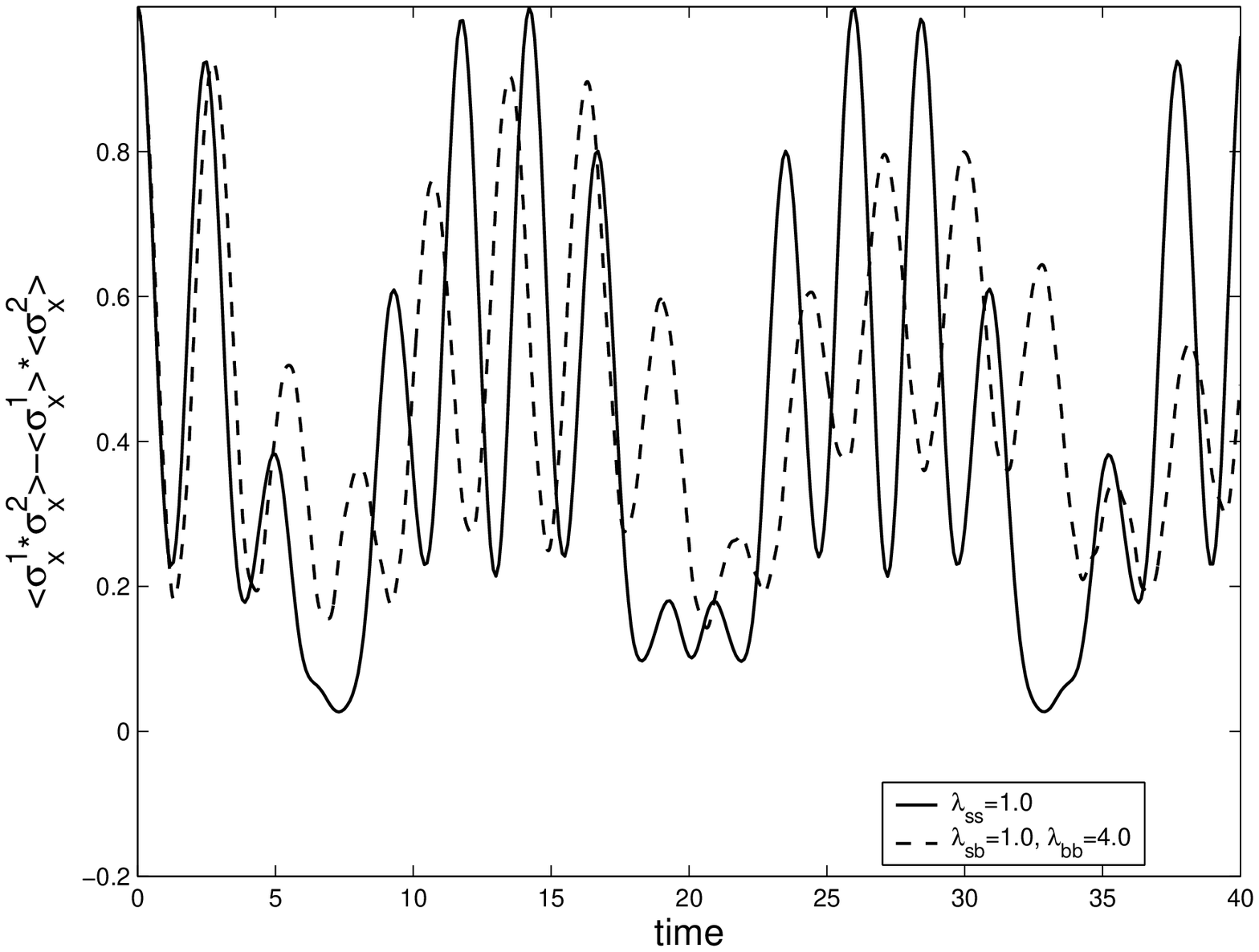}}
\\
\subfigure[$\lambda_{bb}=6.0$]{ \label{fig:1xx6}
\includegraphics[width=3in]{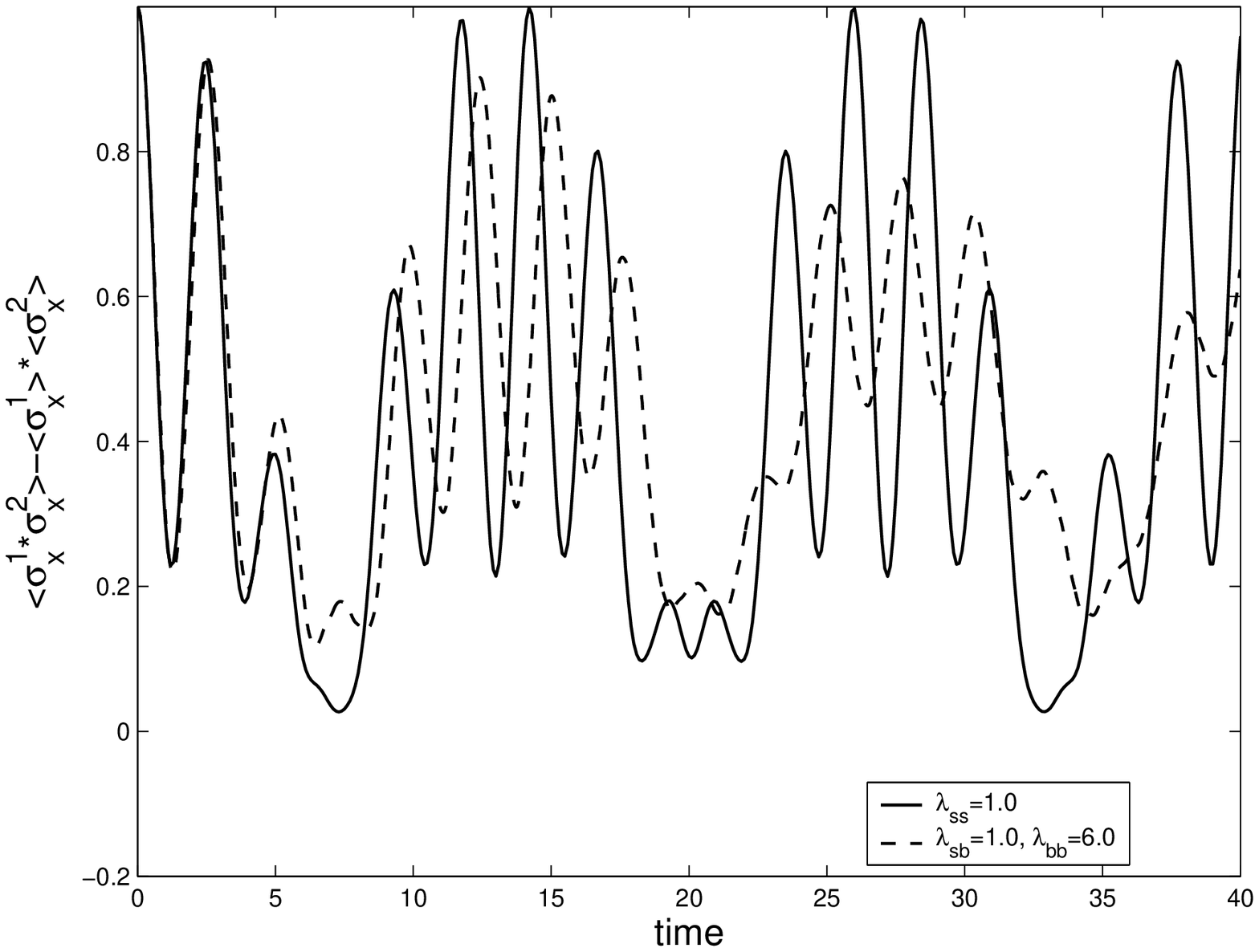}}
\subfigure[$\lambda_{bb}=8.0$]{ \label{fig:1xx8}
\includegraphics[width=3in]{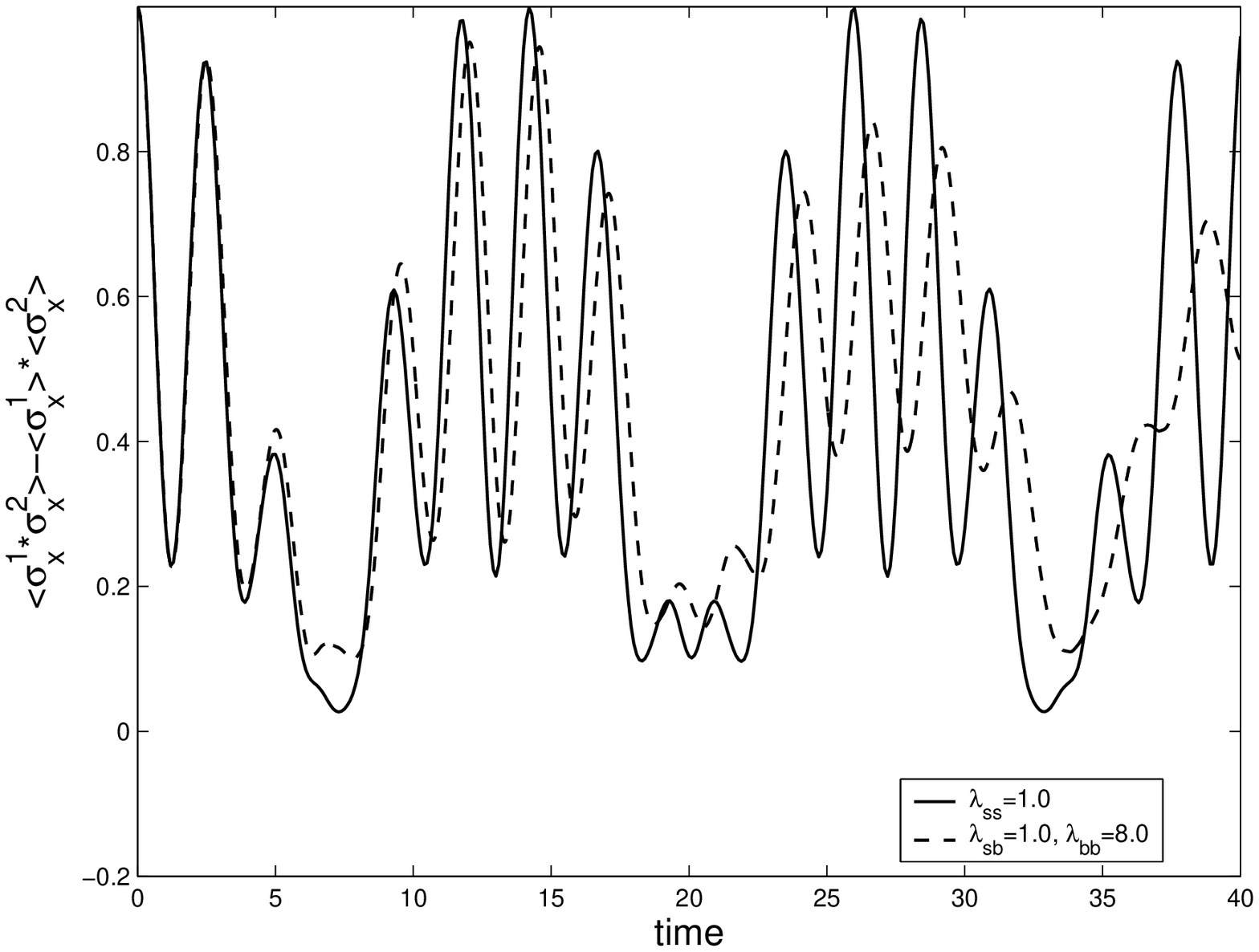}}
  \caption{Evolution for polarization correlation along
$\vec{x}$ direction of the open subsystem,
$\langle\sigma_x^{1}\sigma_x^{2}\rangle
-\langle\sigma_x^{1}\rangle\langle\sigma_x^{2}\rangle$. The initial
state of the  subsystem is $1/\sqrt{2}(|00\rangle+|11\rangle)$.}
\label{fig:1xx}
\end{figure}

\begin{figure}[htbp]
\centering \subfigure[$\lambda_{bb}=2.0$]{ \label{fig:1yy2}
\includegraphics[width=3in]{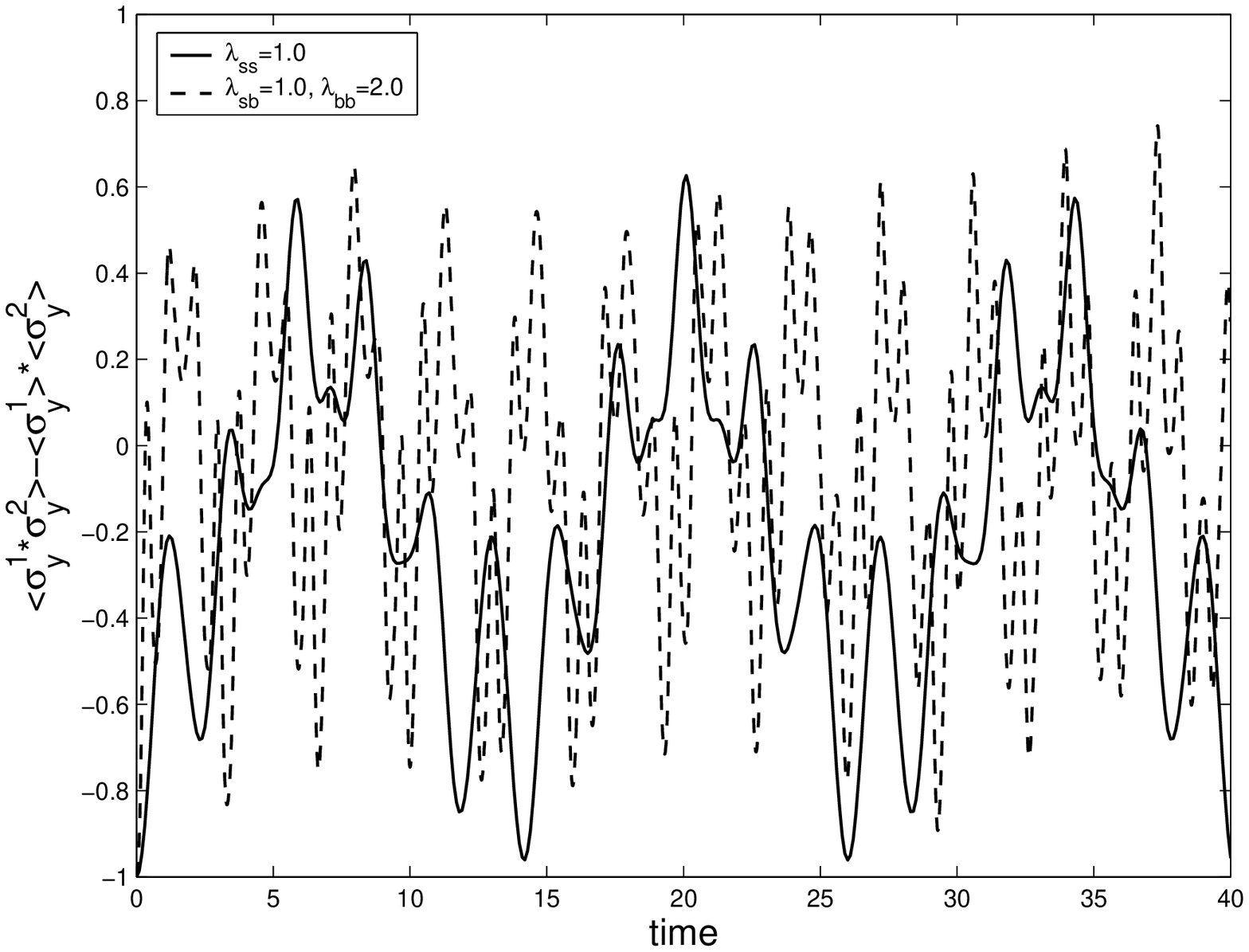}}
\subfigure[$\lambda_{bb}=4.0$]{ \label{fig:1yy4}
\includegraphics[width=3in]{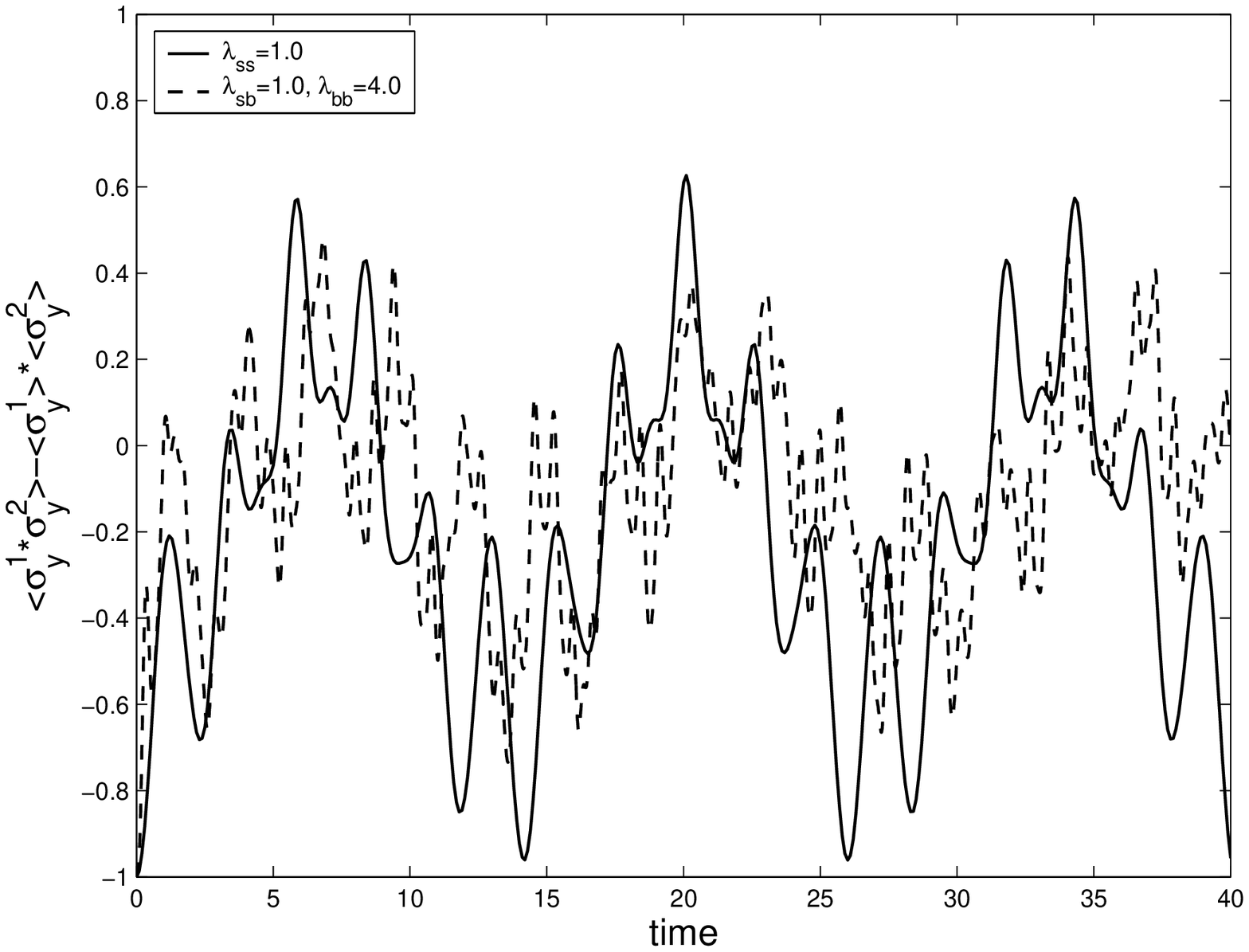}}
\\
\subfigure[$\lambda_{bb}=6.0$]{ \label{fig:1yy6}
\includegraphics[width=3in]{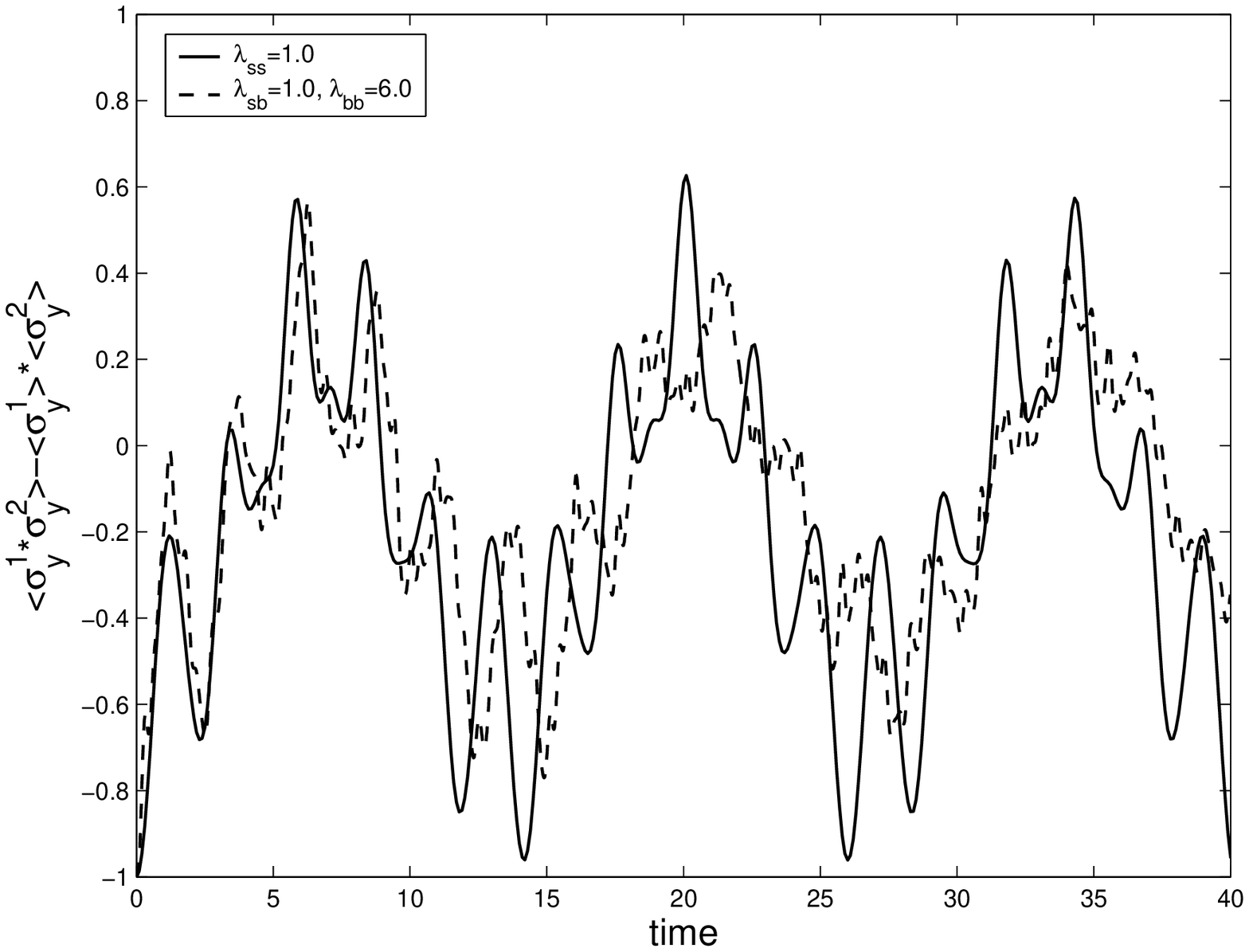}}
\subfigure[$\lambda_{bb}=8.0$]{ \label{fig:1yy8}
\includegraphics[width=3in]{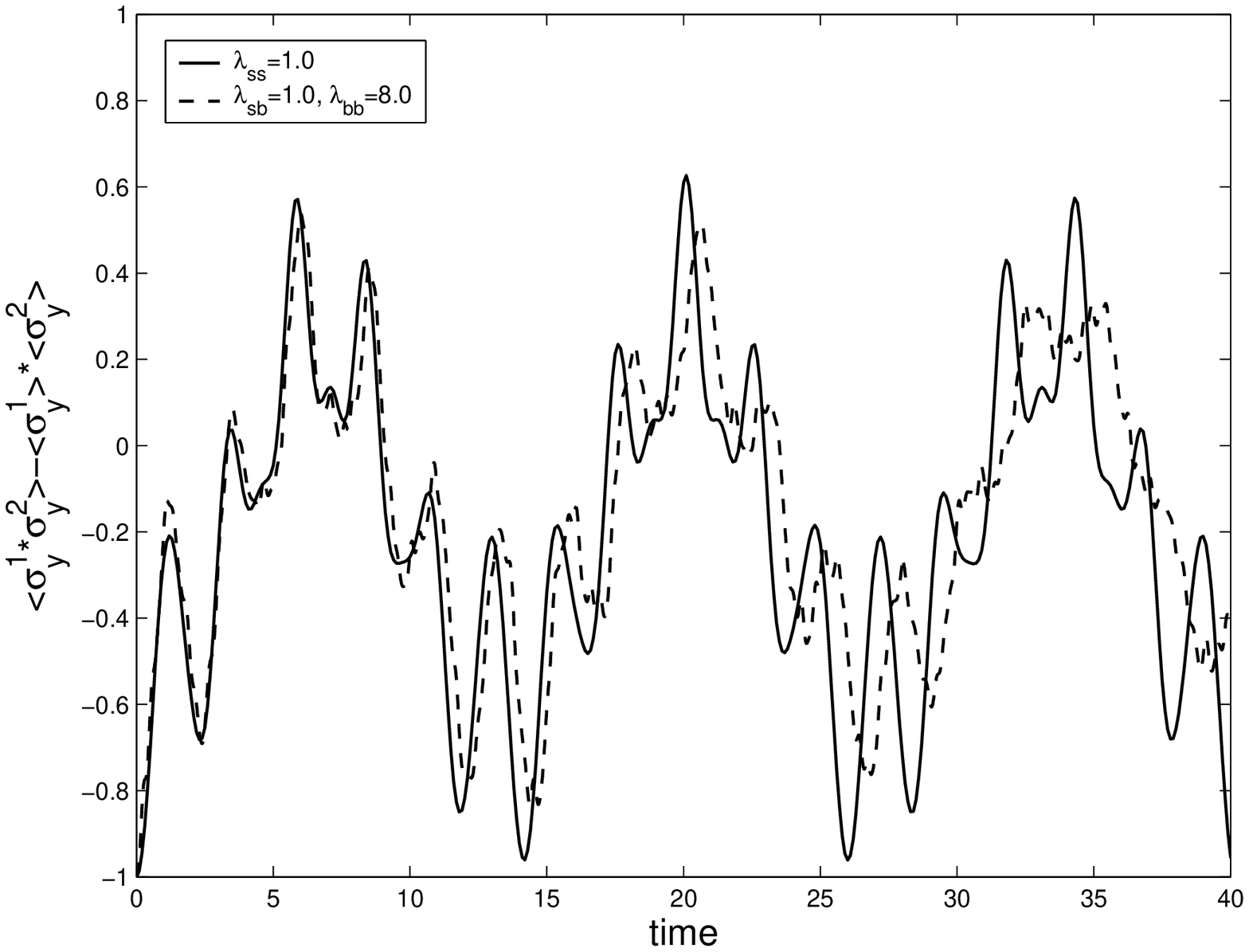}}
  \caption{Evolution for polarization correlation along
$\vec{y}$ direction of the open subsystem,
$\langle\sigma_y^{1}\sigma_y^{2}\rangle
-\langle\sigma_y^{1}\rangle\langle\sigma_y^{2}\rangle$. The initial
state of the subsystem is $1/\sqrt{2}(|00\rangle+|11\rangle)$.}
\label{fig:1yy}
\end{figure}

\begin{figure}[htbp]
\centering \subfigure[$\lambda_{bb}=2.0$]{ \label{fig:1zz2}
\includegraphics[width=3in]{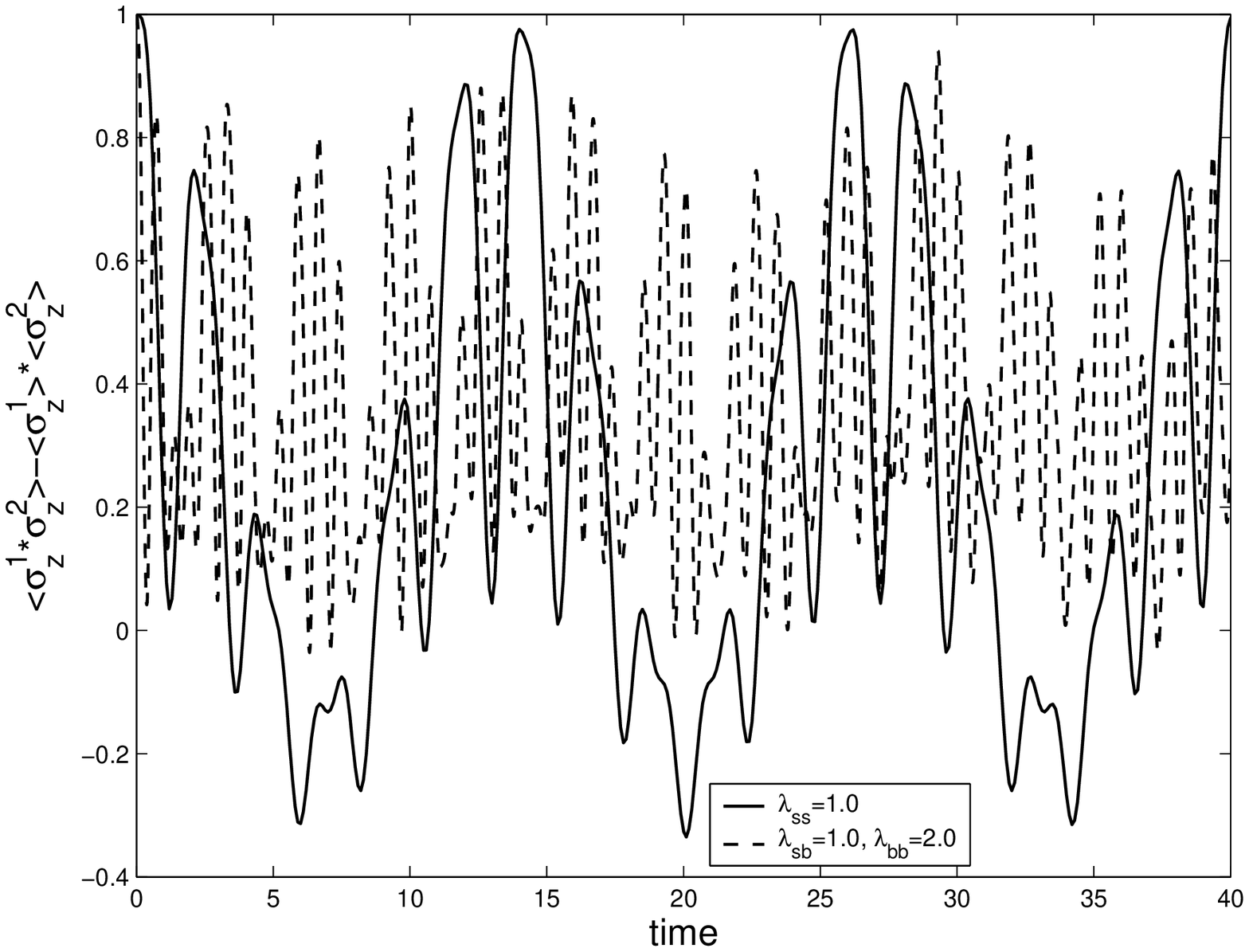}}
\subfigure[$\lambda_{bb}=4.0$]{ \label{fig:1zz4}
\includegraphics[width=3in]{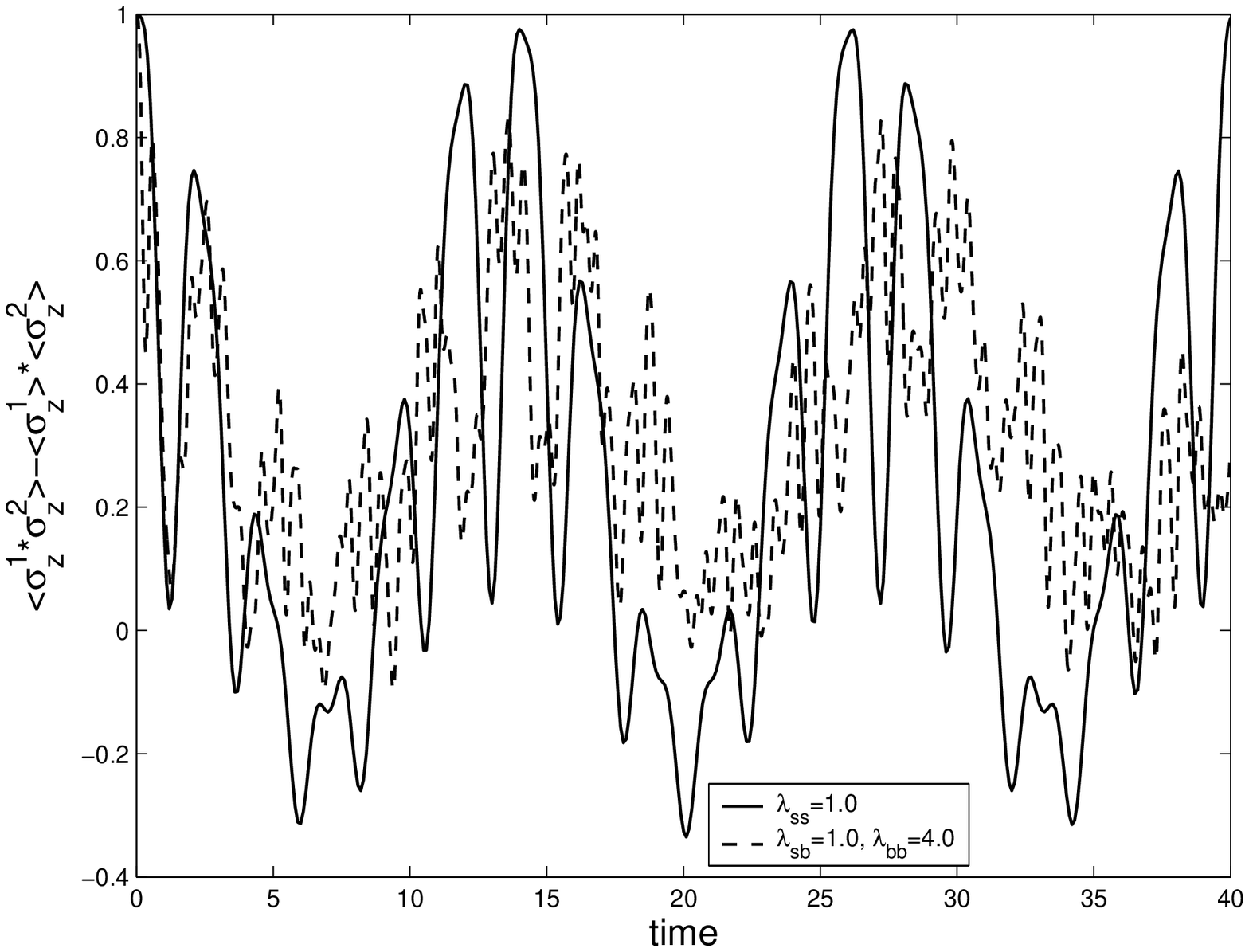}}
\\
\subfigure[$\lambda_{bb}=6.0$]{ \label{fig:1zz6}
\includegraphics[width=3in]{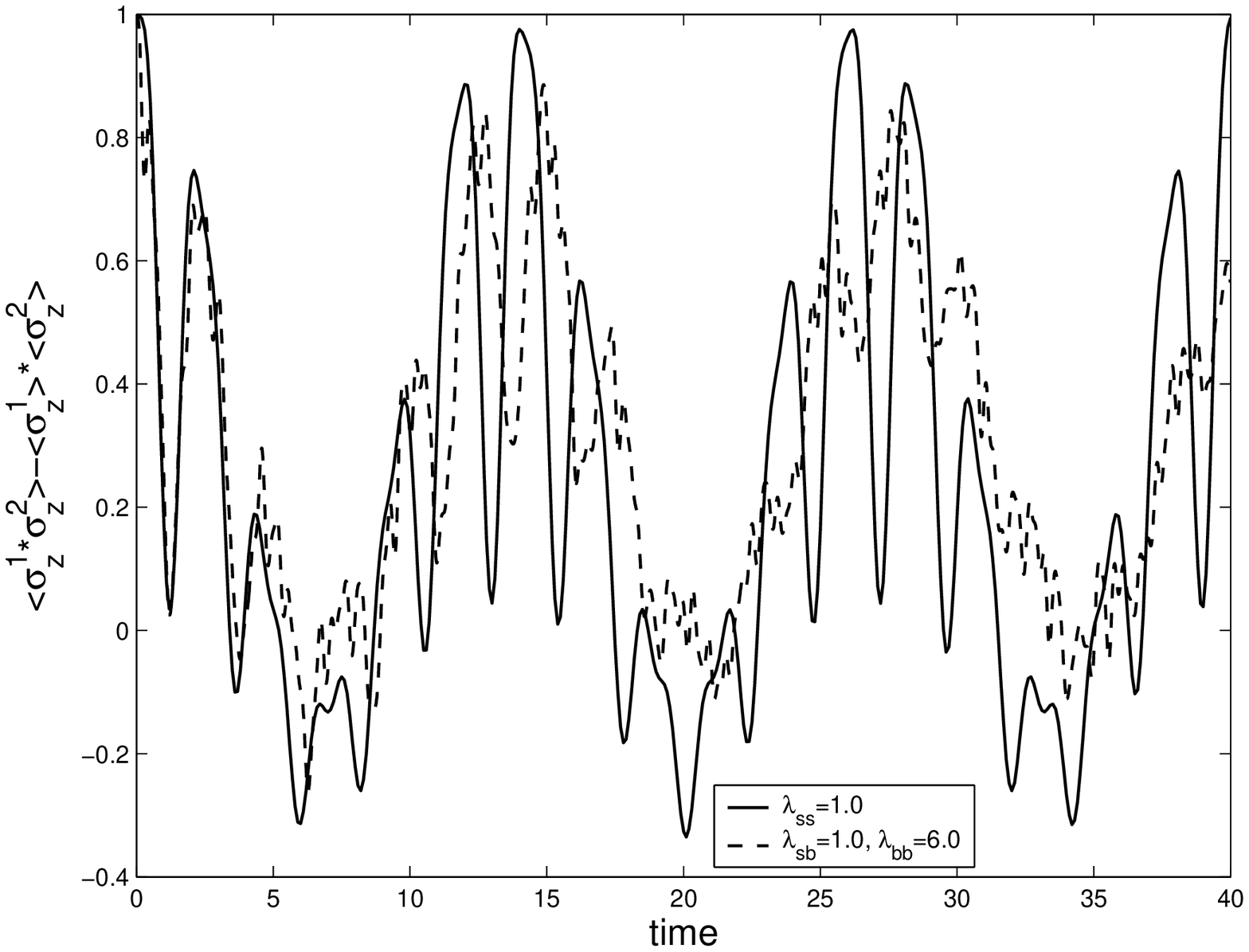}}
\subfigure[$\lambda_{bb}=8.0$]{ \label{fig:1zz8}
\includegraphics[width=3in]{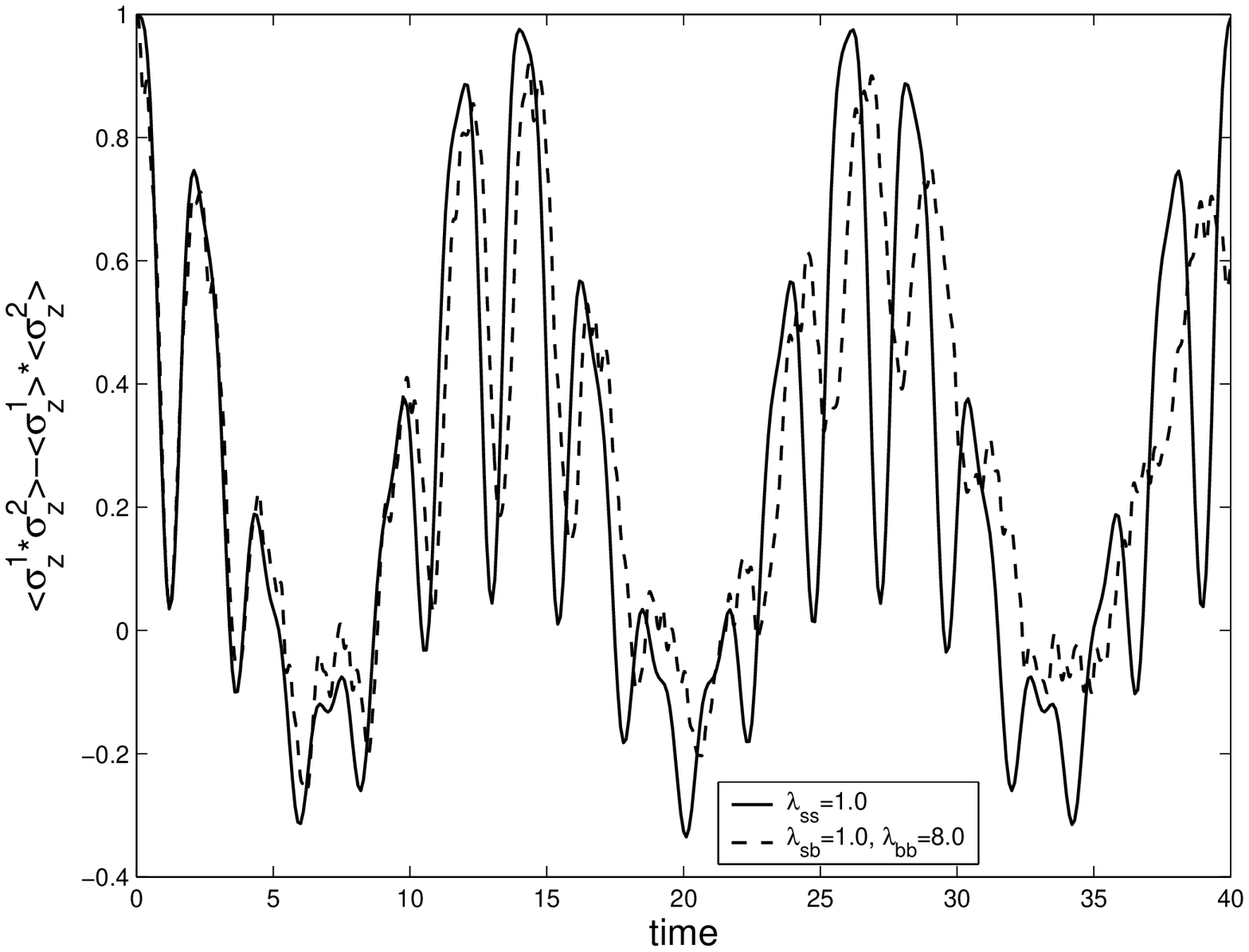}}
  \caption{Evolution for polarization correlation along
$\vec{z}$ direction of the open subsystem,
$\langle\sigma_z^{1}\sigma_z^{2}\rangle
-\langle\sigma_z^{1}\rangle\langle\sigma_z^{2}\rangle$. The initial
state of the  subsystem  is $1/\sqrt{2}(|00\rangle+|11\rangle)$.}
\label{fig:1zz}
\end{figure}

\begin{figure}[htbp]
\centering \subfigure[$\lambda_{bb}=2.0$]{ \label{fig:2xx2}
\includegraphics[width=3in]{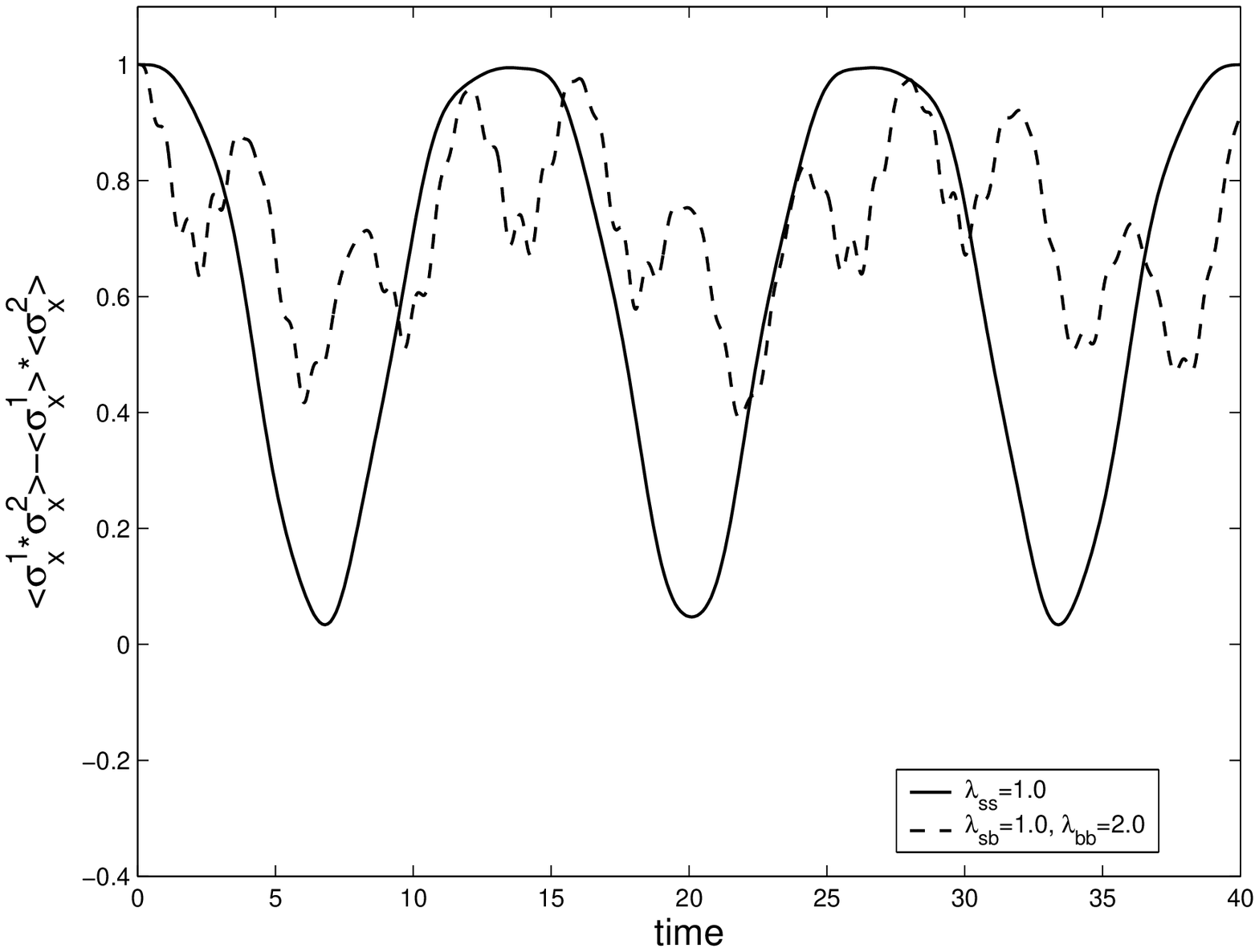}}
\subfigure[$\lambda_{bb}=4.0$]{ \label{fig:2xx4}
\includegraphics[width=3in]{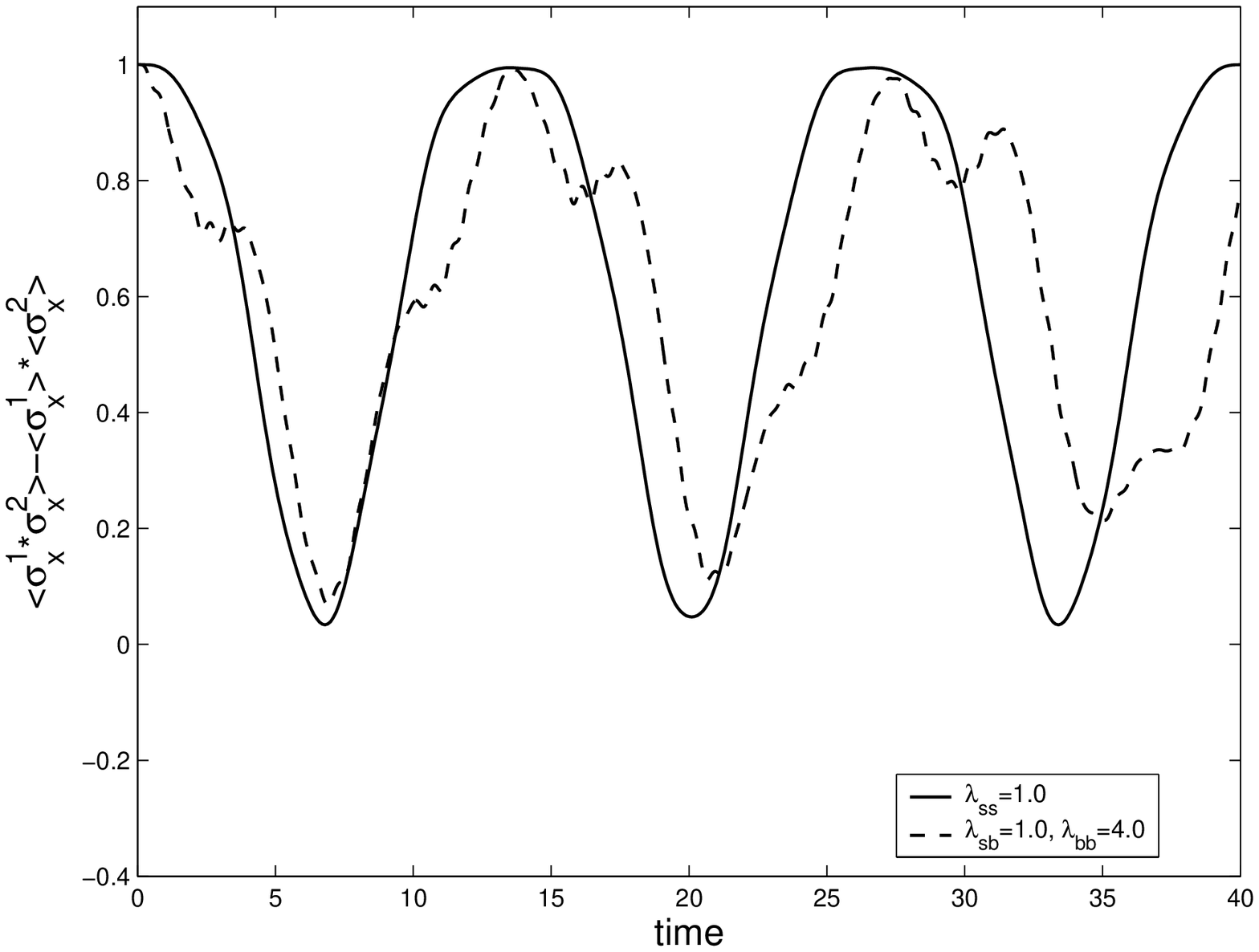}}
\\
\subfigure[$\lambda_{bb}=6.0$]{ \label{fig:2xx6}
\includegraphics[width=3in]{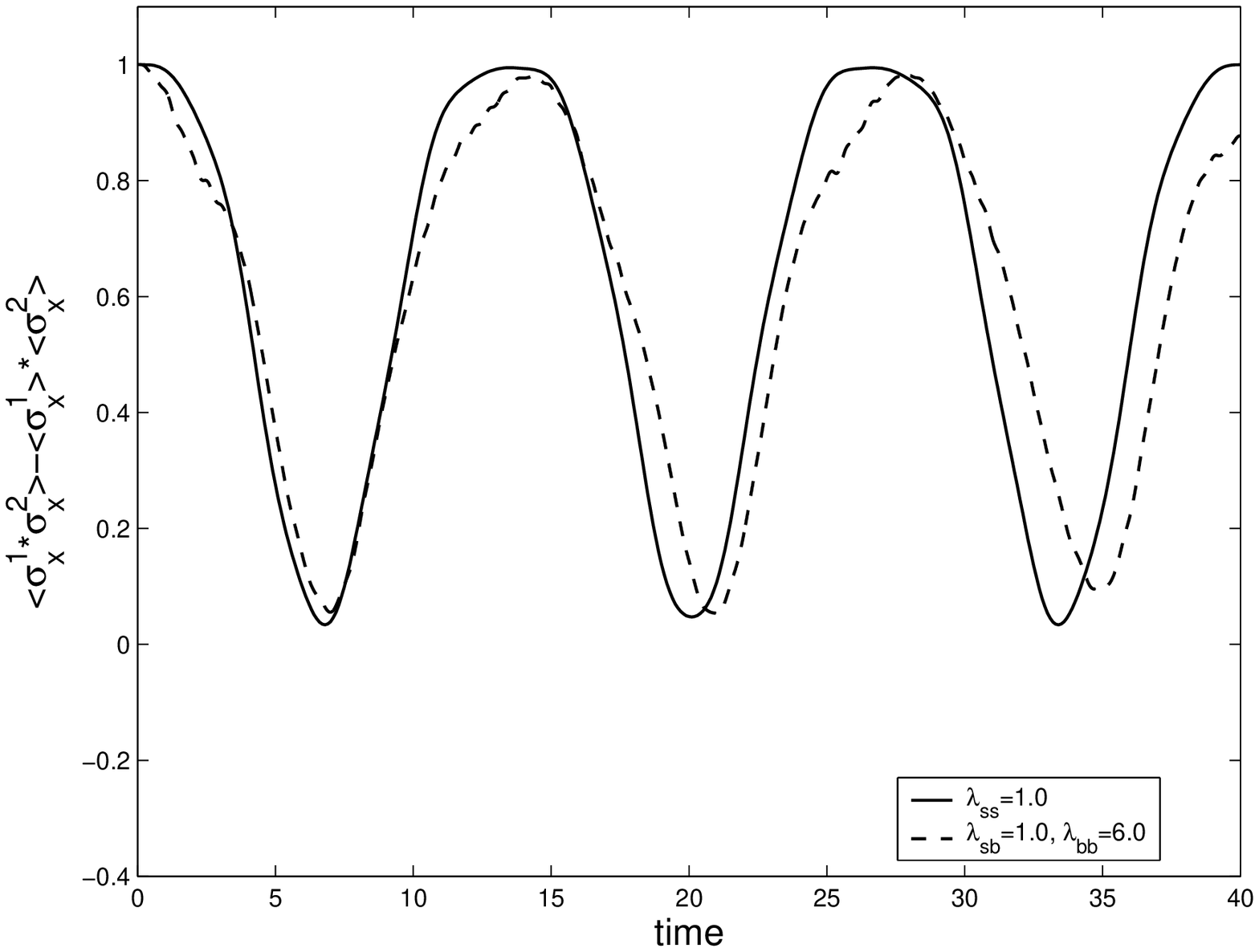}}
\subfigure[$\lambda_{bb}=8.0$]{ \label{fig:2xx8}
\includegraphics[width=3in]{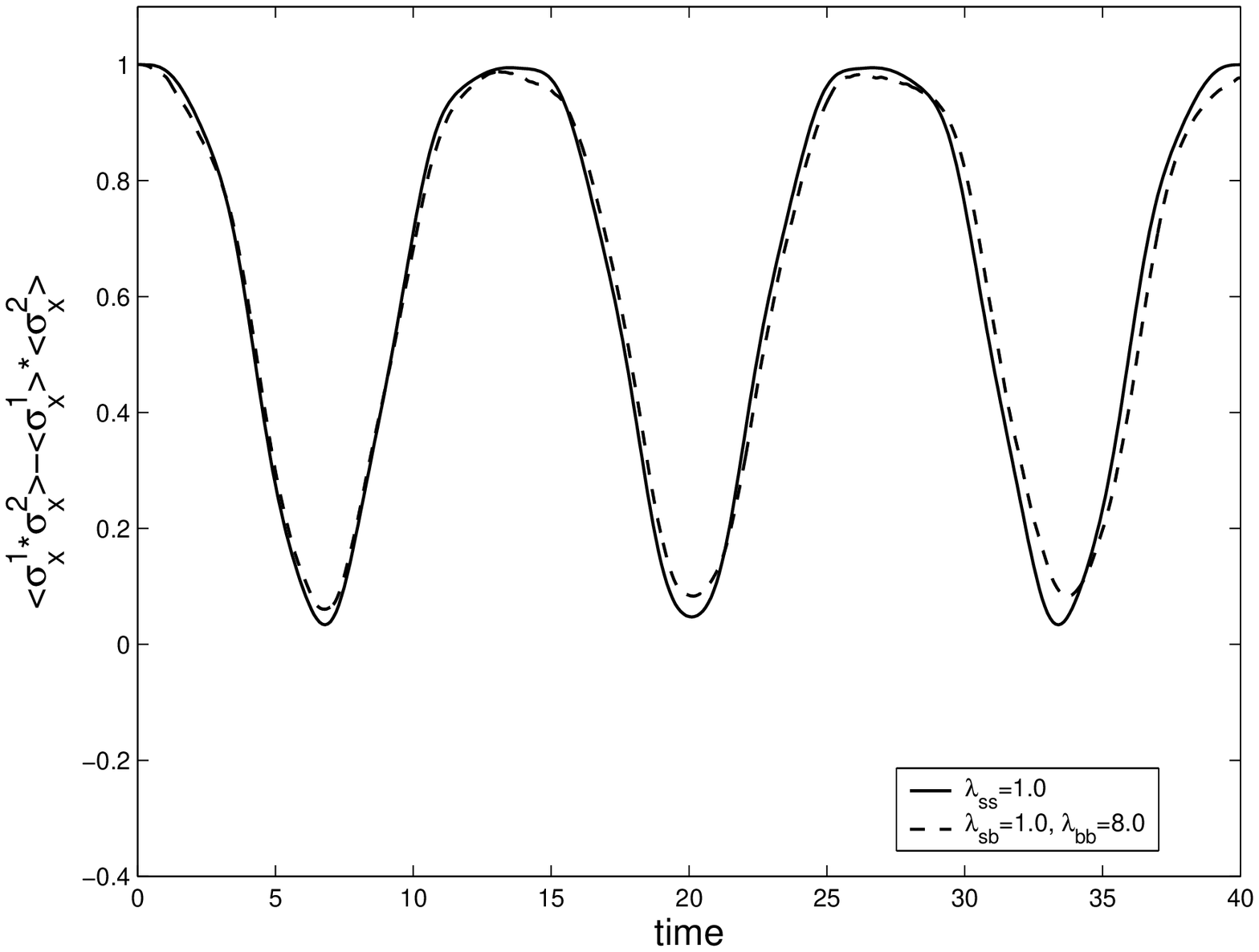}}
  \caption{Evolution for polarization correlation along
$\vec{x}$ direction of the open subsystem,
$\langle\sigma_x^{1}\sigma_x^{2}\rangle
-\langle\sigma_x^{1}\rangle\langle\sigma_x^{2}\rangle$. The initial
state of the subsystem is $1/\sqrt{2}(|01\rangle+|10\rangle)$.}
\label{fig:2xx}
\end{figure}

\begin{figure}[htbp]
\centering \subfigure[$\lambda_{bb}=2.0$]{ \label{fig:2yy2}
\includegraphics[width=3in]{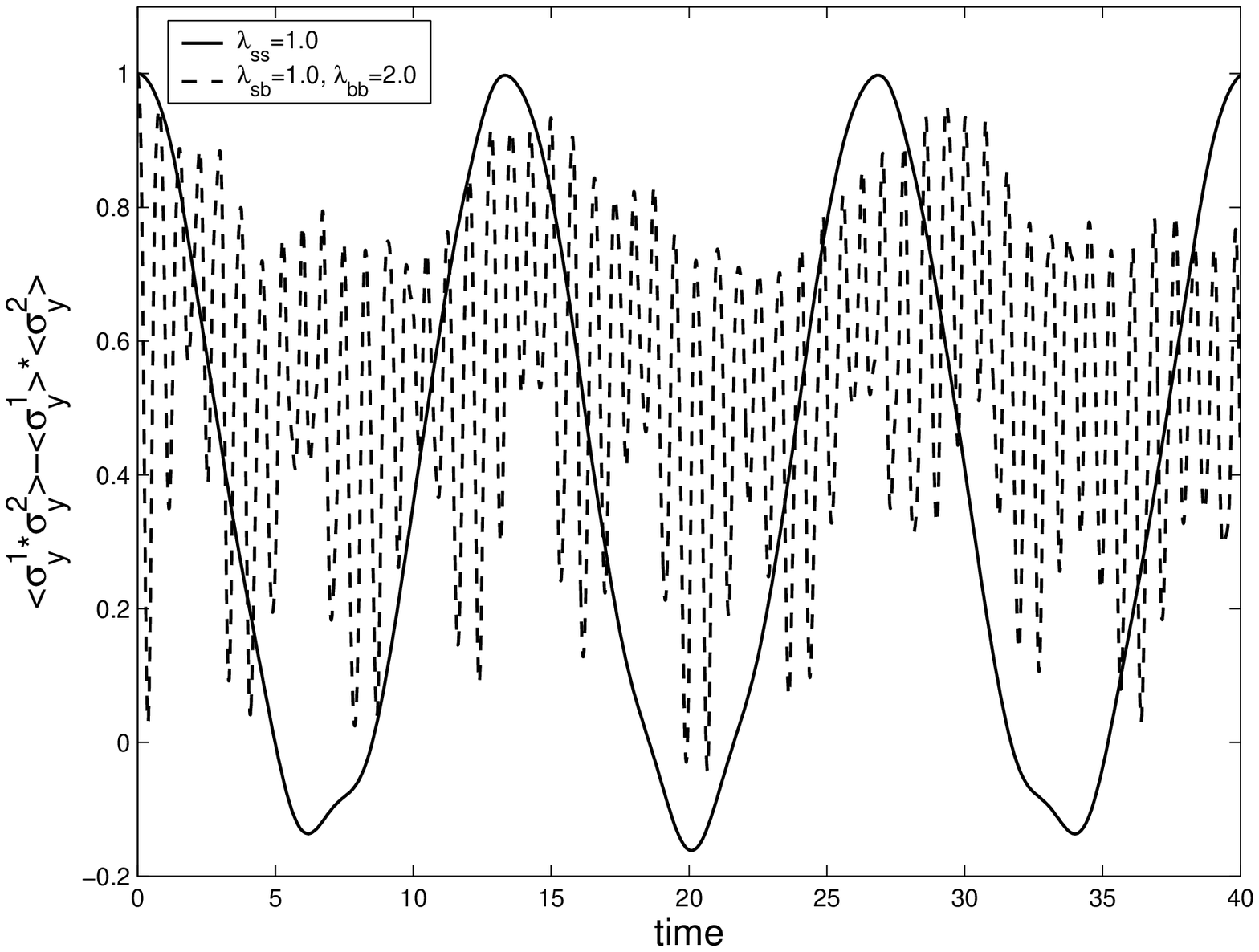}}
\subfigure[$\lambda_{bb}=4.0$]{ \label{fig:2yy4}
\includegraphics[width=3in]{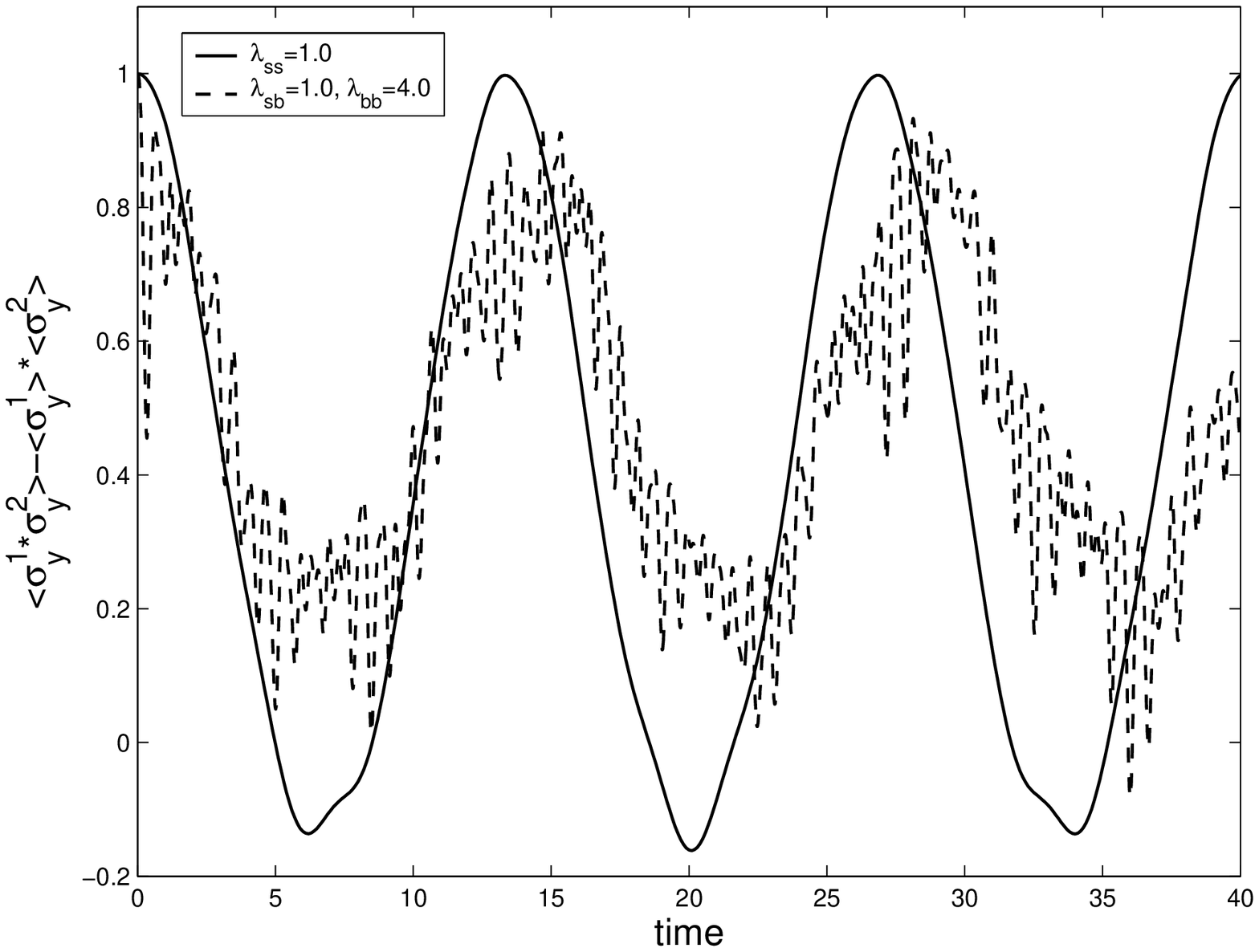}}
\\
\subfigure[$\lambda_{bb}=6.0$]{ \label{fig:2yy6}
\includegraphics[width=3in]{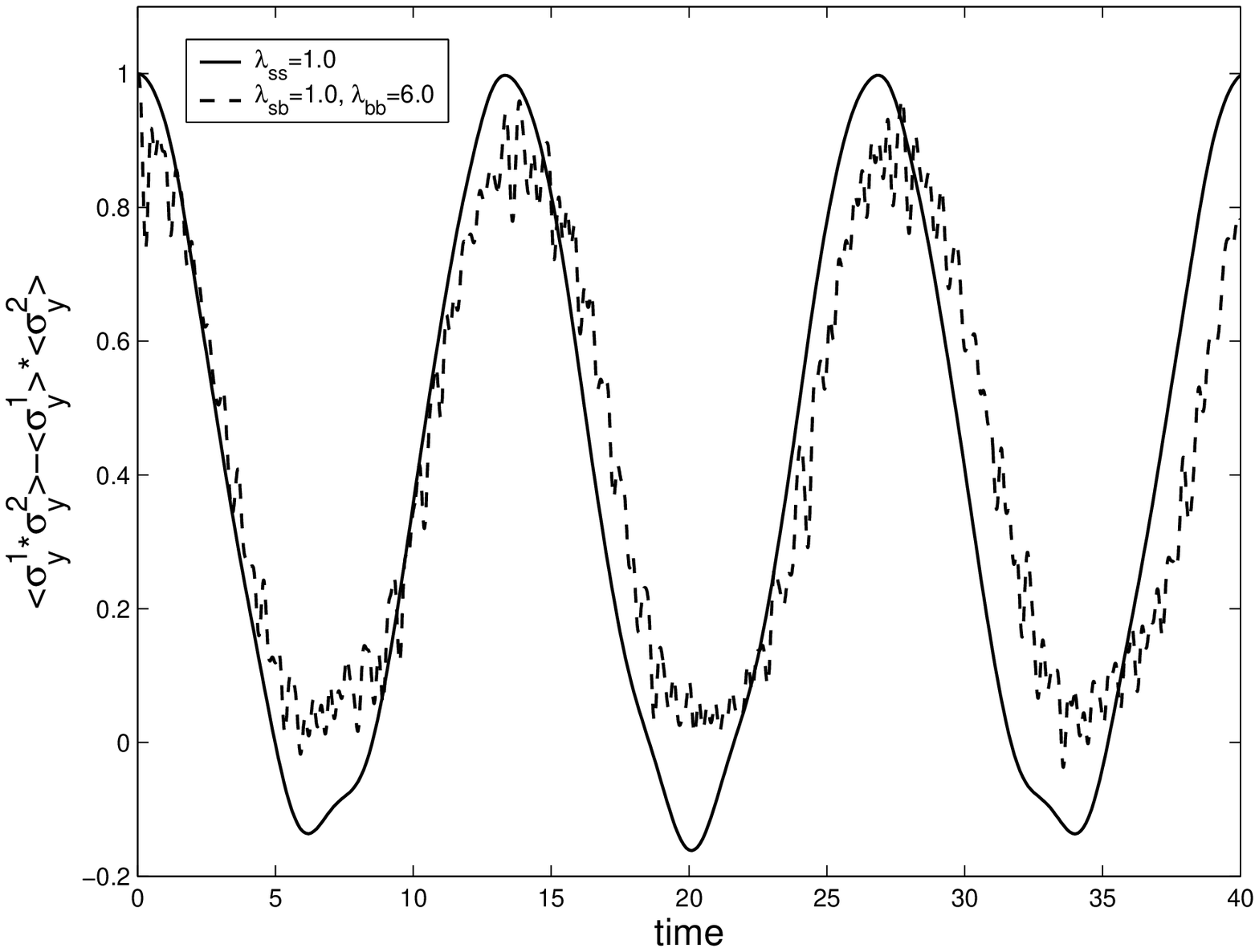}}
\subfigure[$\lambda_{bb}=8.0$]{ \label{fig:2yy8}
\includegraphics[width=3in]{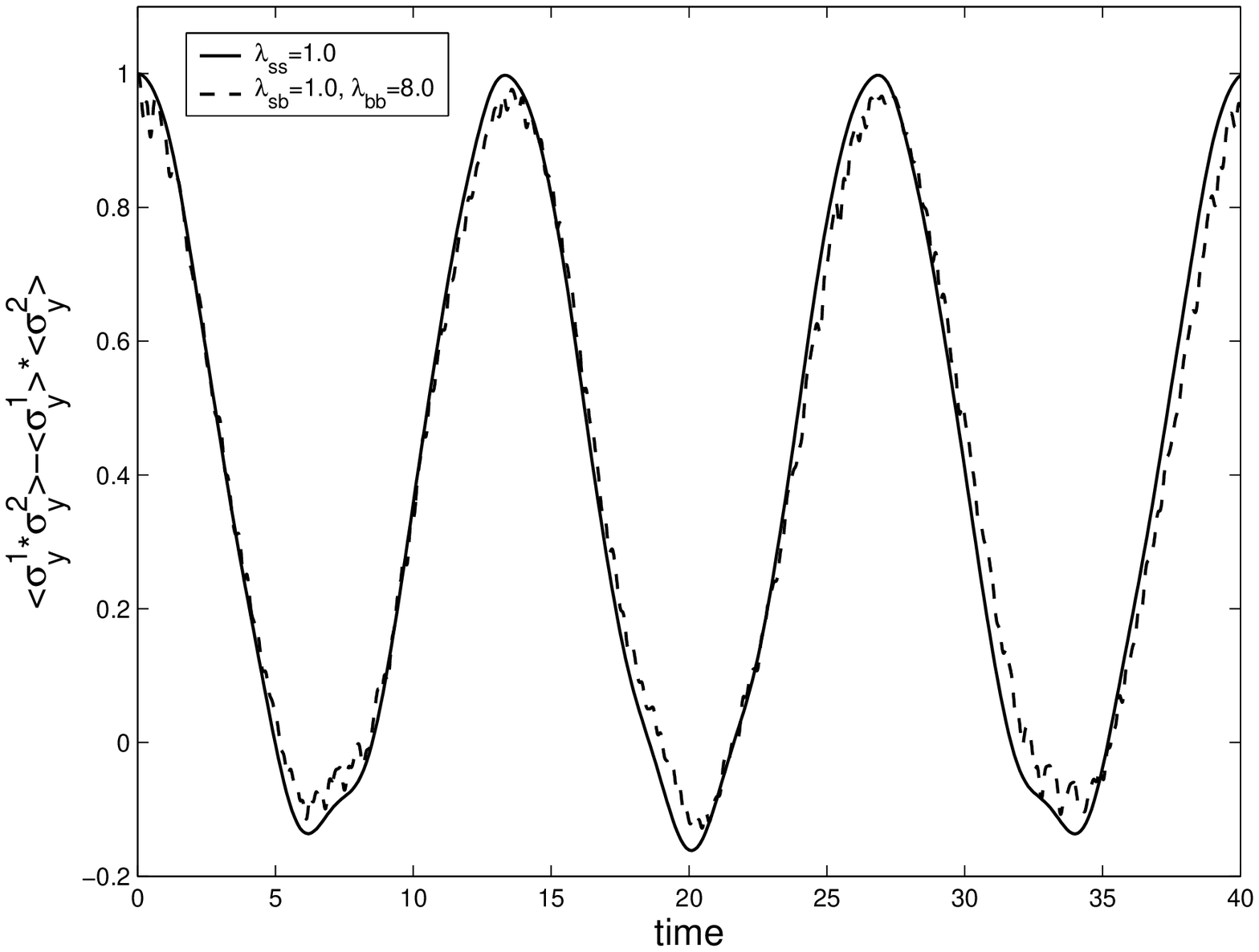}}
  \caption{Evolution for polarization correlation along
$\vec{y}$ direction of the open subsystem,
$\langle\sigma_y^{1}\sigma_y^{2}\rangle
-\langle\sigma_y^{1}\rangle\langle\sigma_y^{2}\rangle$. The  initial
state of the  subsystem  is $1/\sqrt{2}(|01\rangle+|10\rangle)$.}
\label{fig:2yy}
\end{figure}

\begin{figure}[htbp]
\centering \subfigure[$\lambda_{bb}=2.0$]{ \label{fig:2zz2}
\includegraphics[width=3in]{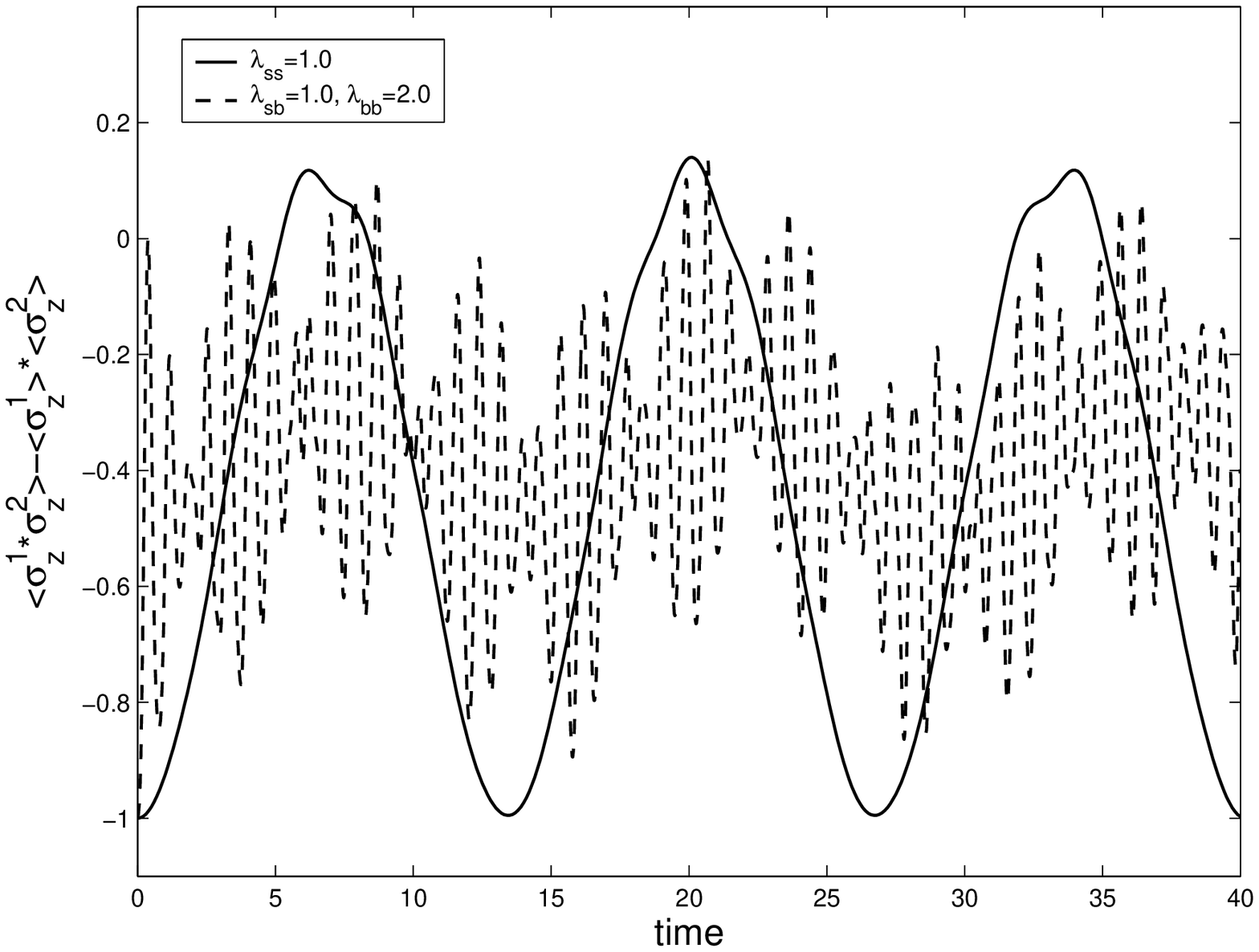}}
\subfigure[$\lambda_{bb}=4.0$]{ \label{fig:2zz4}
\includegraphics[width=3in]{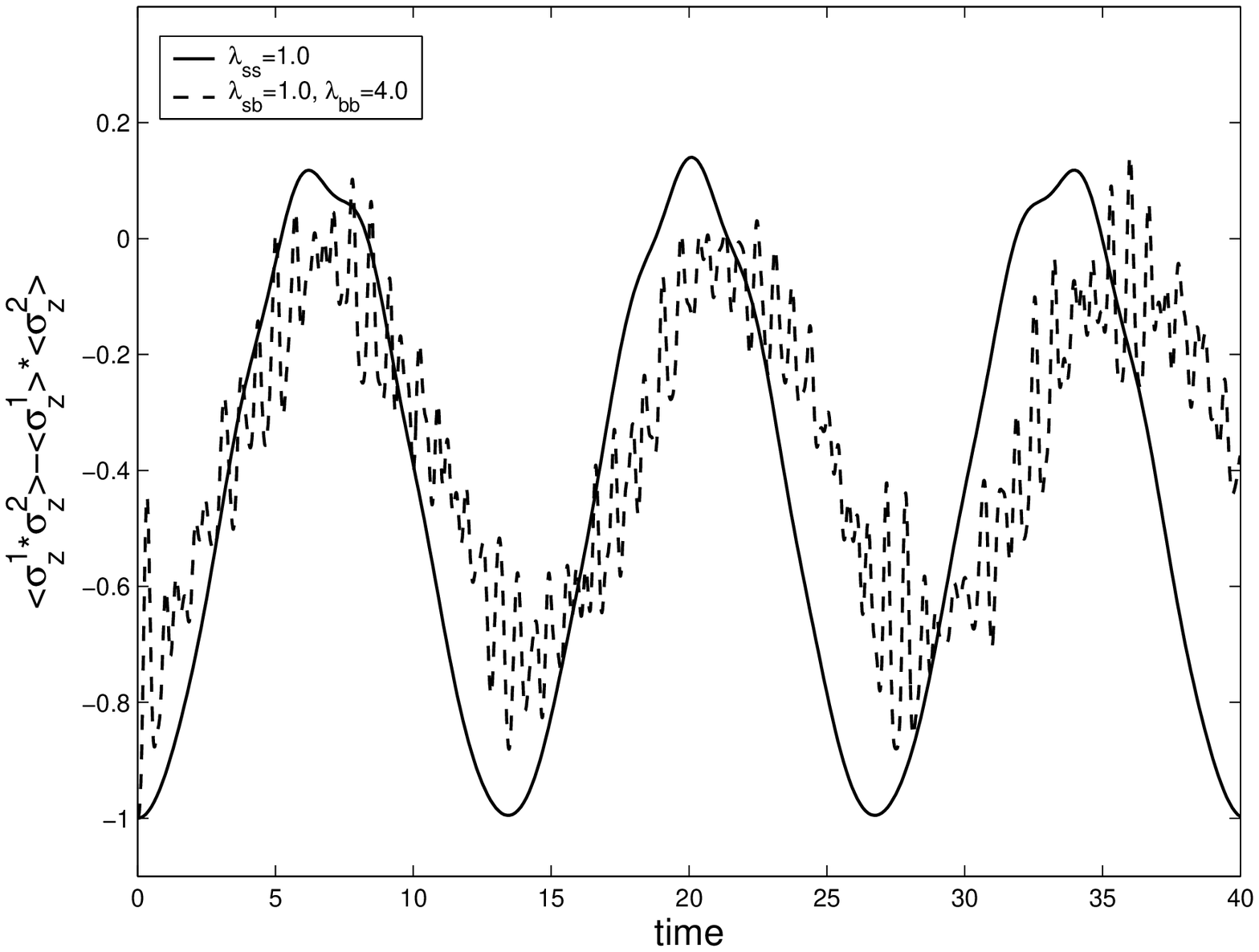}}
\\
\subfigure[$\lambda_{bb}=6.0$]{ \label{fig:2zz6}
\includegraphics[width=3in]{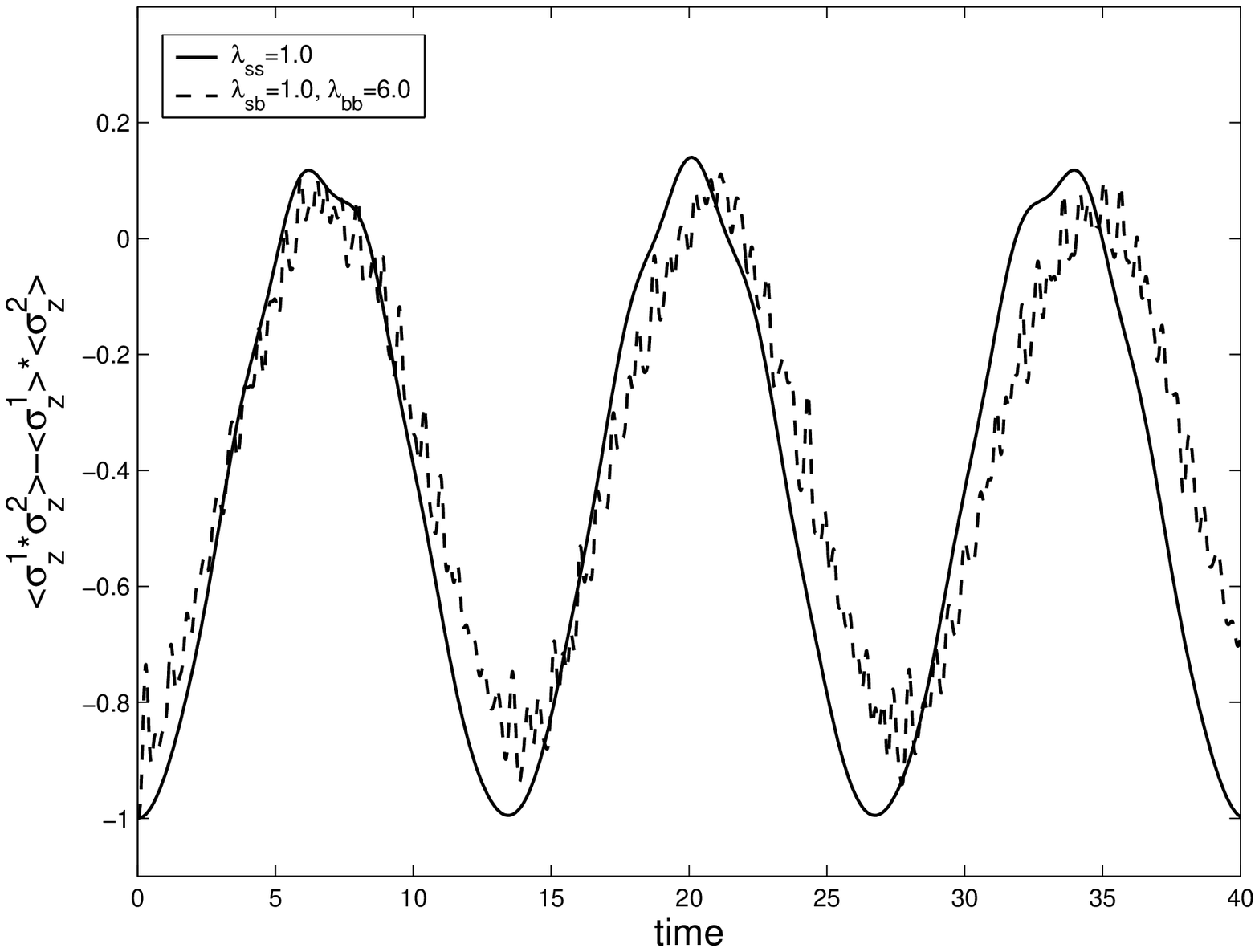}}
\subfigure[$\lambda_{bb}=8.0$]{ \label{fig:2zz8}
\includegraphics[width=3in]{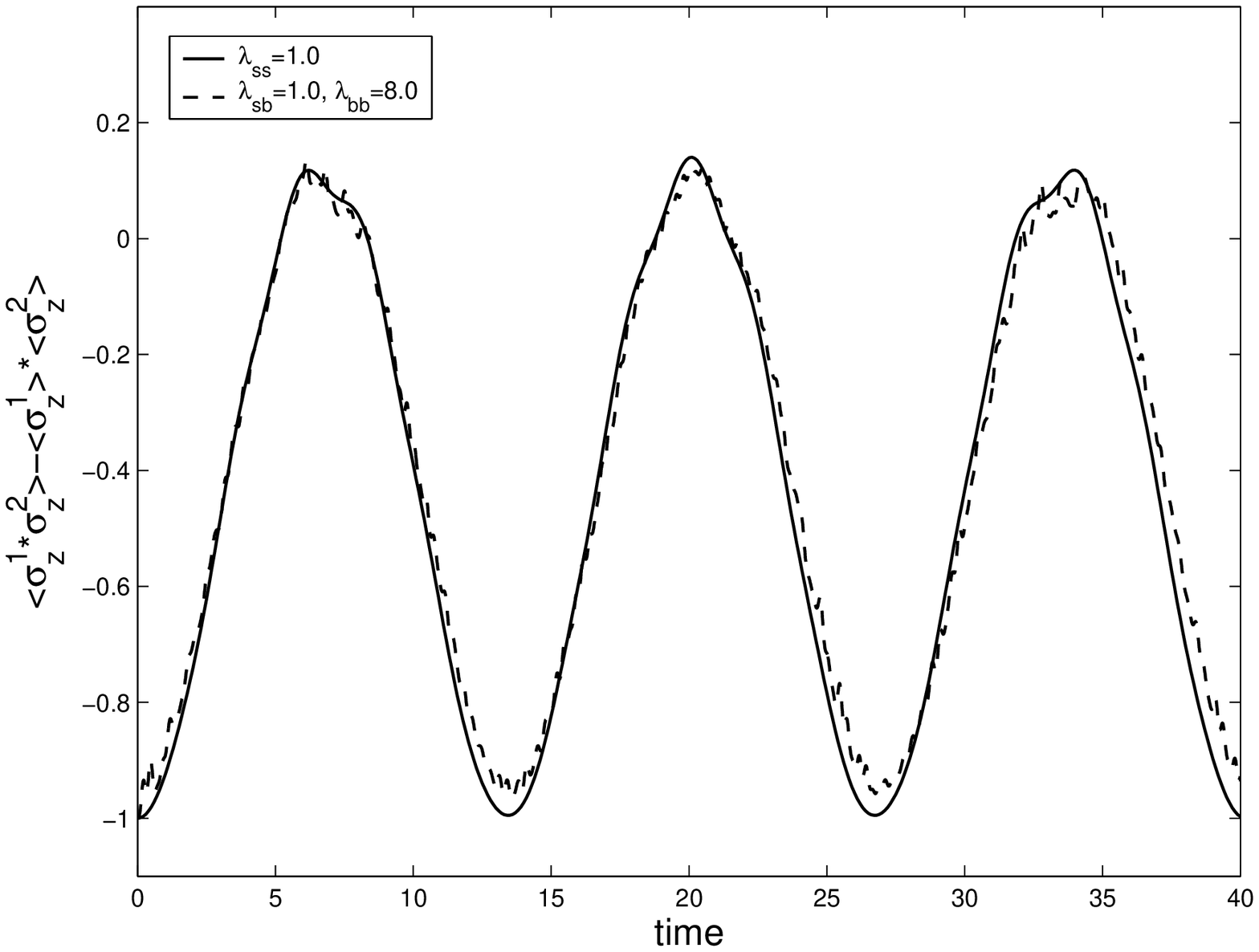}}
  \caption{Evolution for polarization correlation along
$\vec{z}$ direction of the open subsystem,
$\langle\sigma_z^{1}\sigma_z^{2}\rangle
-\langle\sigma_z^{1}\rangle\langle\sigma_z^{2}\rangle$. The initial
state of the  subsystem  is $1/\sqrt{2}(|01\rangle+|10\rangle)$.}
\label{fig:2zz}
\end{figure}

\begin{figure}[htbp]
\centering \subfigure[$\lambda_{bb}=2.0$]{ \label{fig:3xx2}
\includegraphics[width=3in]{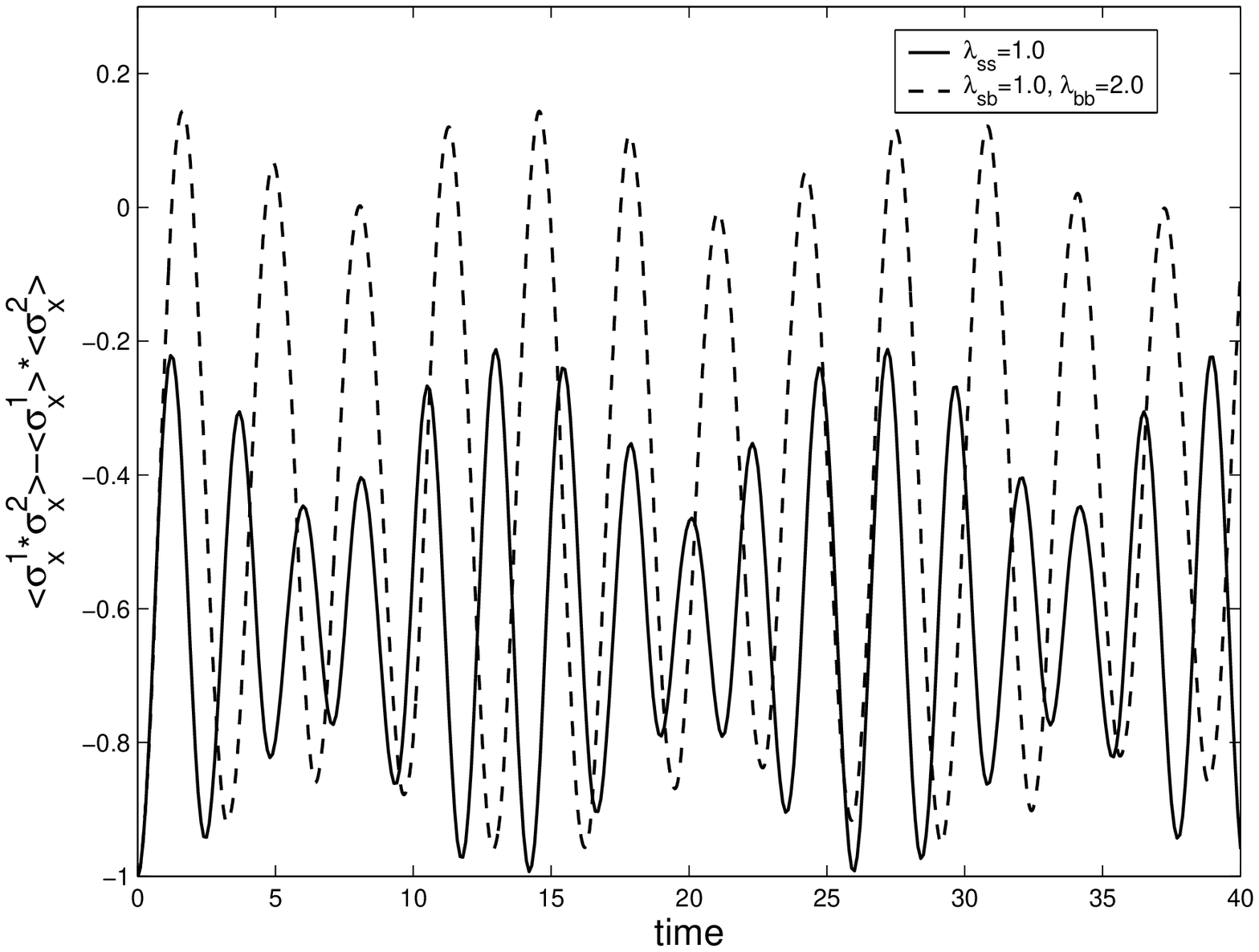}}
\subfigure[$\lambda_{bb}=4.0$]{ \label{fig:3xx4}
\includegraphics[width=3in]{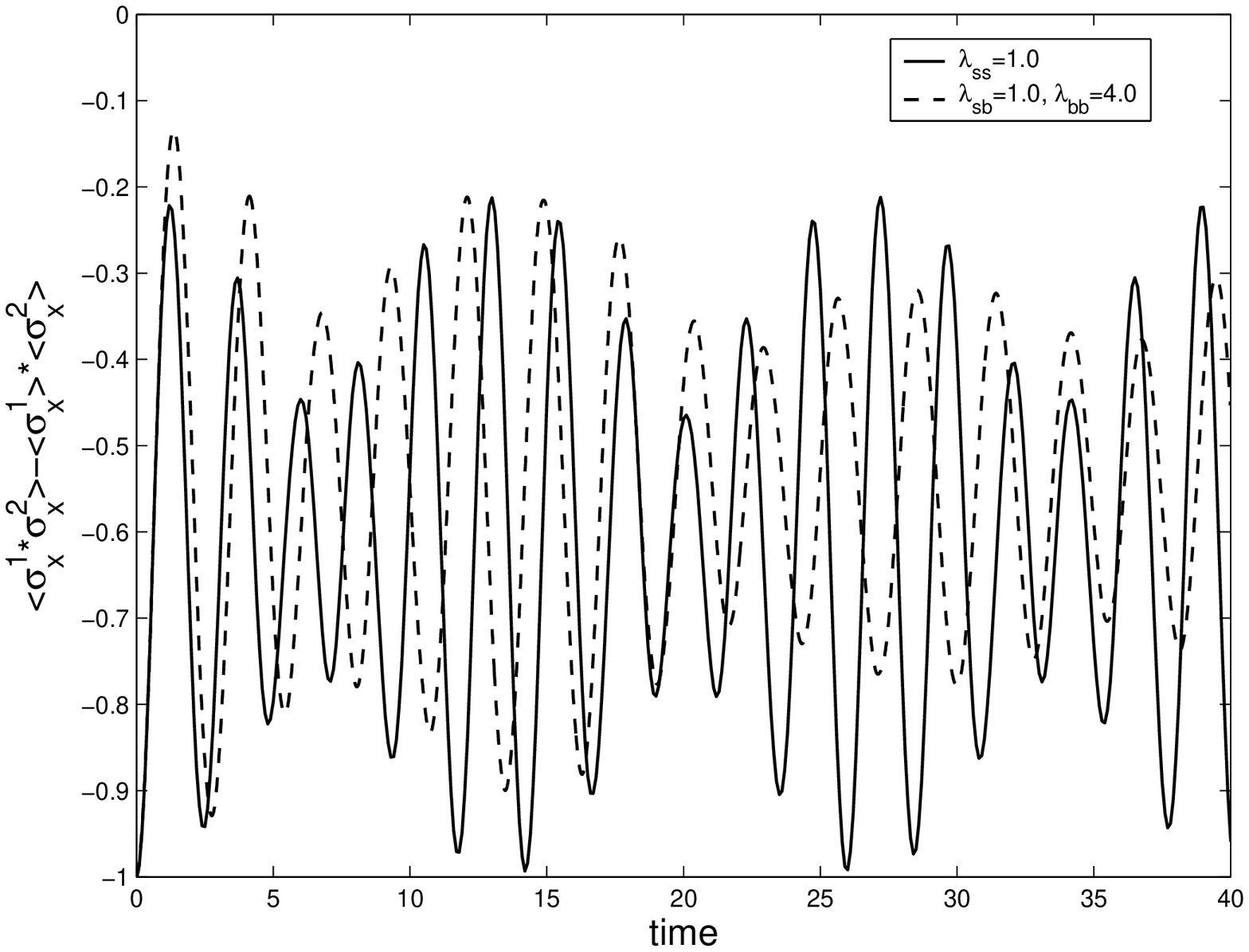}}
\\
\subfigure[$\lambda_{bb}=6.0$]{ \label{fig:3xx6}
\includegraphics[width=3in]{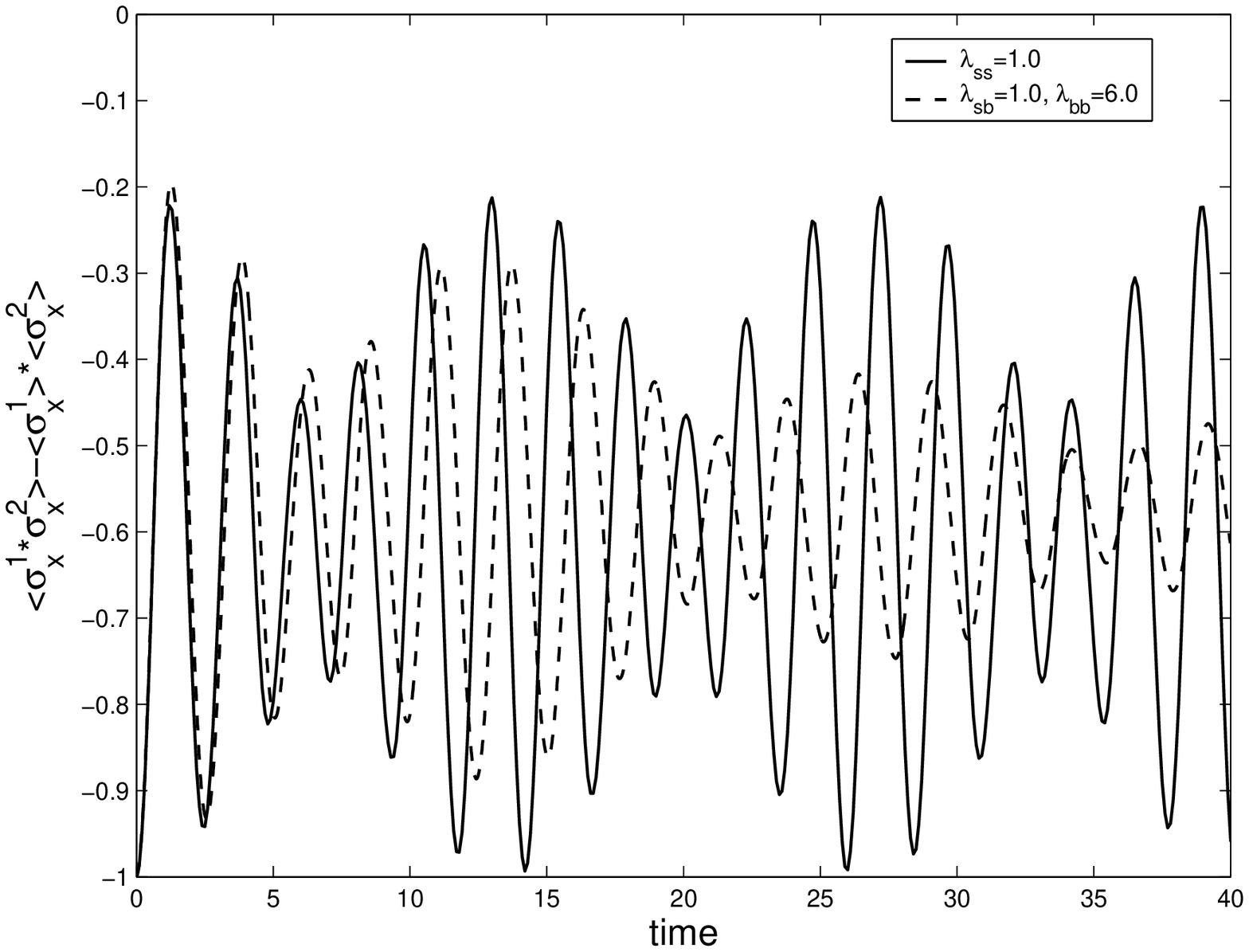}}
\subfigure[$\lambda_{bb}=8.0$]{ \label{fig:3xx8}
\includegraphics[width=3in]{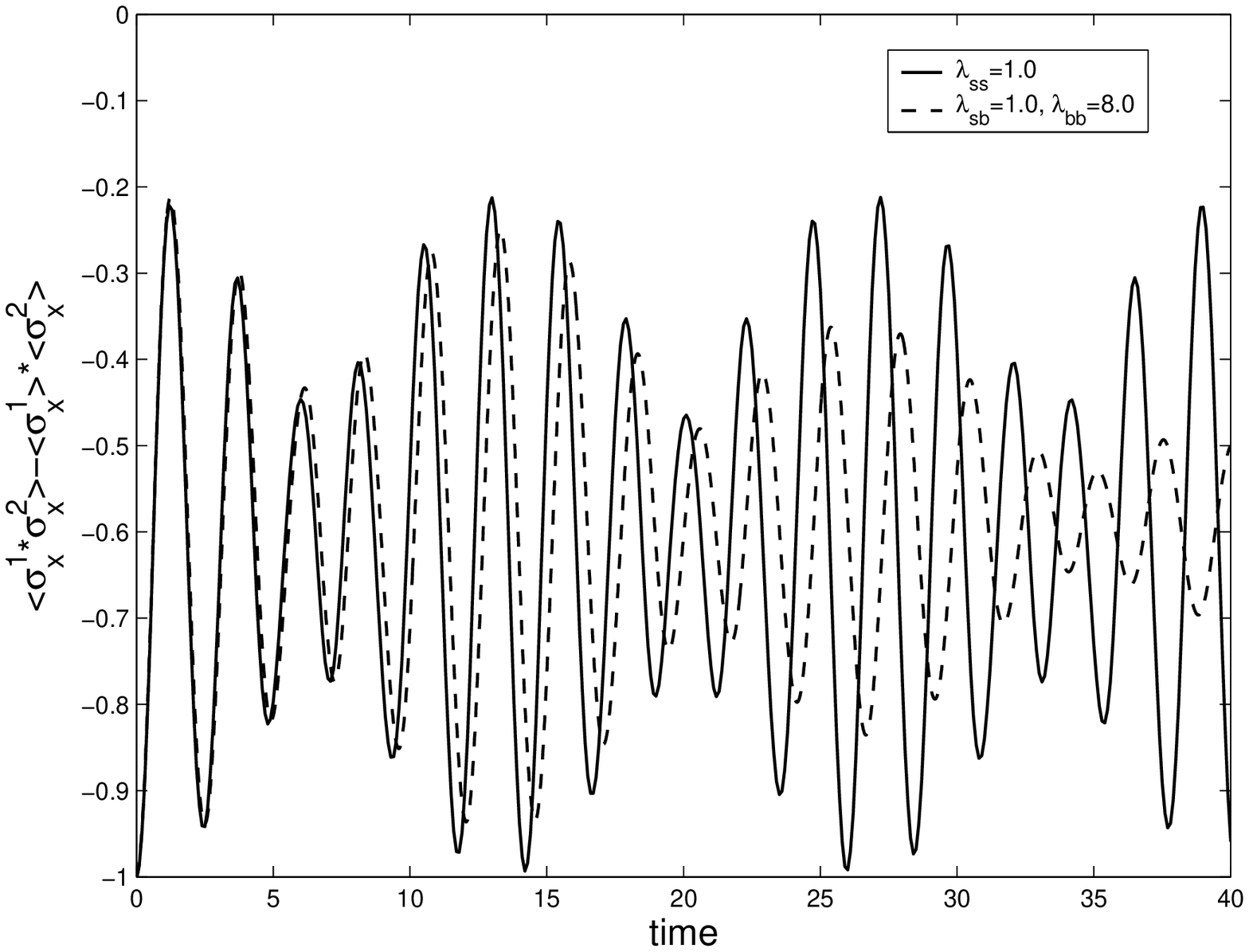}}
  \caption{Evolution for polarization correlation along
$\vec{x}$ direction of the open subsystem,
$\langle\sigma_x^{1}\sigma_x^{2}\rangle
-\langle\sigma_x^{1}\rangle\langle\sigma_x^{2}\rangle$. The  initial
state of the  subsystem  is $1/\sqrt{2}(|11\rangle-|00\rangle)$.}
\label{fig:3xx}
\end{figure}

\begin{figure}[htbp]
\centering \subfigure[$\lambda_{bb}=2.0$]{ \label{fig:3yy2}
\includegraphics[width=3in]{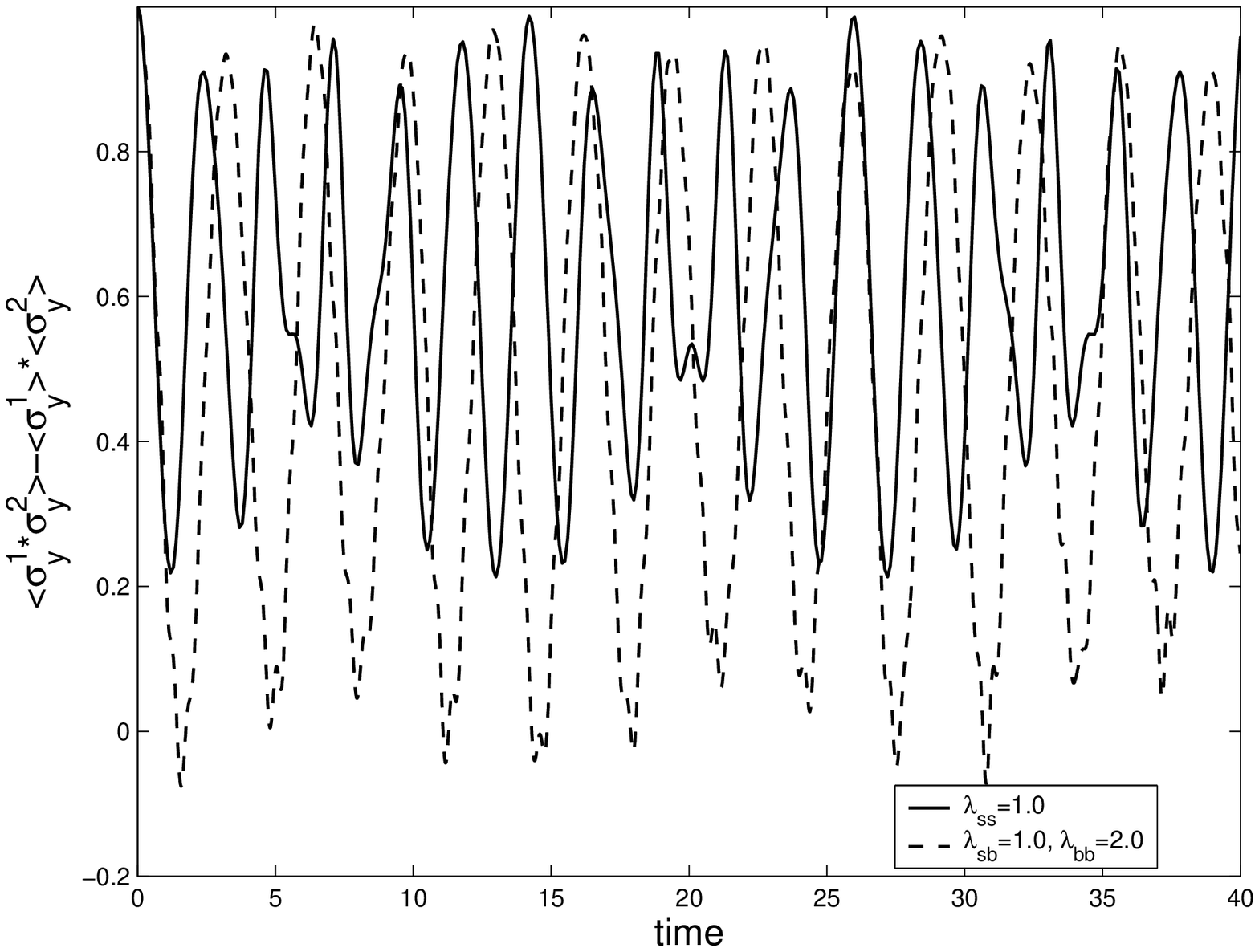}}
\subfigure[$\lambda_{bb}=4.0$]{ \label{fig:3yy4}
\includegraphics[width=3in]{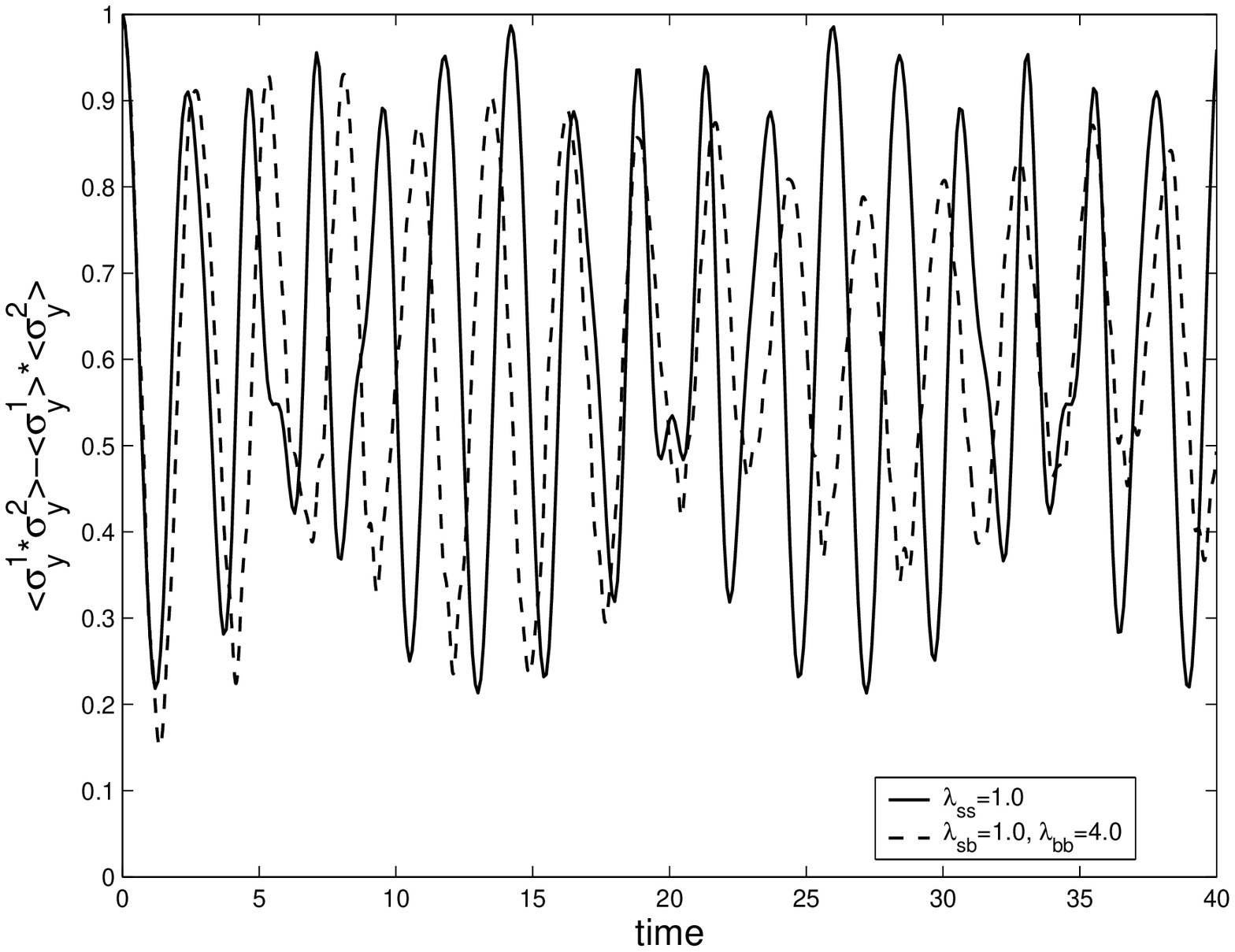}}
\\
\subfigure[$\lambda_{bb}=6.0$]{ \label{fig:3yy6}
\includegraphics[width=3in]{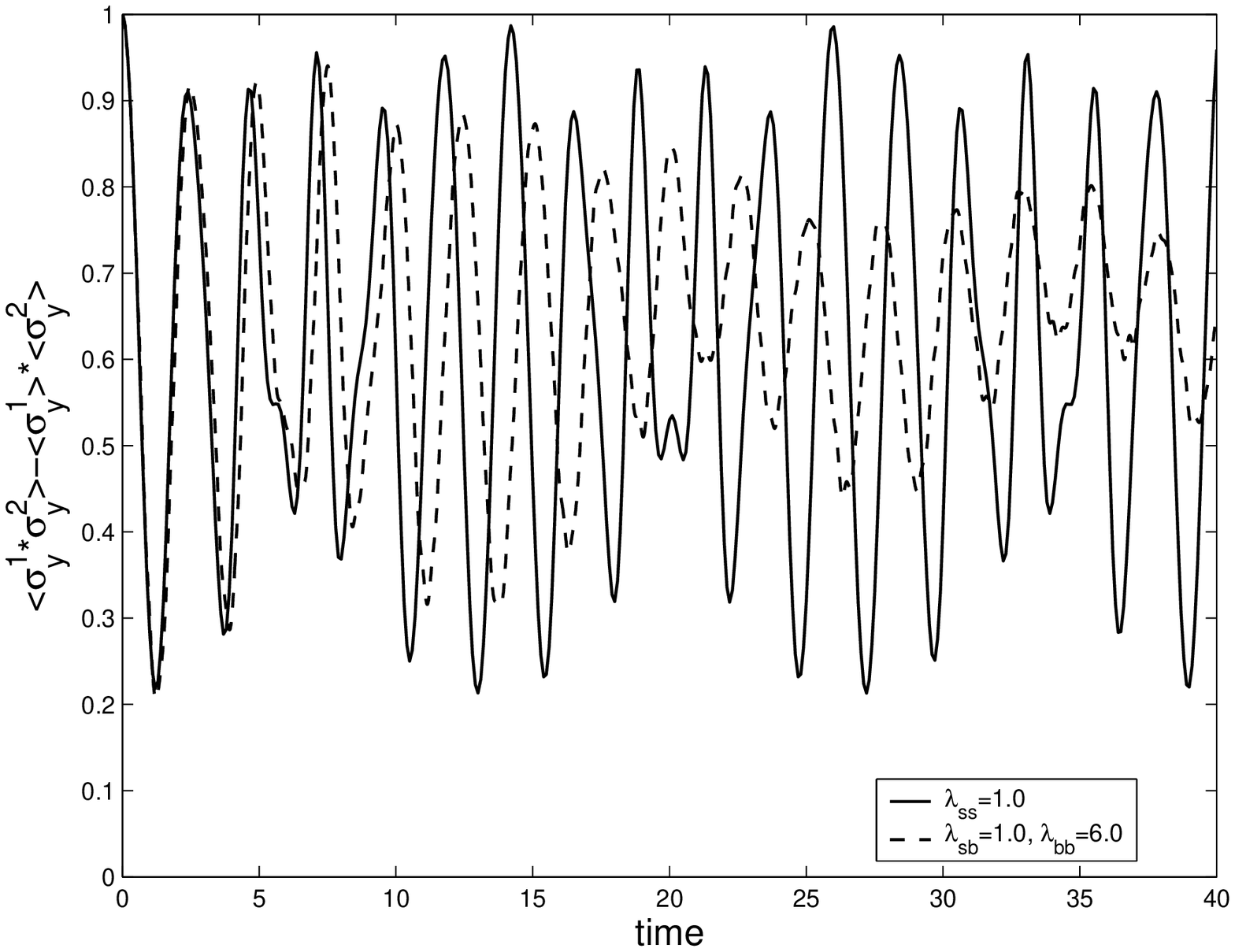}}
\subfigure[$\lambda_{bb}=8.0$]{ \label{fig:3yy8}
\includegraphics[width=3in]{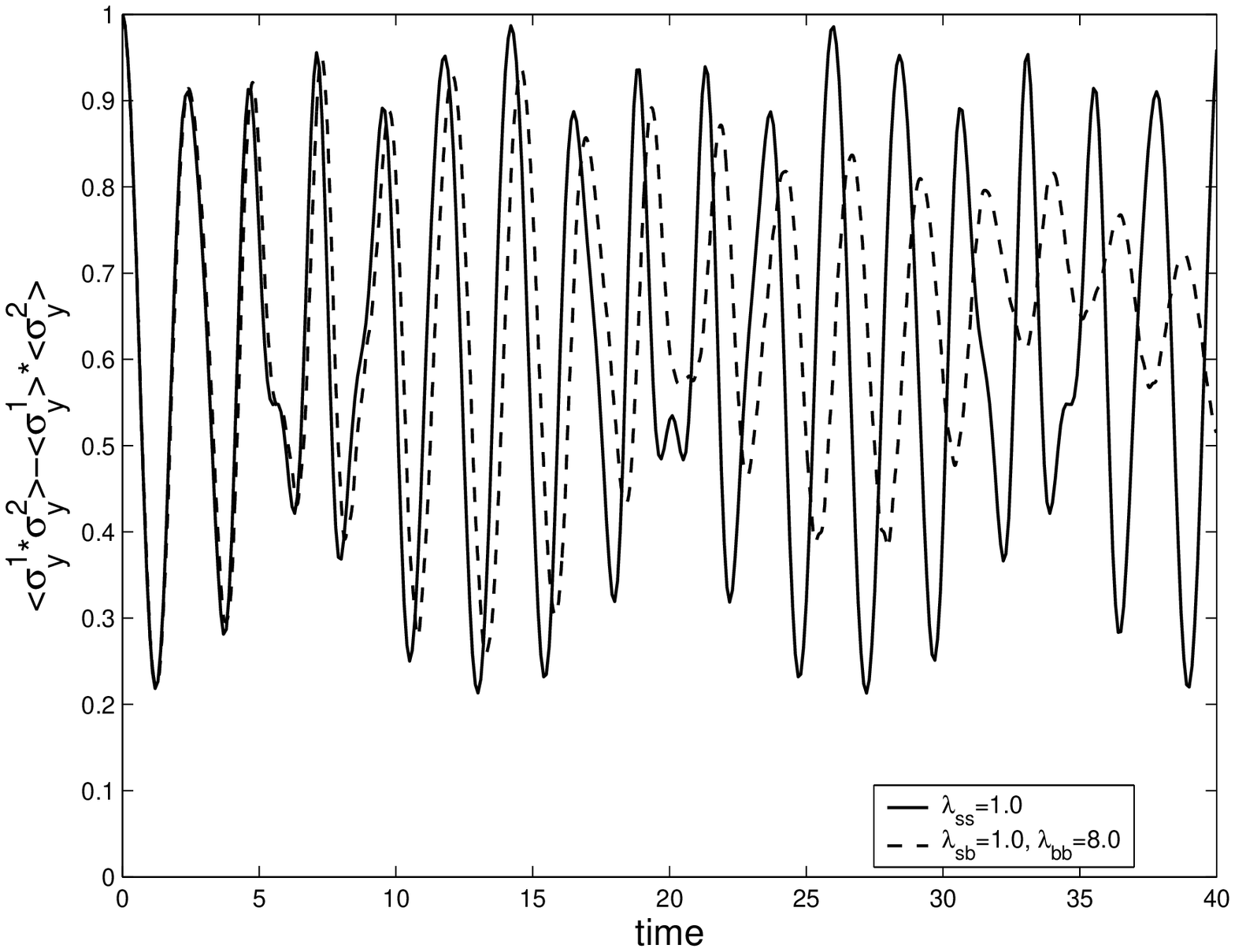}}
\caption{Evolution for polarization correlation along $\vec{y}$
direction of the open subsystem,
$\langle\sigma_y^{1}\sigma_y^{2}\rangle
-\langle\sigma_y^{1}\rangle\langle\sigma_y^{2}\rangle$. The  initial
state of the  subsystem  is $1/\sqrt{2}(|11\rangle-|00\rangle)$.}
\label{fig:3yy}
\end{figure}

\begin{figure}[htbp]
\centering \subfigure[$\lambda_{bb}=2.0$]{ \label{fig:3zz2}
\includegraphics[width=3in]{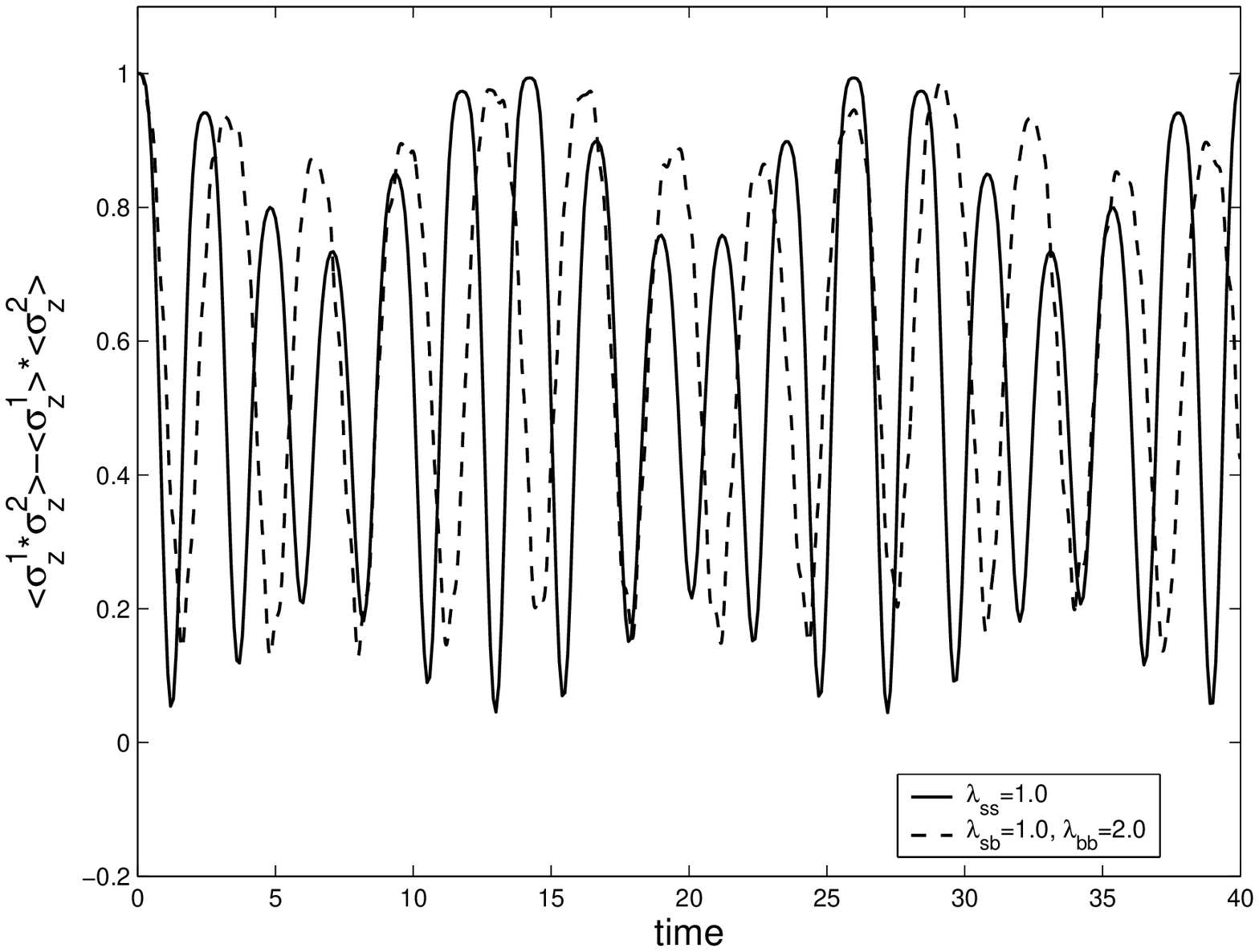}}
\subfigure[$\lambda_{bb}=4.0$]{ \label{fig:3zz4}
\includegraphics[width=3in]{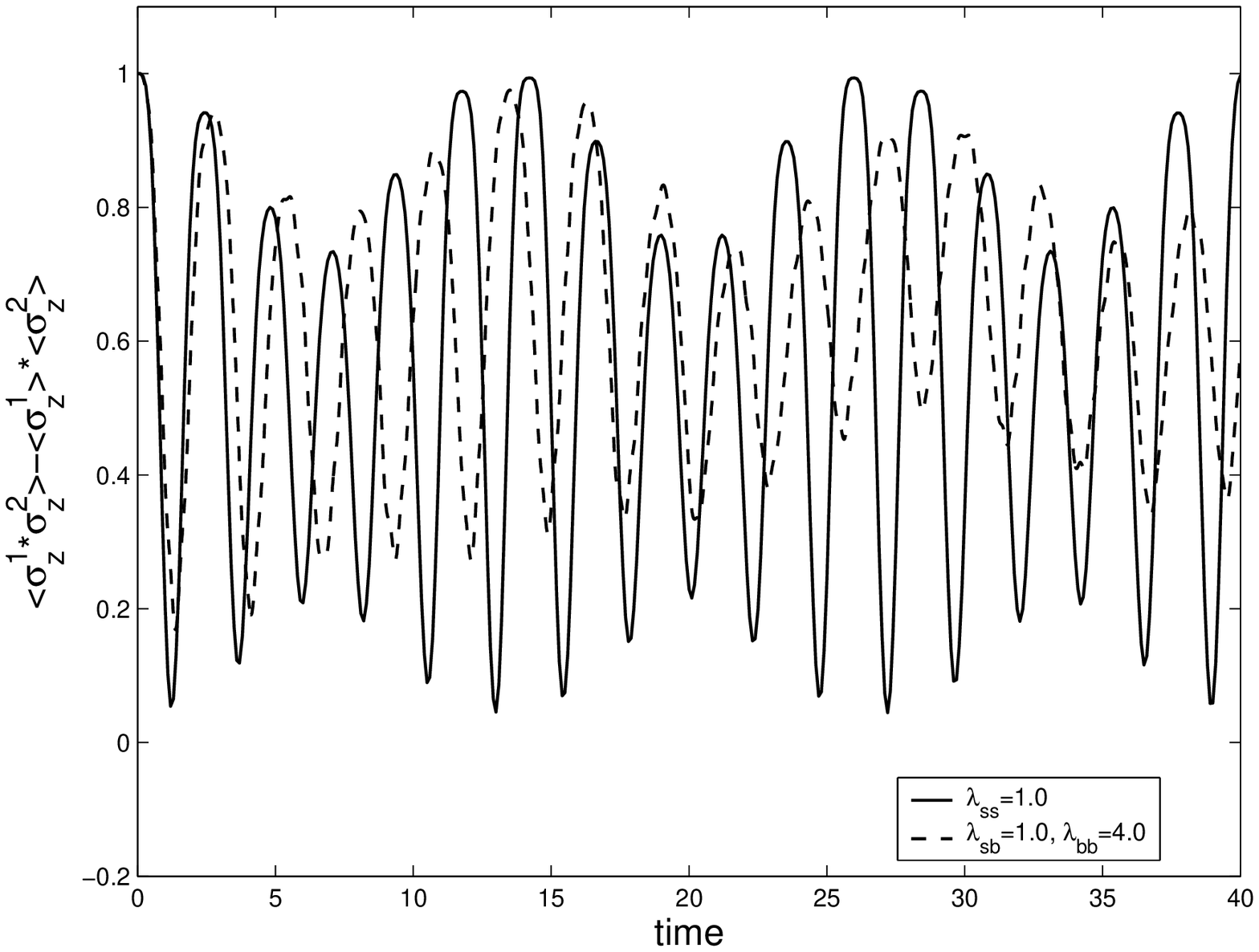}}
\\
\subfigure[$\lambda_{bb}=6.0$]{ \label{fig:3zz6}
\includegraphics[width=3in]{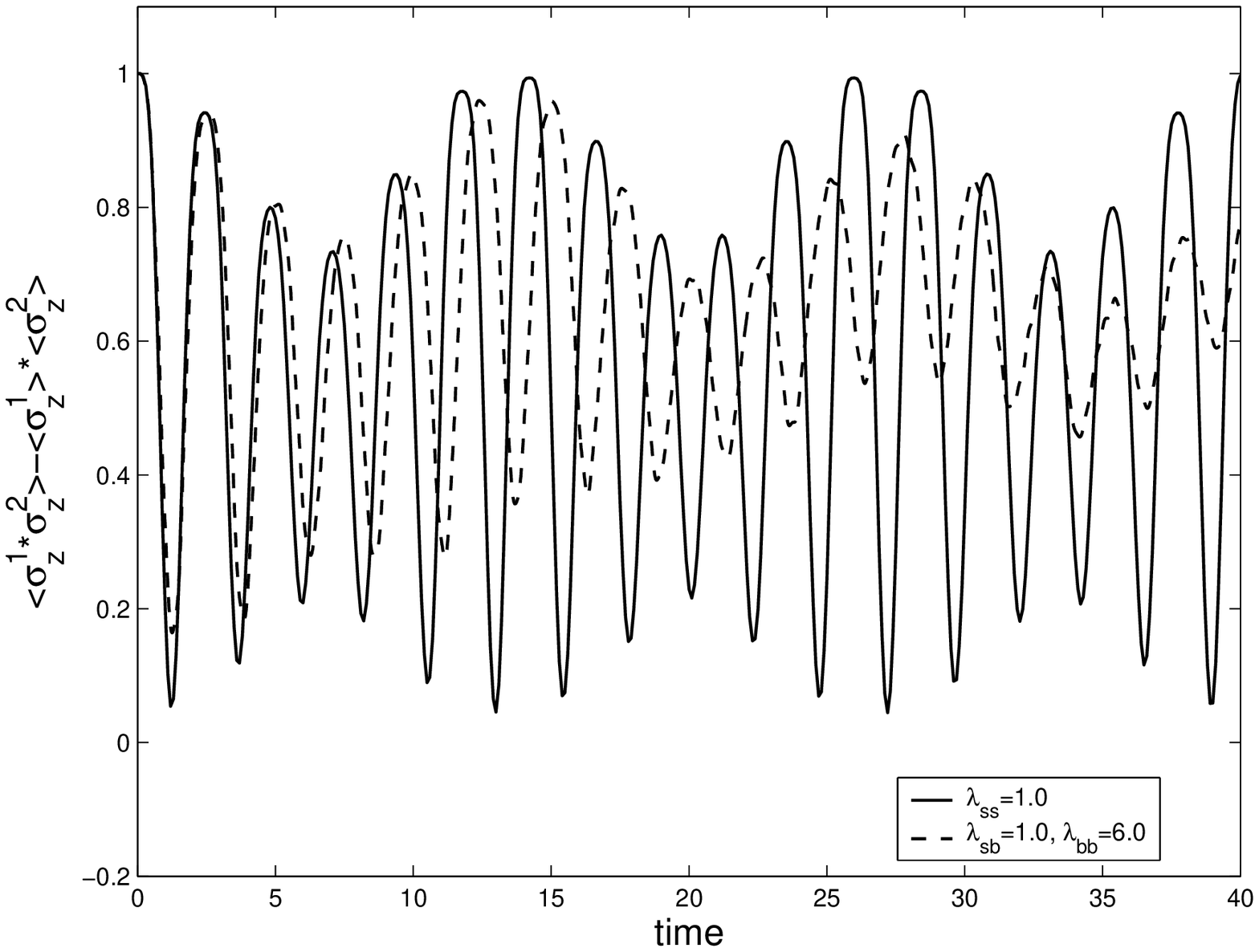}}
\subfigure[$\lambda_{bb}=8.0$]{ \label{fig:3zz8}
\includegraphics[width=3in]{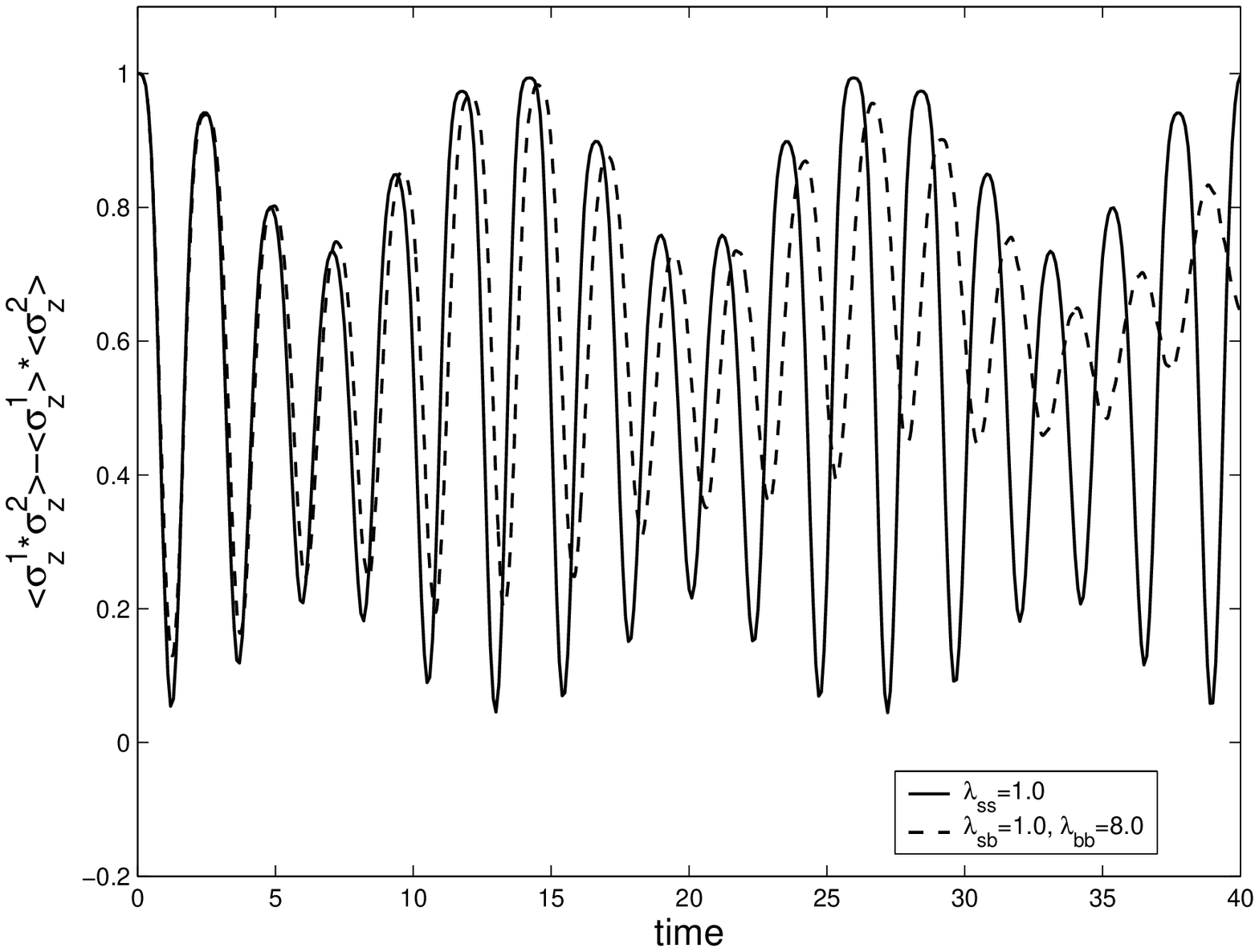}}
\caption{Evolution for polarization correlation along $\vec{z}$
direction of the open subsystem,
$\langle\sigma_z^{1}\sigma_z^{2}\rangle
-\langle\sigma_z^{1}\rangle\langle\sigma_z^{2}\rangle$. The  initial
state of the  subsystem  is $1/\sqrt{2}(|11\rangle-|00\rangle)$.}
\label{fig:3zz}
\end{figure}

\begin{figure}[htbp]
\centering \subfigure[$\langle\sigma_x^{1}\sigma_x^{2}\rangle
  -\langle\sigma_x^{1}\rangle\langle\sigma_x^{2}\rangle$,
$\lambda_{bb}=6.0$]{ \label{fig:N6:x6}
\includegraphics[width=3in]{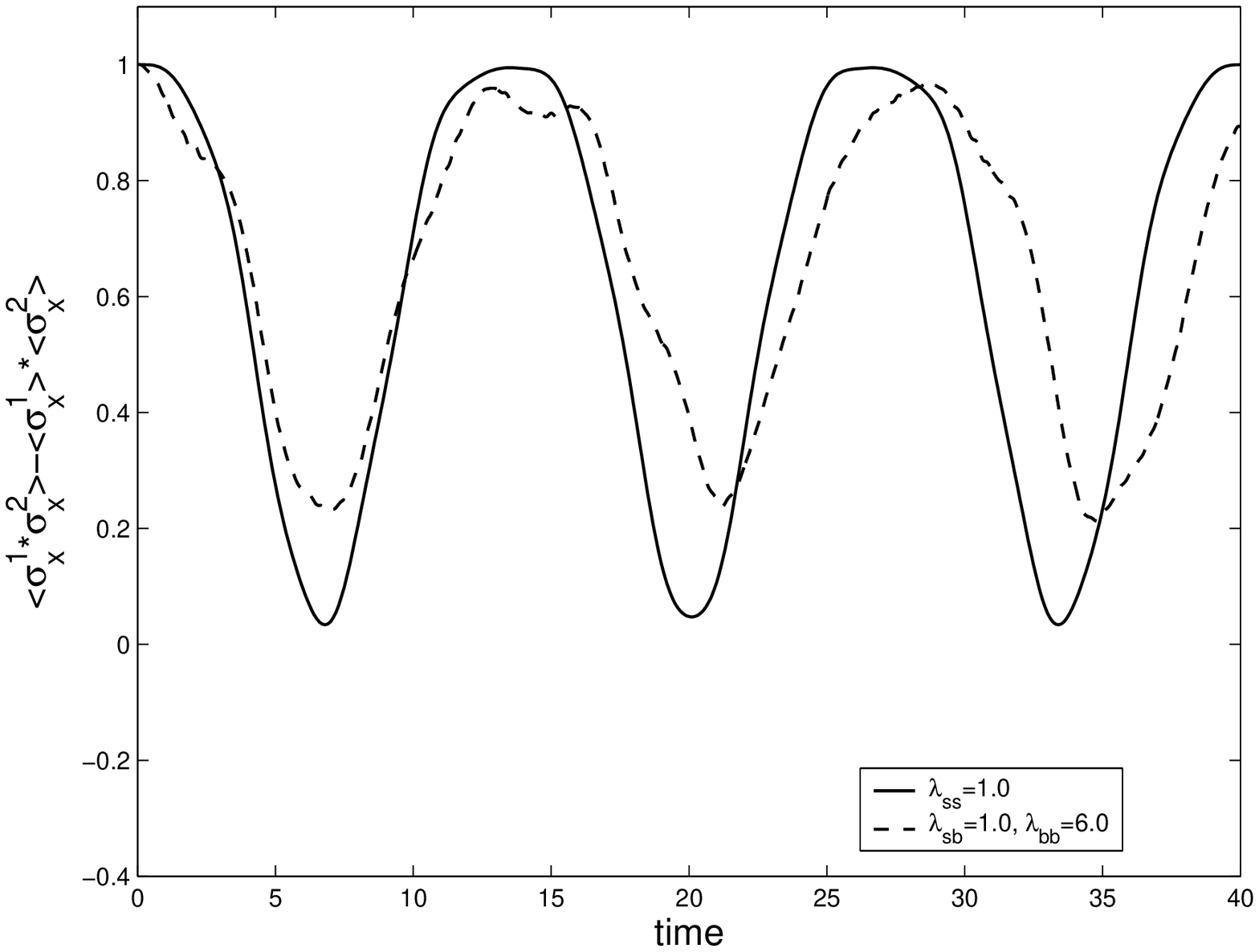}}
\subfigure[$\langle\sigma_x^{1}\sigma_x^{2}\rangle
  -\langle\sigma_x^{1}\rangle\langle\sigma_x^{2}\rangle$, $\lambda_{bb}=10.0$]{
\label{fig:N6:x10}
\includegraphics[width=3in]{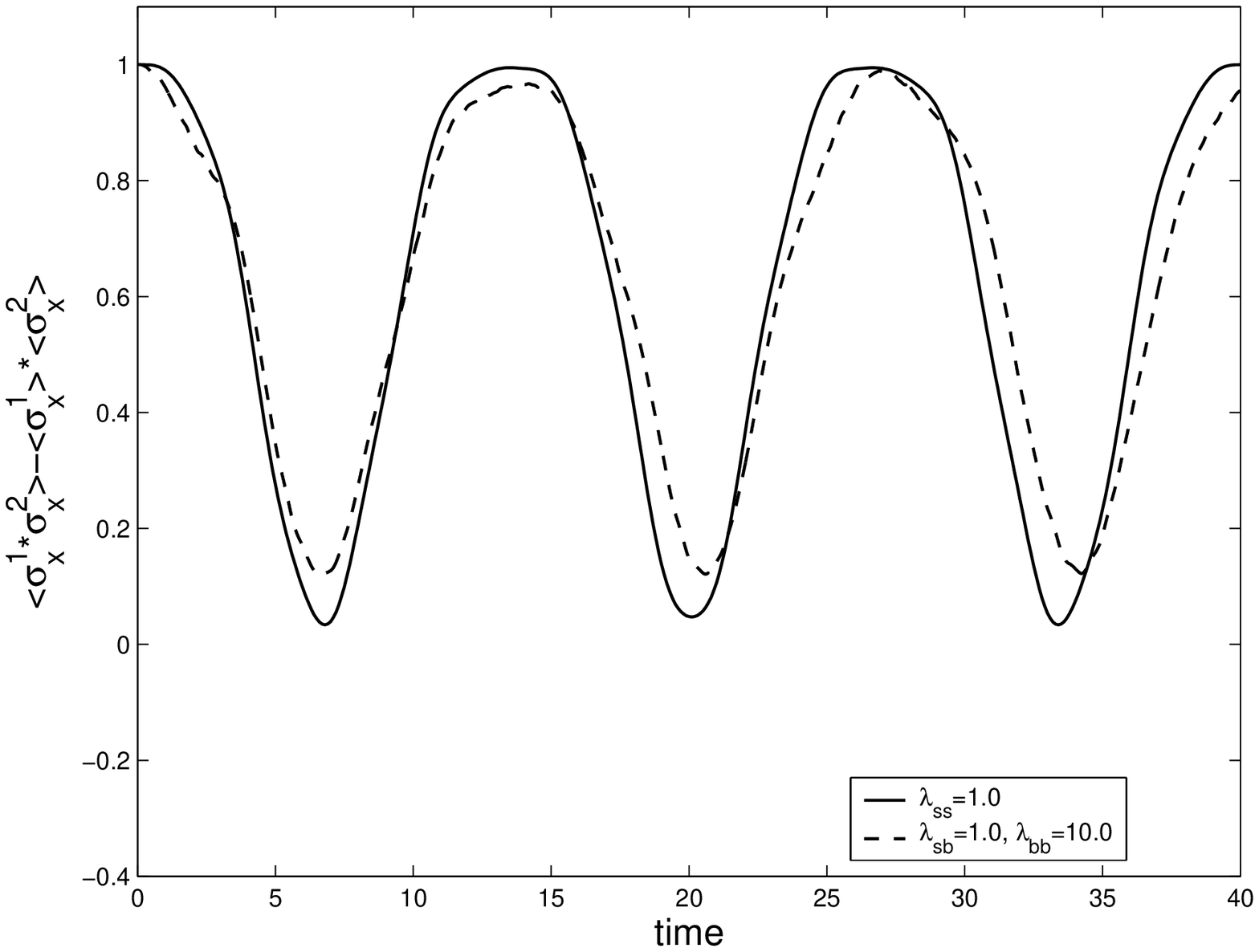}}
\subfigure[$\langle\sigma_y^{1}\sigma_y^{2}\rangle
  -\langle\sigma_y^{1}\rangle\langle\sigma_y^{2}\rangle$, $\lambda_{bb}=6.0$]{
\label{fig:N6:y6}
\includegraphics[width=3in]{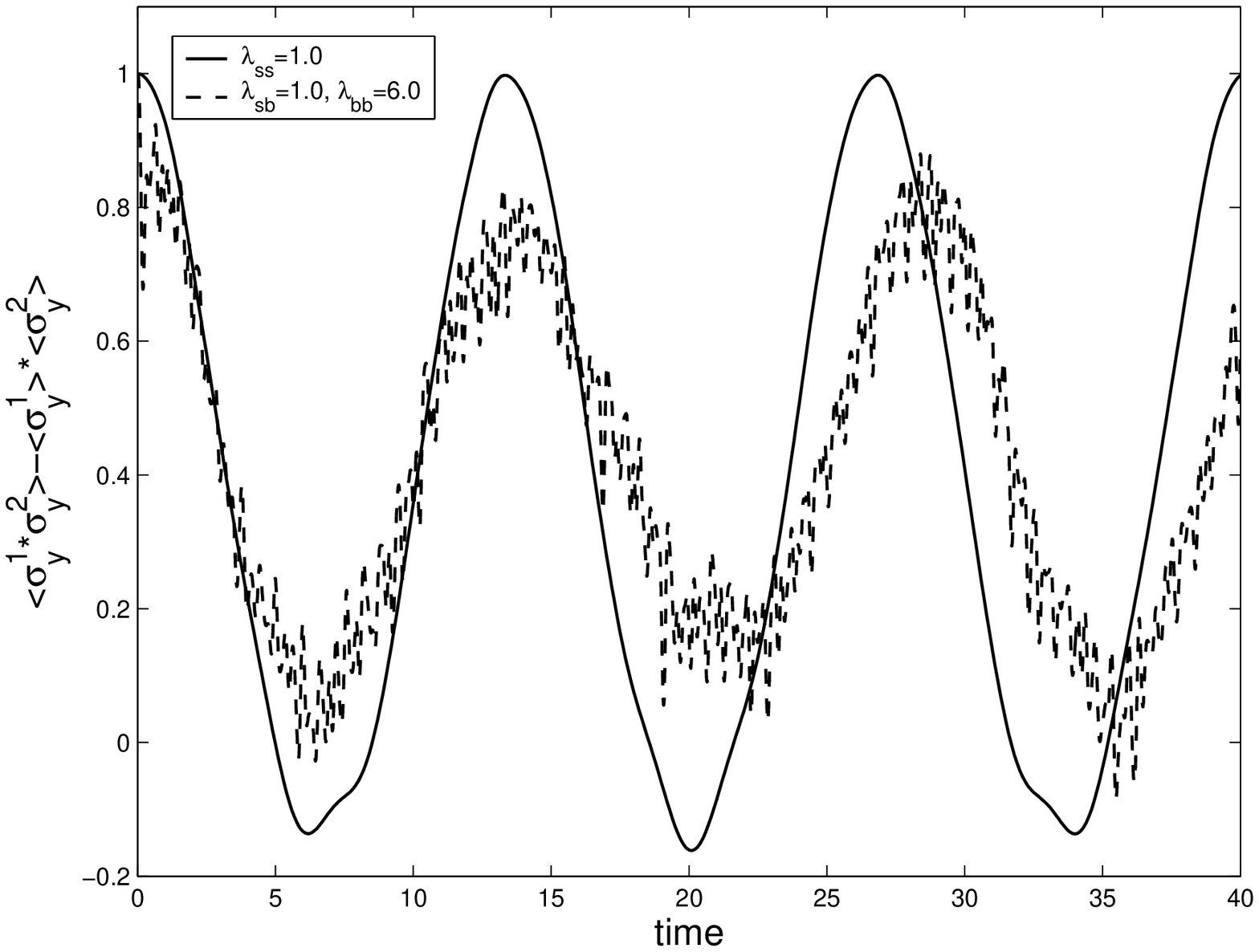}}
\subfigure[$\langle\sigma_y^{1}\sigma_y^{2}\rangle
  -\langle\sigma_y^{1}\rangle\langle\sigma_y^{2}\rangle$, $\lambda_{bb}=10.0$]{
\label{fig:N6:y10}
\includegraphics[width=3in]{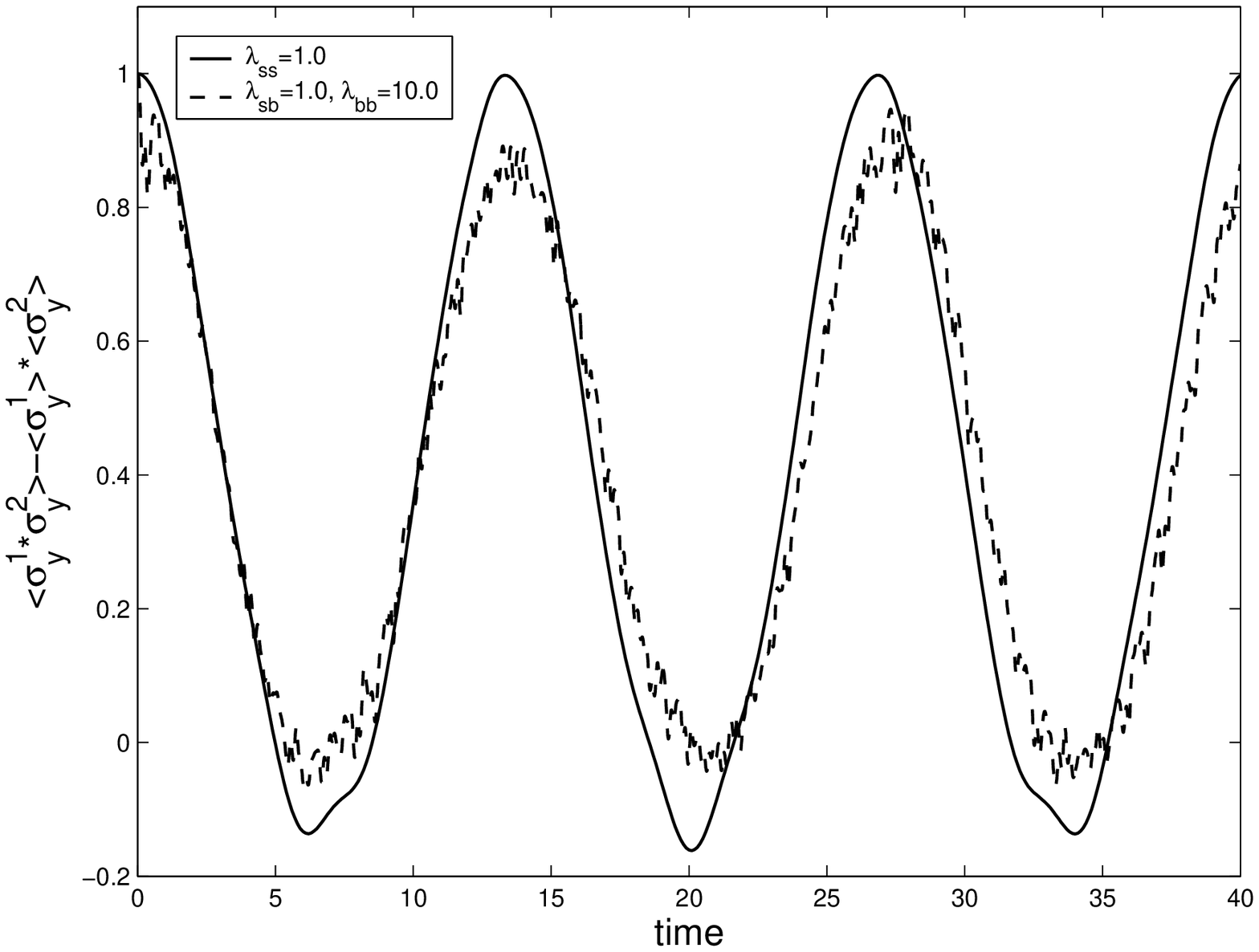}}
\subfigure[$\langle\sigma_z^{1}\sigma_z^{2}\rangle
  -\langle\sigma_z^{1}\rangle\langle\sigma_z^{2}\rangle$, $\lambda_{bb}=6.0$]{
\label{fig:N6:z6}
\includegraphics[width=3in]{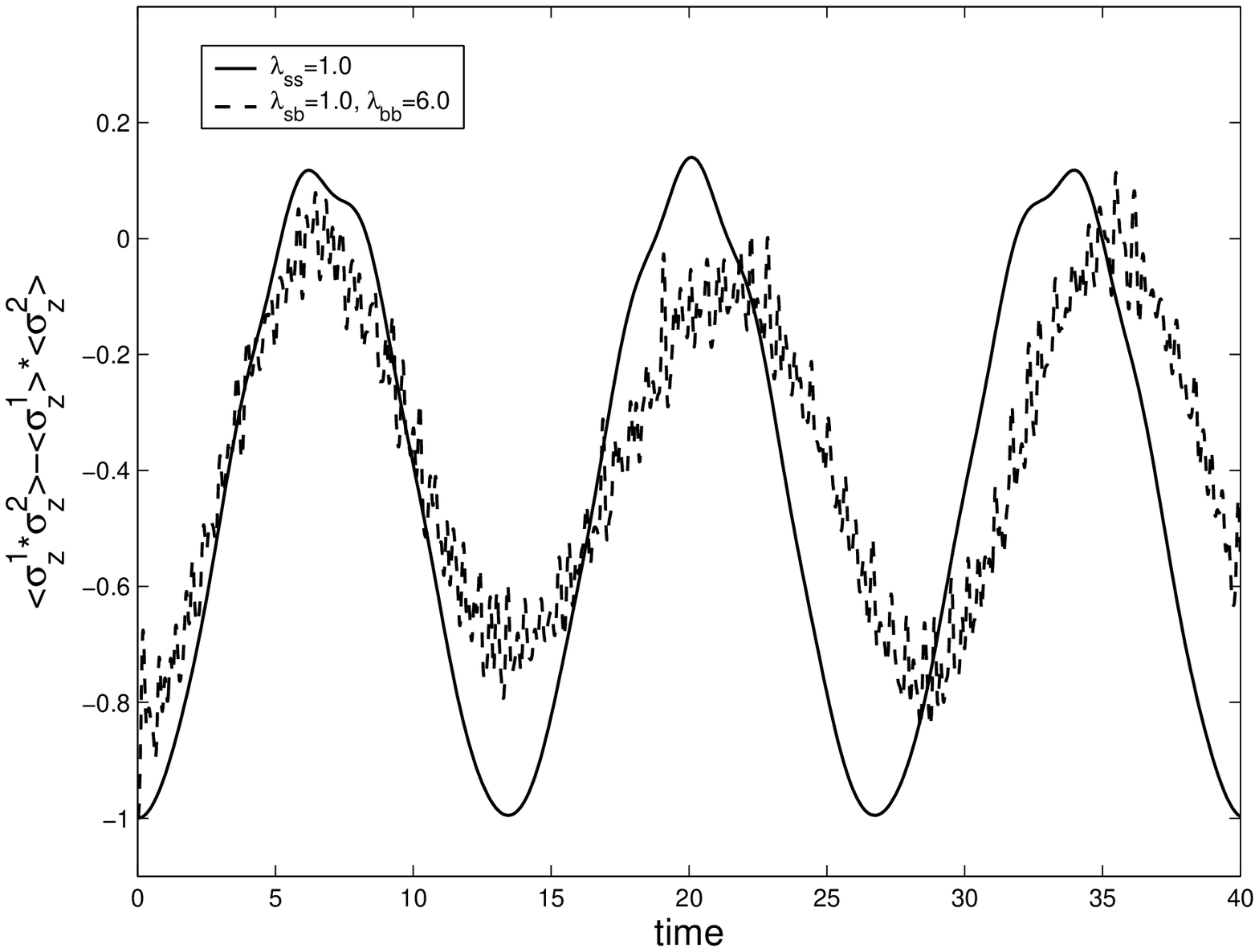}}
\subfigure[$\langle\sigma_z^{1}\sigma_z^{2}\rangle
  -\langle\sigma_z^{1}\rangle\langle\sigma_z^{2}\rangle$, $\lambda_{bb}=10.0$]{
\label{fig:N6:z10}
\includegraphics[width=3in]{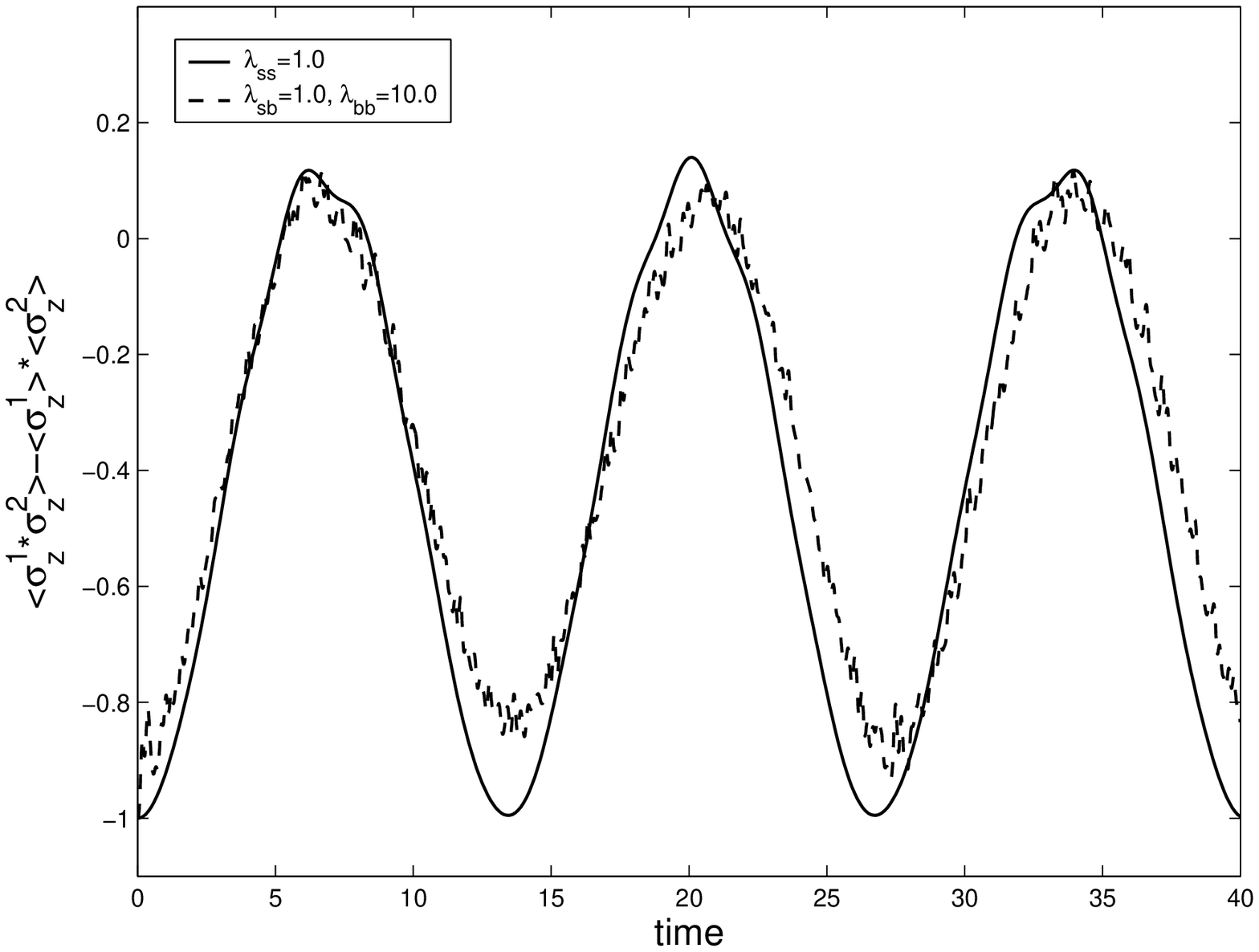}}
\caption{Evolution for polarization correlations along three
directions of the open subsystem. Where there are $6$ spins in the
bath and the initial state of the subsystem is
$1/\sqrt{2}(|01\rangle+|10\rangle)$.} \label{fig:N6}
\end{figure}

Figure \ref{fig:1xx} through figure \ref{fig:3zz} are the
polarization correlations when the subsystem coupled to the bath
with different $\lambda_{bb}$. The corresponding results for
isolated systems are also drawn on each figure as solid lines. In
these figures, we have $m=4$ spins in the bath, though the number of
bath spins is small, it still can give some information on the
influence of bath coupling to the system. We also did calculations
with more bath spins, $m=6$, show in figure \ref{fig:N6}.\\

It is clear from the figures that when the subsystem coupled to the
bath, all the polarization correlations are different from the
isolated systems. The difference reflects the influences of the bath
and decoherence of the subsystem. However, as pointed out in
reference \cite{TWmodel} and \cite{Milburn} for the case of a single
spin system, the coupling strength among environment spins can
suppress the decoherence and partially restore the subsystem
coherence. Figure \ref{fig:1xx} through figure \ref{fig:3zz} are
plots of the polarization correlations as function of the time with
different initial states and increasing intra-bath coupling
strength. At weak intra-bath coupling, the results are drastically
different from the isolated systems and similar with figure
\ref{fig2:N60}; as the intra-bath coupling is increased, the
difference between the open subsystem and the isolated system
becomes smaller; and at very strong strengths, the difference
becomes very small especially in the case of $|\psi_S(0)\rangle^2$.
The other two initial Bell states will be dissipated when time is
elongated.

\begin{figure}[htbp]
\centering \subfigure[$\langle\sigma_x^{1}\sigma_x^{2}\rangle
  -\langle\sigma_x^{1}\rangle\langle\sigma_x^{2}\rangle$,
$\lambda_{bb}=4.0$]{ \label{fig:N8:x4}
\includegraphics[width=3in]{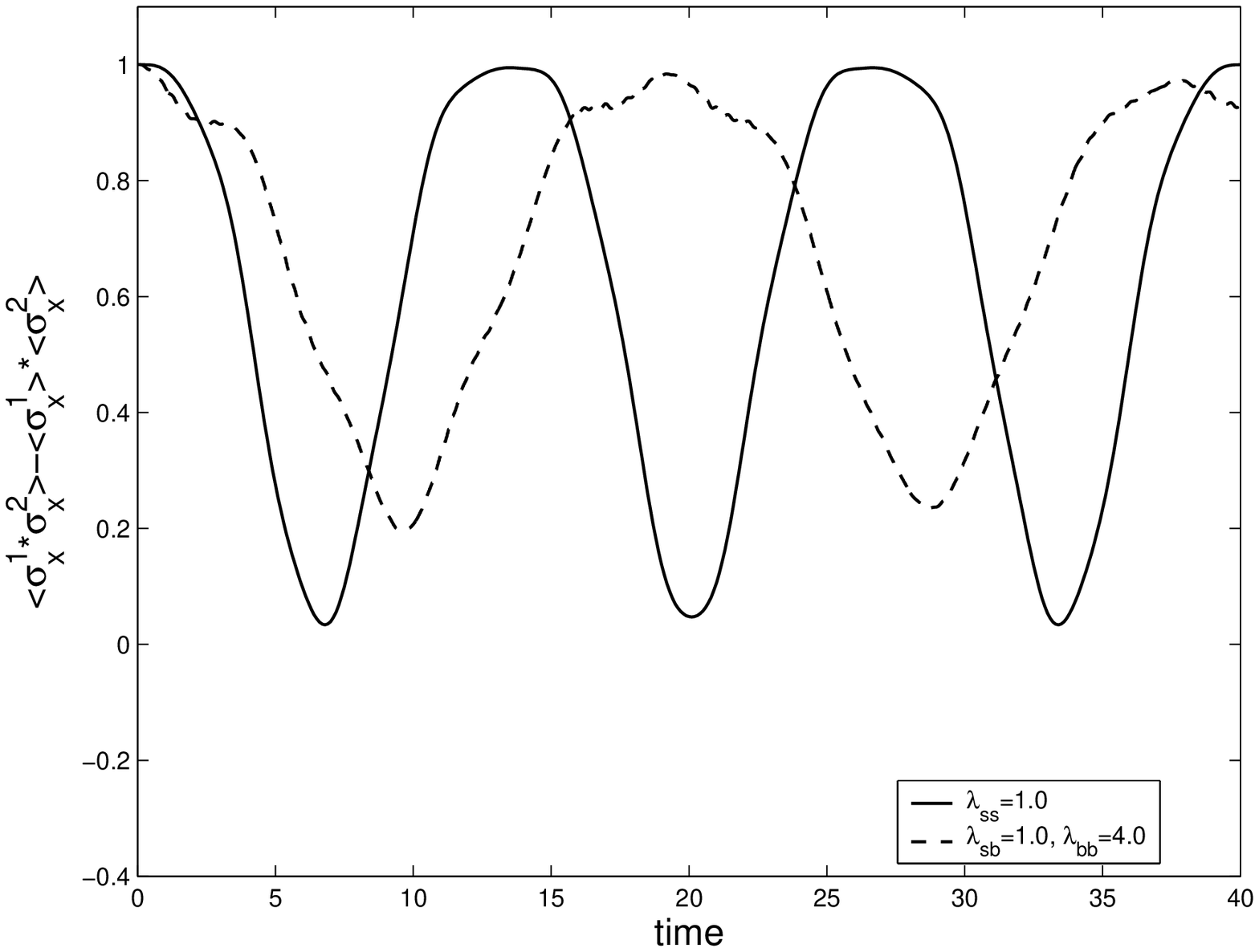}}
\subfigure[$\langle\sigma_x^{1}\sigma_x^{2}\rangle
  -\langle\sigma_x^{1}\rangle\langle\sigma_x^{2}\rangle$, $\lambda_{bb}=24.0$]{
\label{fig:N8:x24}
\includegraphics[width=3in]{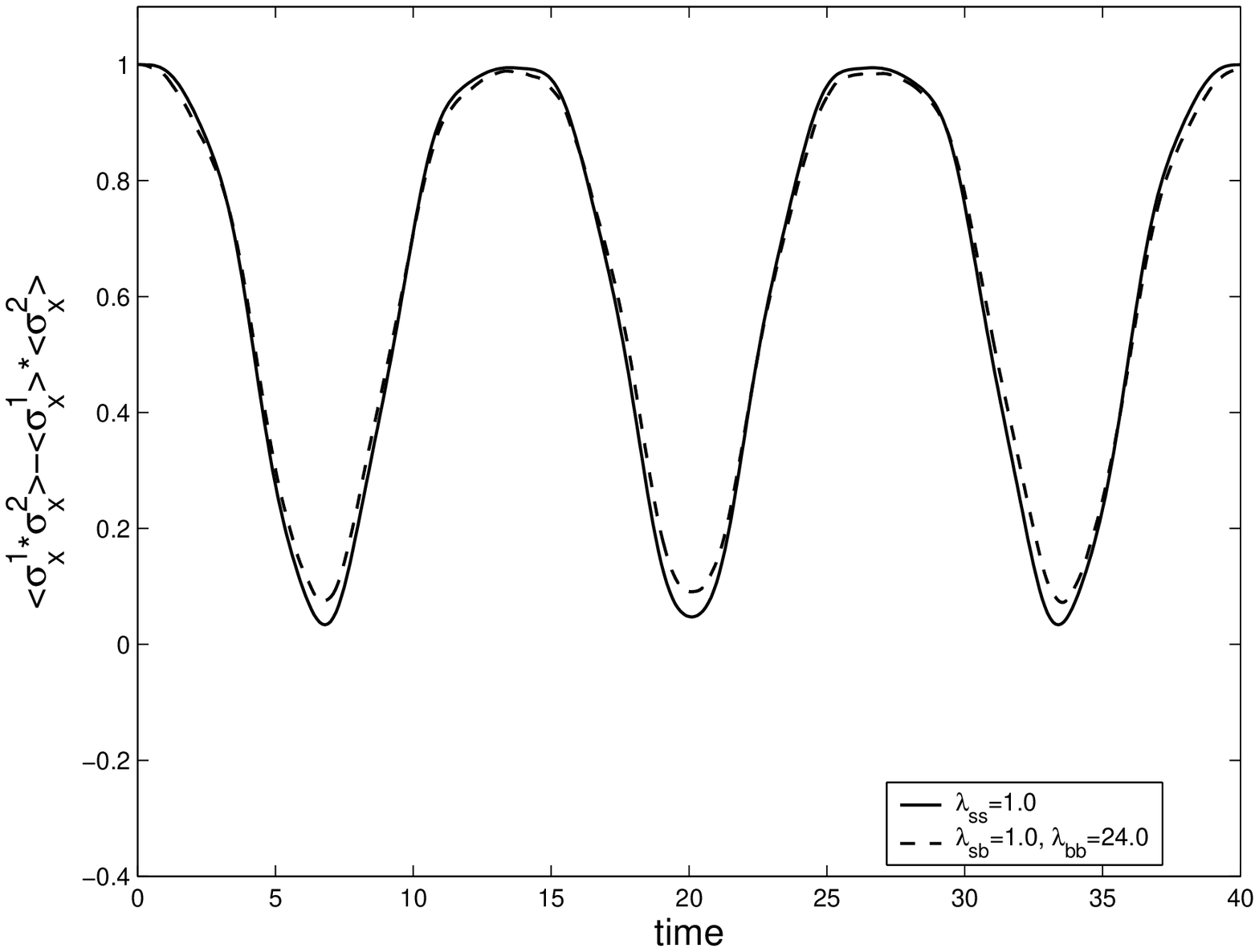}}
\subfigure[$\langle\sigma_y^{1}\sigma_y^{2}\rangle
  -\langle\sigma_y^{1}\rangle\langle\sigma_y^{2}\rangle$, $\lambda_{bb}=4.0$]{
\label{fig:N8:y4}
\includegraphics[width=3in]{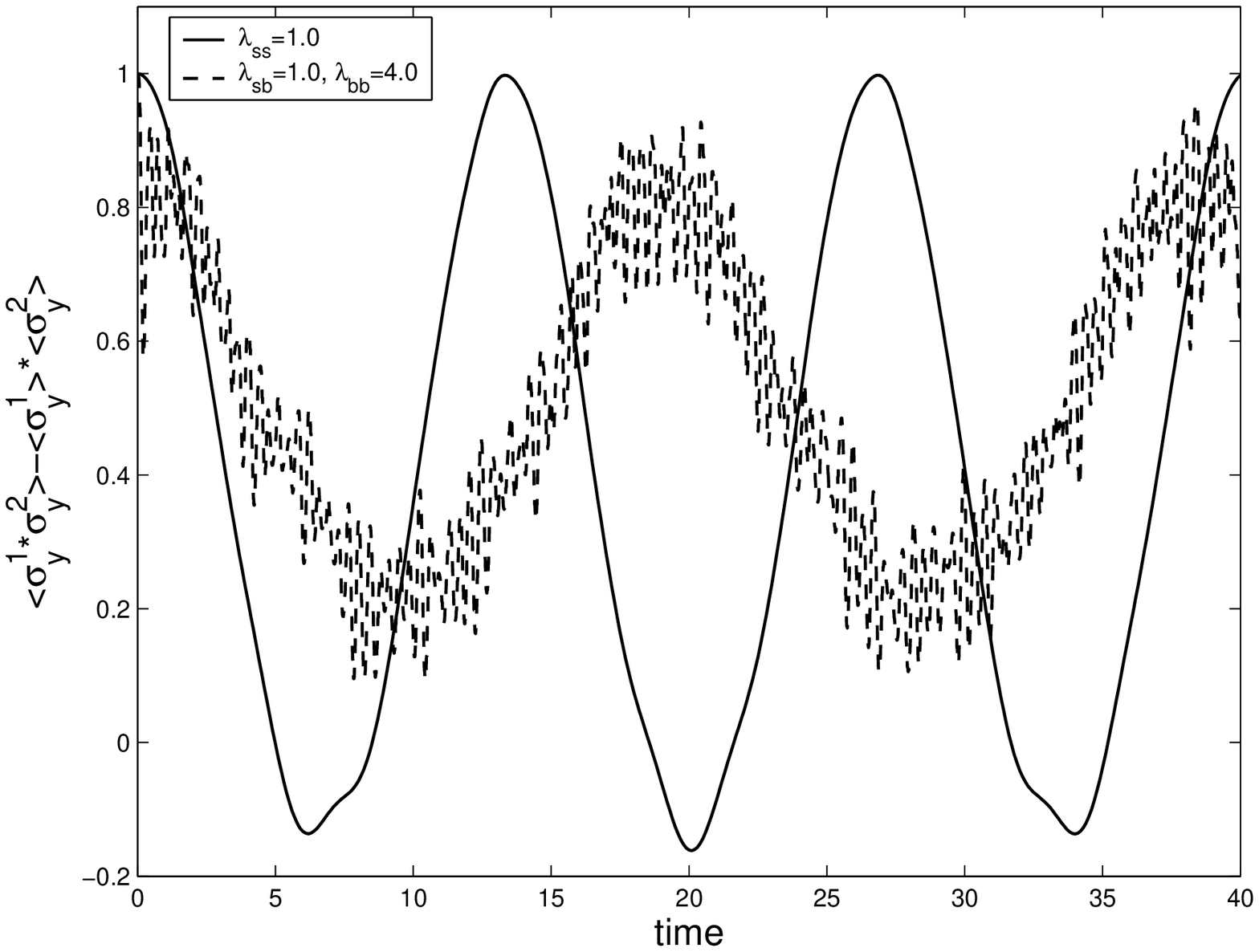}}
\subfigure[$\langle\sigma_y^{1}\sigma_y^{2}\rangle
  -\langle\sigma_y^{1}\rangle\langle\sigma_y^{2}\rangle$, $\lambda_{bb}=24.0$]{
\label{fig:N8:y24}
\includegraphics[width=3in]{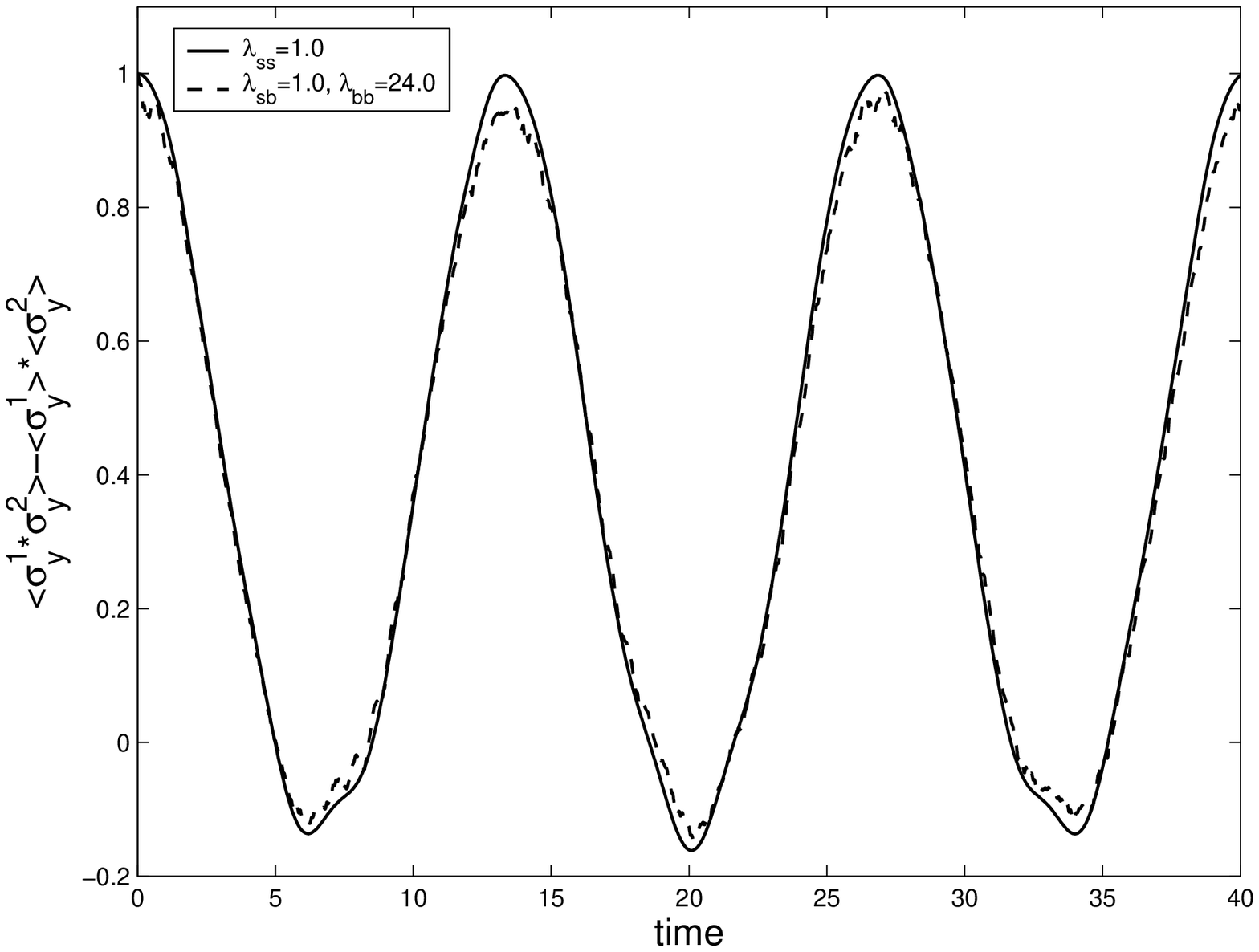}}
\subfigure[$\langle\sigma_z^{1}\sigma_z^{2}\rangle
  -\langle\sigma_z^{1}\rangle\langle\sigma_z^{2}\rangle$, $\lambda_{bb}=4.0$]{
\label{fig:N8:z4}
\includegraphics[width=3in]{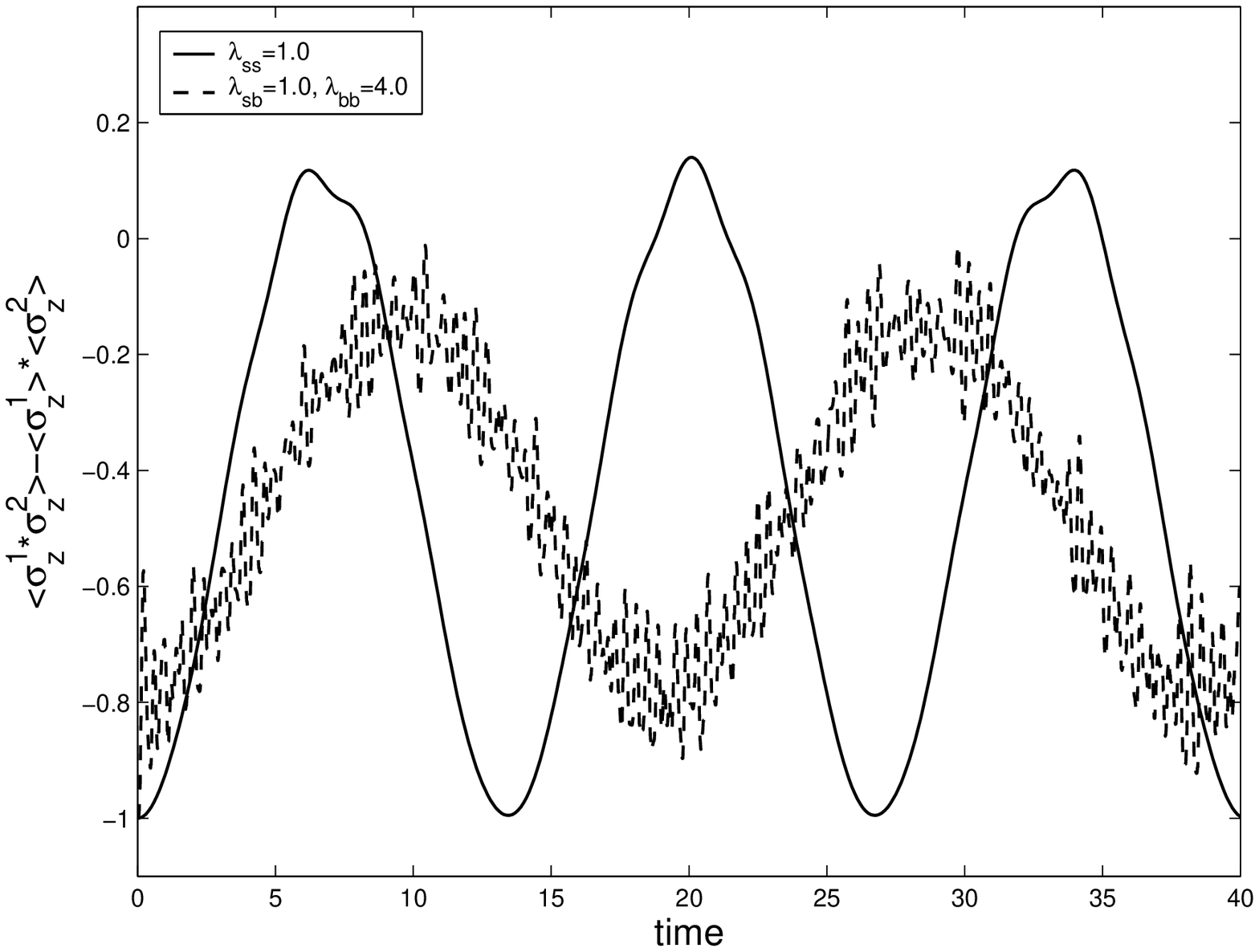}}
\subfigure[$\langle\sigma_z^{1}\sigma_z^{2}\rangle
  -\langle\sigma_z^{1}\rangle\langle\sigma_z^{2}\rangle$, $\lambda_{bb}=24.0$]{
\label{fig:N8:z24}
\includegraphics[width=3in]{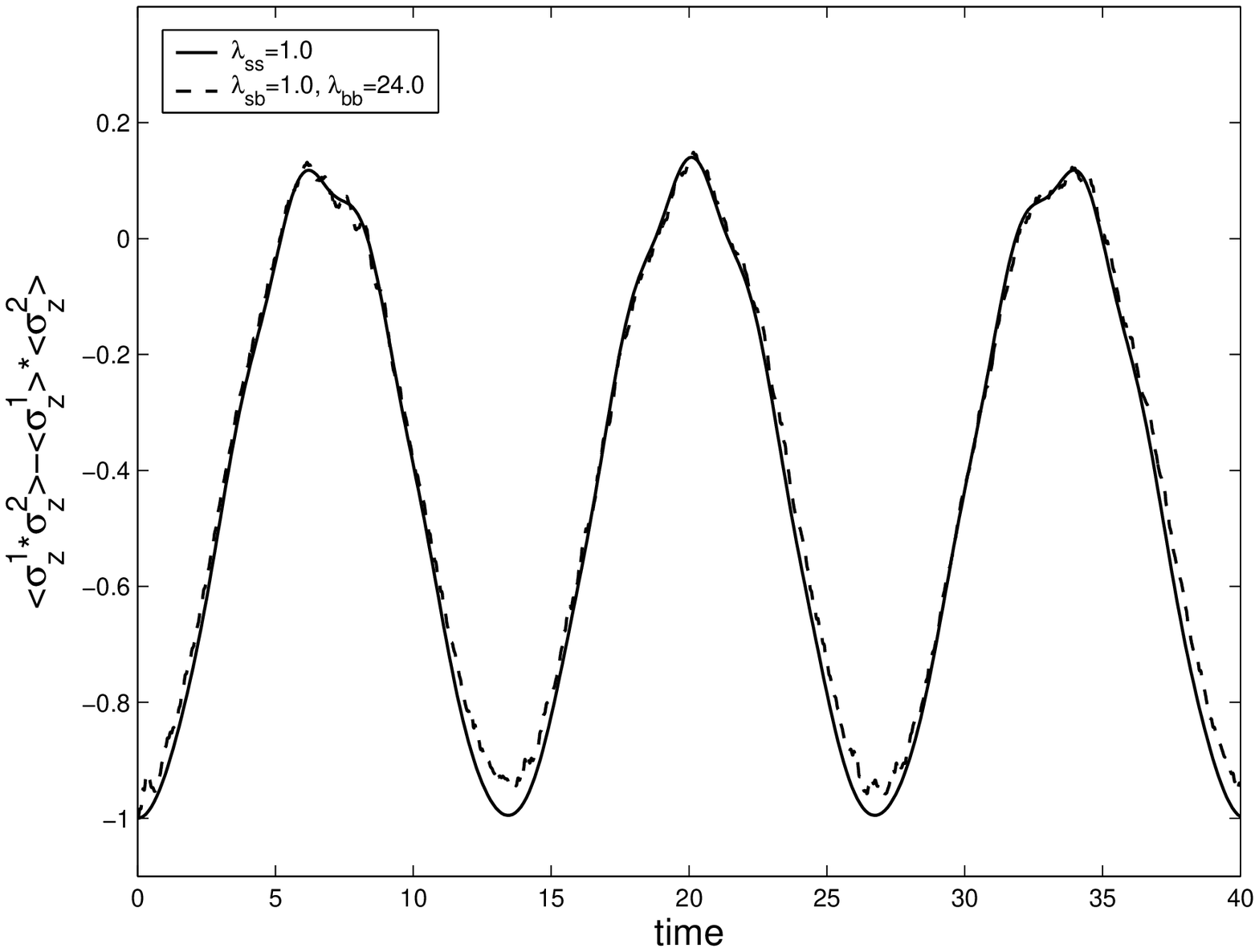}}
\caption{Evolution for polarization correlations along three
directions of the open subsystem. Where there are $8$ spins in the
bath and the initial state of the subsystem is
$1/\sqrt{2}(|01\rangle+|10\rangle)$.} \label{fig:N8}
\end{figure}

These results suggest that by changing the bath coupling strength,
one can effectively control the subsystem coherence properties. The
suppression of decoherence effect is also depends on the number of
bath spins, it is found that the more bath spins, the less effective
of suppression of the decoherence with the same bath coupling
strength. To show this, we provide the results of $\lambda_{bb}=6.0$
and $\lambda_{bb}=10.0$ on the Bell state of
$|\psi_S(0)\rangle^2=1/\sqrt{2}(|01\rangle+|10\rangle)$ in figure
\ref{fig:N6}, where there are $m=6$ bath spins. The values of other
parameters are the same as before. Through the comparison of figure
\ref{fig:N6:x6} with figure \ref{fig:2xx6}, we see that when $m=4$,
$\lambda_{bb}=6.0$ is almost sufficient to recover the isolated case
except a slightly elongated period (see figure \ref{fig:2xx6}), but
when $m$ is increased to $6$, the coupling strength need to be
increased at least to $\lambda_{bb}=10.0$ to get the same effect
(see figure \ref{fig:N6:x10}). The same feature also happens to the
evolution of $C_{yy}$ and $C_{zz}$ which can be seen by comparing
figure \ref{fig:N6:y6} with figure \ref{fig:2yy6} and figure
\ref{fig:N6:z6} with figure \ref{fig:2zz6}. If we go on to increase
the coupling strength $\lambda_{bb}$, we can suppress the
decoherence and disentanglement effect due to the bath with more
spins. In figure \ref{fig:N8}, we compare the results of the cases
of $\lambda_{bb}=4.0$ and $\lambda_{bb}=24.0$, in which the
subsystem seems to be decoupled from the bath; the initial state is
also $1/\sqrt{2}(|01\rangle+|10\rangle)$.

\FloatBarrier
\subsection{Concurrence}

\begin{figure}[htbp]
  \centering
  \includegraphics[scale=0.6]{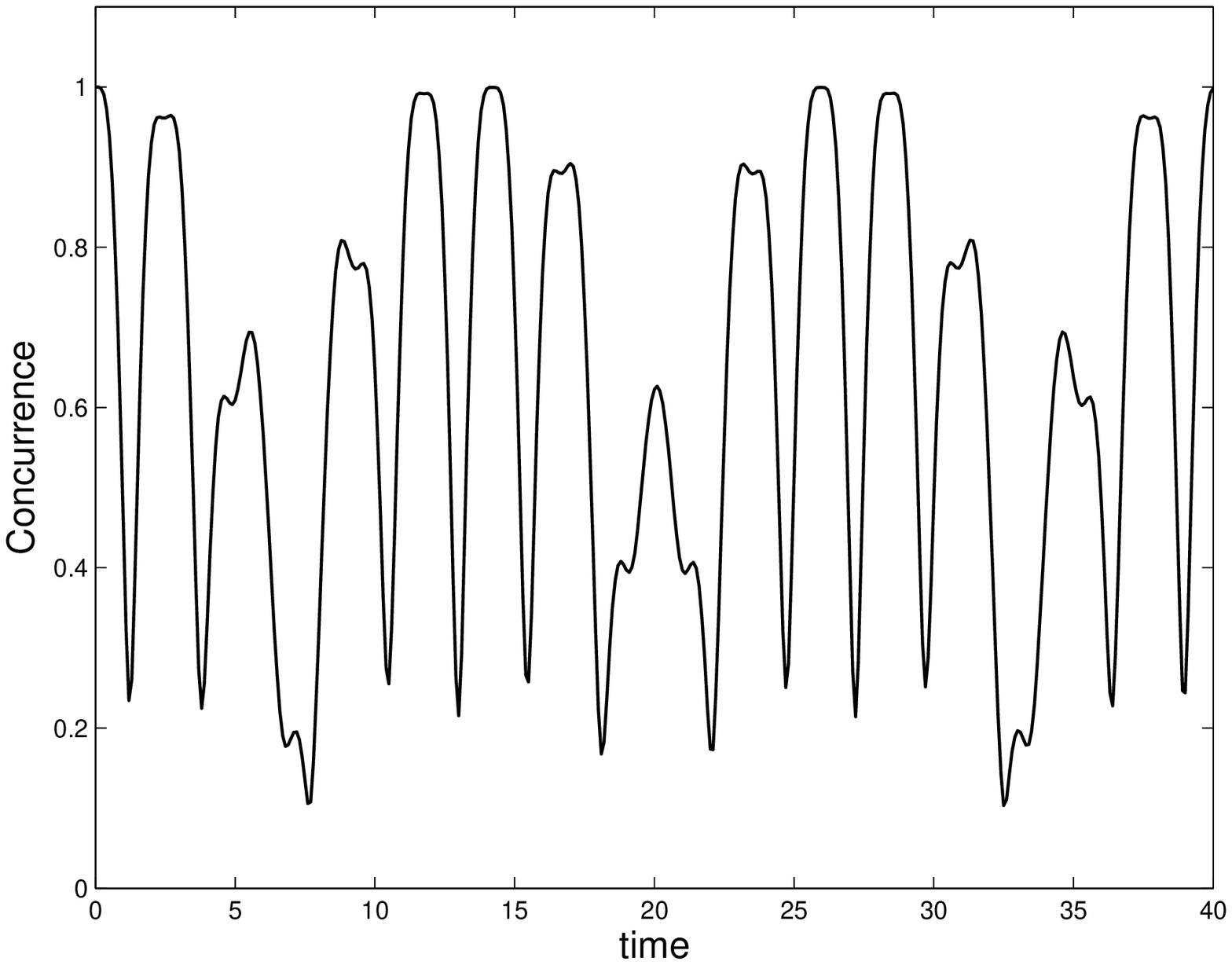}
  \caption{Evolution of the concurrence of the two
subsystem spins for the isolated subsystem. The initial state of the
subsystem is $1/\sqrt{2}(|00\rangle+|11\rangle)$.} \label{fig:Con1}
\end{figure}

\begin{figure}[htbp]
  \centering
  \includegraphics[scale=0.6]{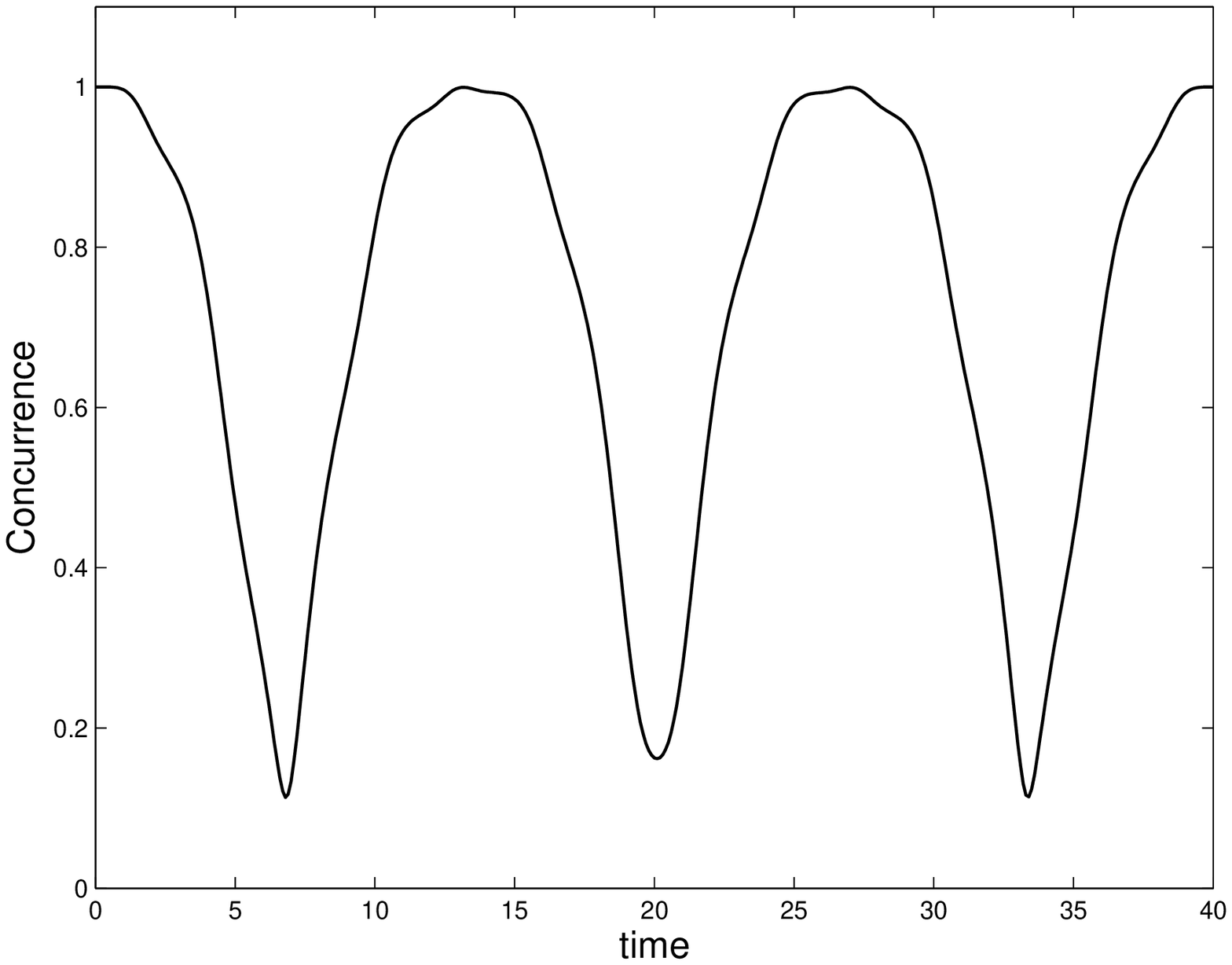}
  \caption{Evolution of the concurrence of the two
subsystem spins for the isolated subsystem. The initial state of the
subsystem  is $1/\sqrt{2}(|01\rangle+|10\rangle)$.} \label{fig:Con2}
\end{figure}

\begin{figure}[htbp]
  \centering
  \includegraphics[scale=0.6]{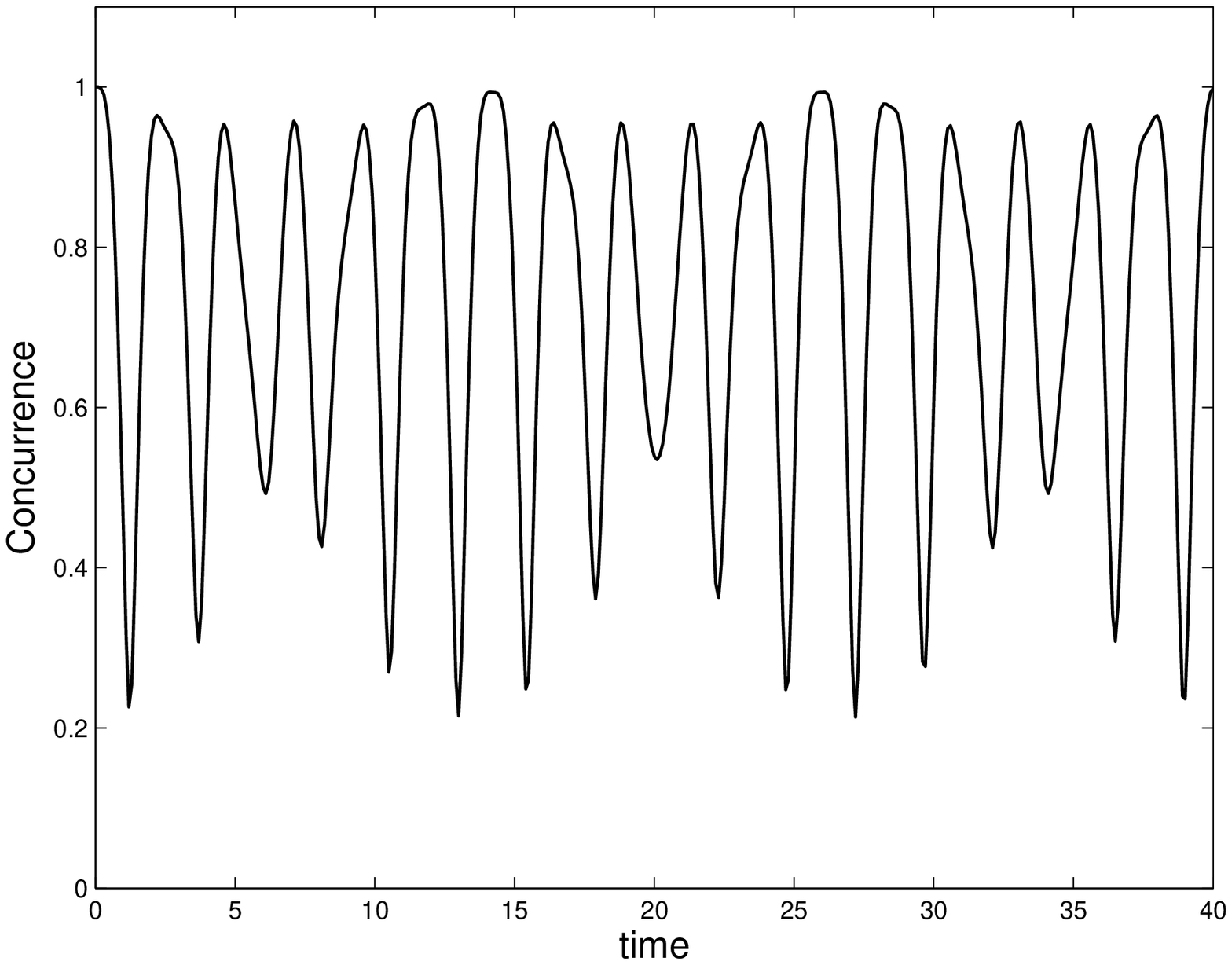}
  \caption{Evolution of the concurrence of the two
subsystem spins for the isolated subsystem. The initial state of the
subsystem is $1/\sqrt{2}(|00\rangle-|11\rangle)$.} \label{fig:Con3}
\end{figure}

\begin{figure}[htbp]
\centering
\includegraphics[scale=0.6]{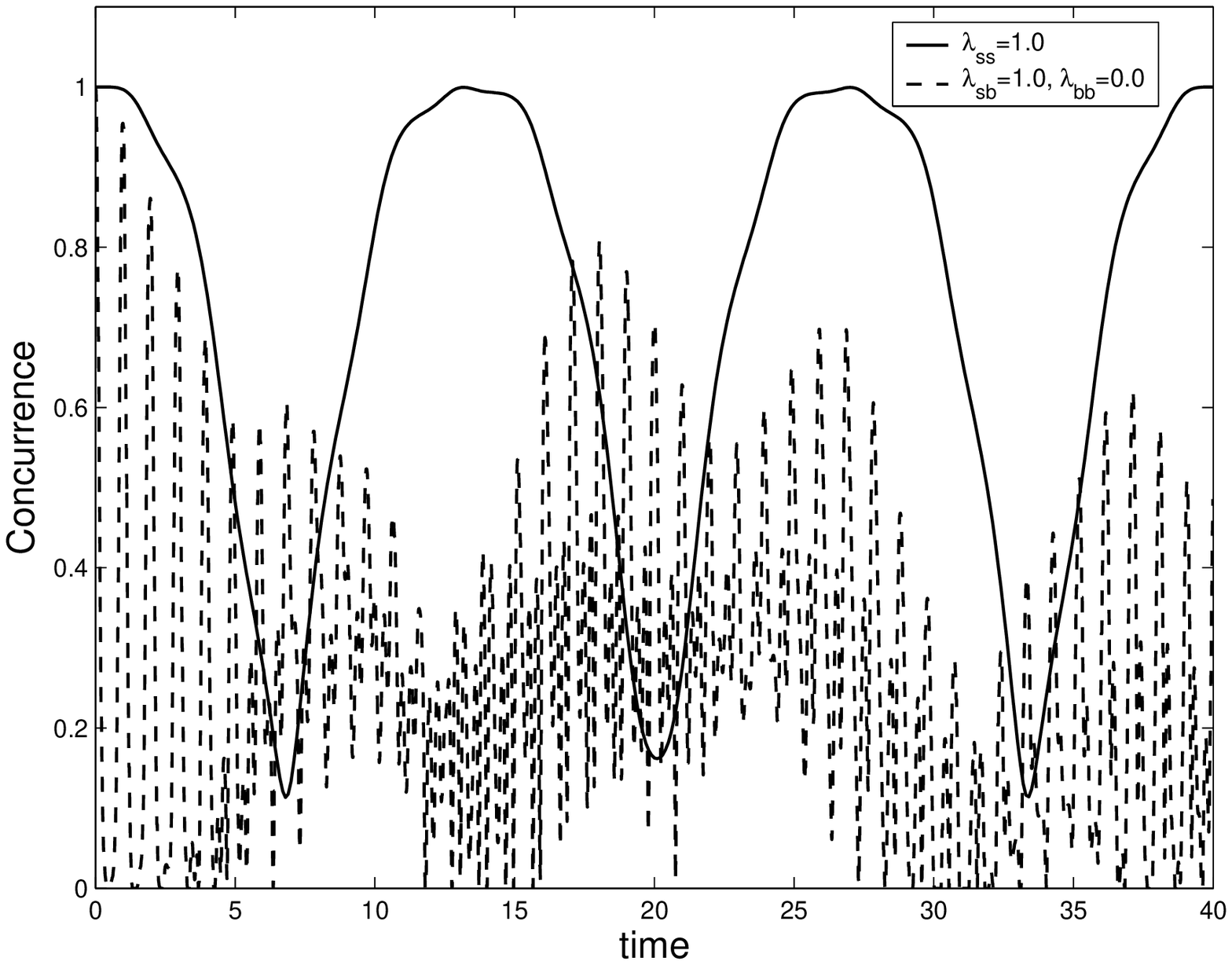}
\caption{The evolution of concurrence of the open subsystem. Where
there are $6$ spins in the bath and the initial state of the
subsystem is $1/\sqrt{2}(|10\rangle+|01\rangle)$ and
$\lambda_{ss}=\lambda_{sb}=1.0$, $\lambda_{bb}=0.0$.}
\label{fig2:N6C0}
\end{figure}

\begin{figure}[htbp]
\centering \subfigure[$\lambda_{bb}=6.0$]{ \label{fig2:N6C:6}
\includegraphics[width=3in]{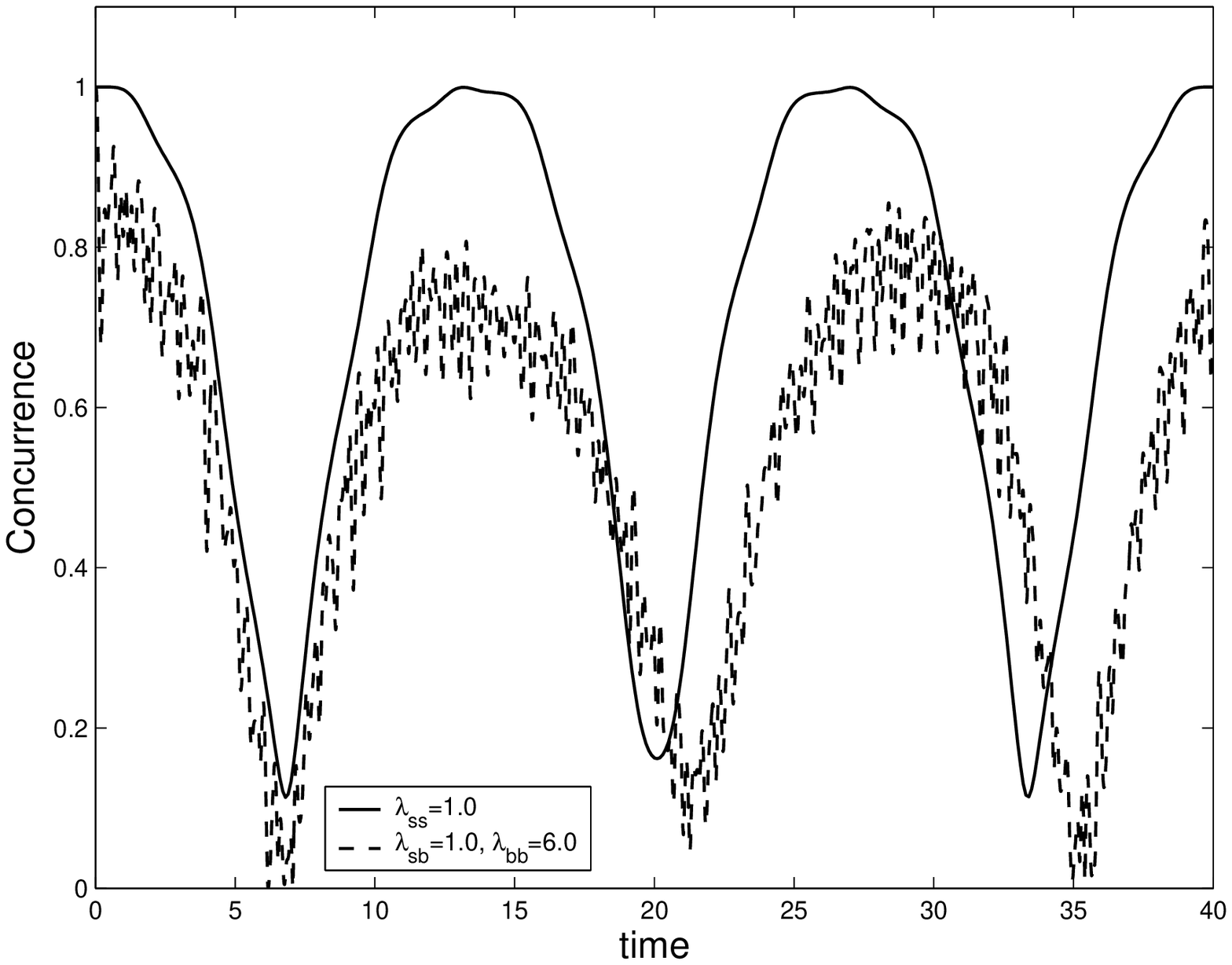}}
\subfigure[$\lambda_{bb}=10.0$]{ \label{fig2:N6C:10}
\includegraphics[width=3in]{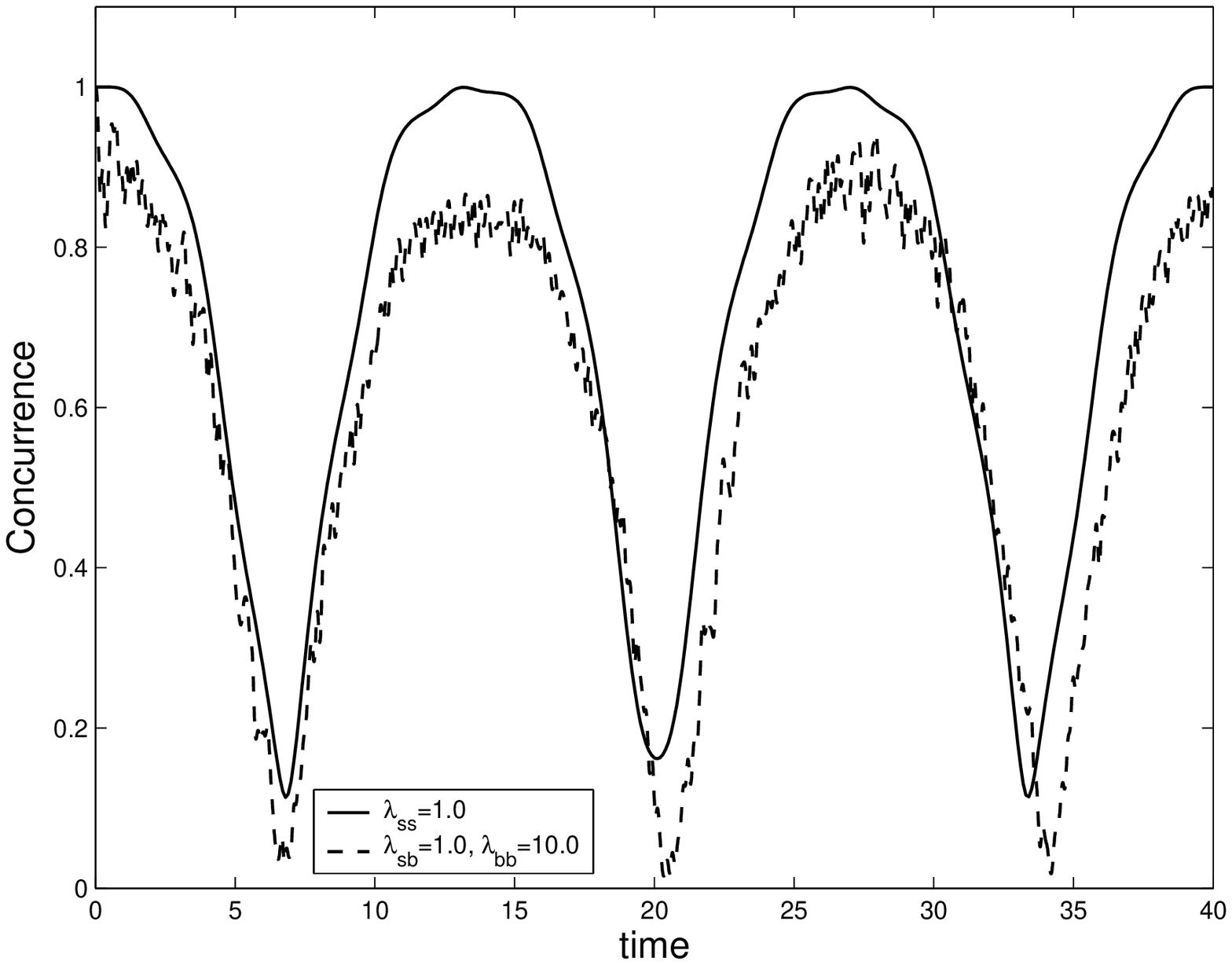}}
\caption{The evolution of concurrence of the open subsystem. Where
there are $6$ spins in the bath and the  the initial state of the
subsystem is $1/\sqrt{2}(|01\rangle+|10\rangle)$ and
$\lambda_{ss}=\lambda_{sb}=1.0$} \label{fig2:N6C}
\end{figure}

The effect of coupling strengths of bath spins on the concurrence
are also studied. In figures \ref{fig:Con1}, \ref{fig:Con2} and
\ref{fig:Con3} we plot the time evolution of the concurrence of
three Bell states for the isolated subsystem. An example that the
subsystem-bath coupling is considered but the intra-bath coupling
strength is zero is showed by figure \ref{fig2:N6C0}, which can be
compared with figure \ref{fig2:N6C} because of the same compaction
condition except $\lambda_{bb}$. \\

\begin{figure}[htbp]
\centering \subfigure[$\lambda_{bb}=6.0$]{ \label{fig:N6C:6}
\includegraphics[width=3in]{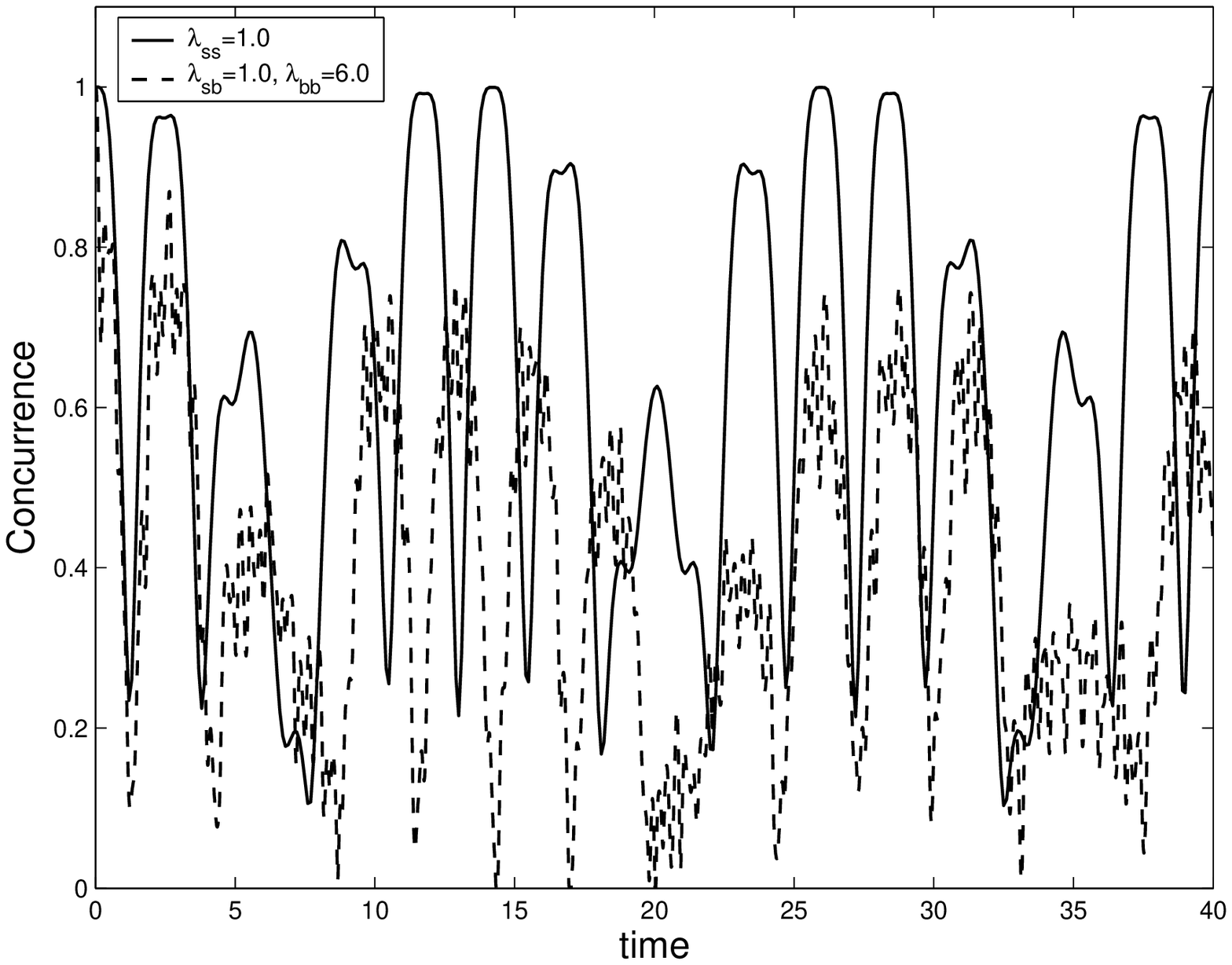}}
\subfigure[$\lambda_{bb}=10.0$]{ \label{fig:N6C:10}
\includegraphics[width=3in]{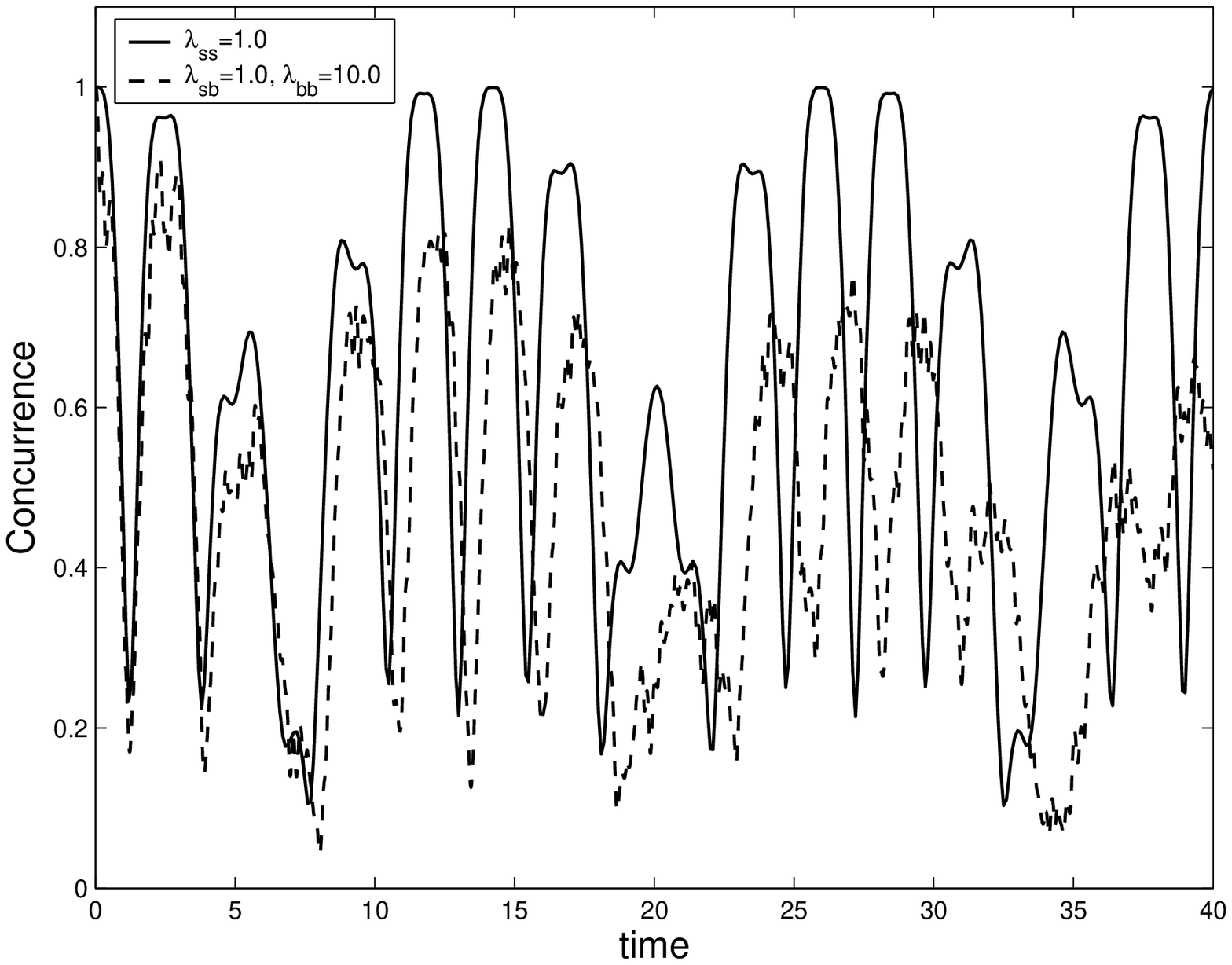}}
\caption{The evolution of concurrence of the open subsystem. Where
there are $6$ spins in the bath and the initial state of the
subsystem is $1/\sqrt{2}(|00\rangle+|11\rangle)$ and
$\lambda_{ss}=\lambda_{sb}=1.0$} \label{fig:N6C}
\end{figure}

\begin{figure}[htbp]
\centering \subfigure[$\lambda_{bb}=6.0$]{ \label{fig3:N6C:6}
\includegraphics[width=3in]{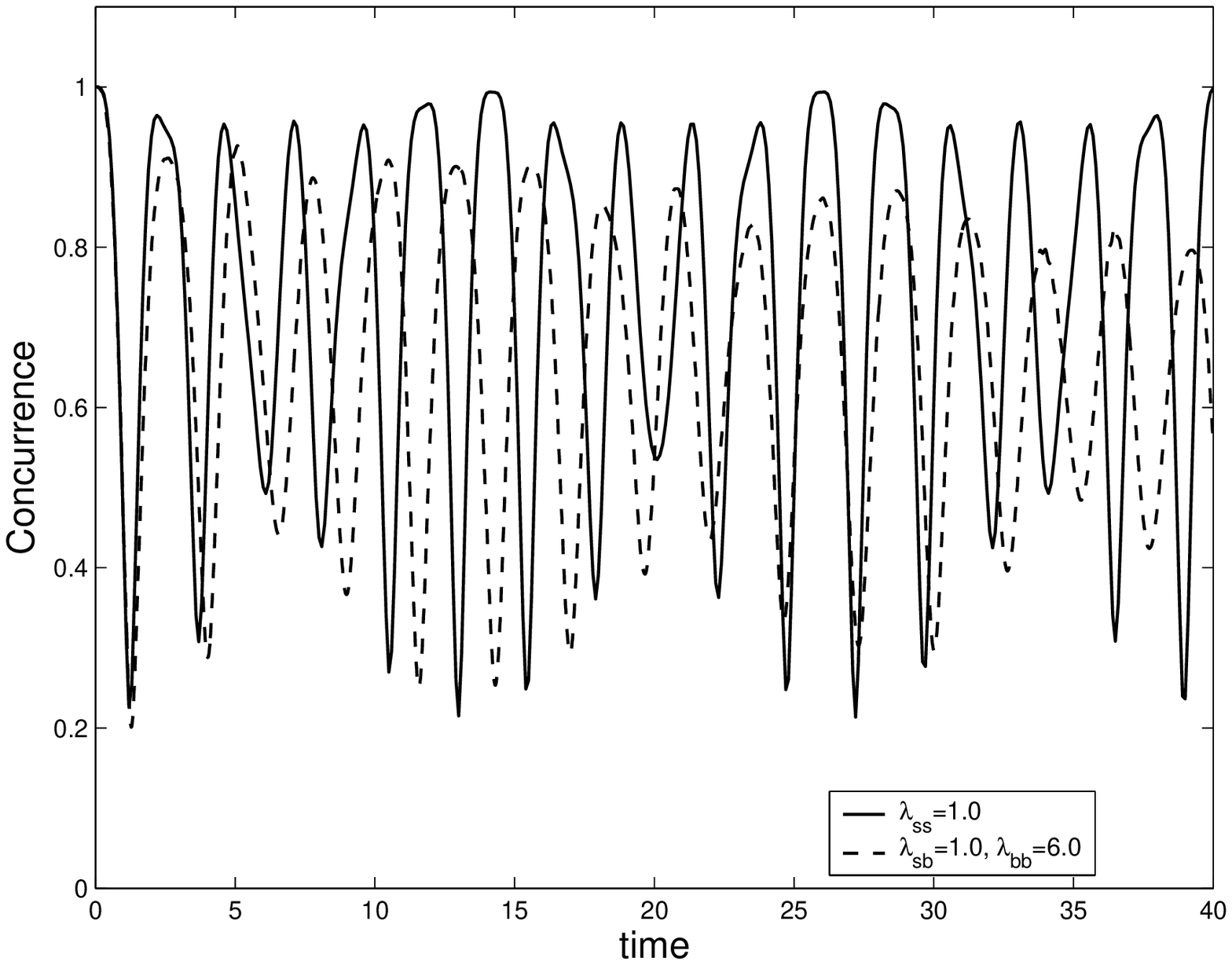}}
\subfigure[$\lambda_{bb}=10.0$]{ \label{fig3:N6C:10}
\includegraphics[width=3in]{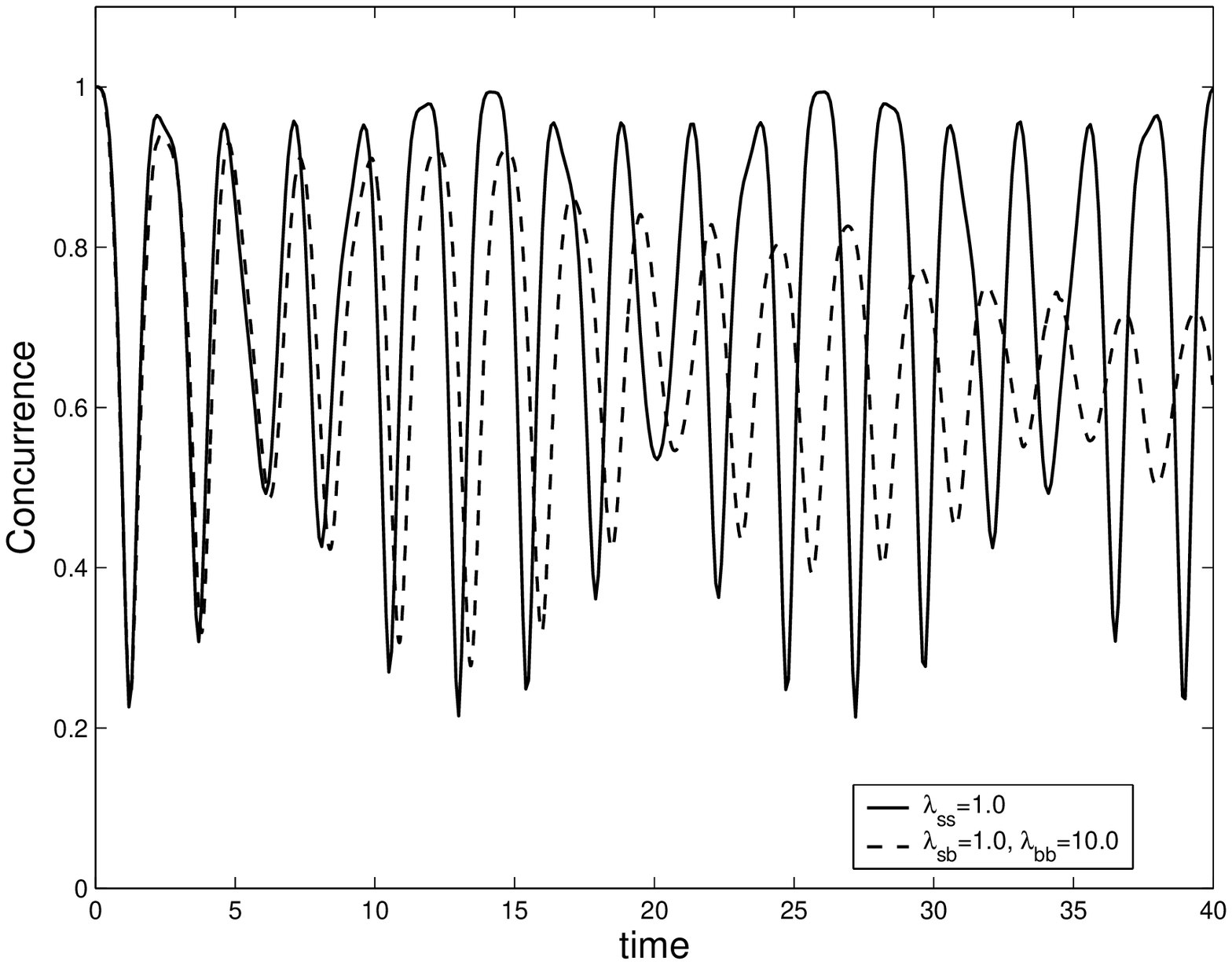}}
\caption{The evolution of concurrence of the open subsystem. Where
there are $6$ spins in the bath and the initial state of the
subsystem is $1/\sqrt{2}(|00\rangle-|11\rangle)$ and
$\lambda_{ss}=\lambda_{sb}=1.0$} \label{fig3:N6C}
\end{figure}

\begin{figure}[htbp]
\centering \subfigure[$\lambda_{bb}=4.0$]{ \label{fig2:N8C:4}
\includegraphics[width=3in]{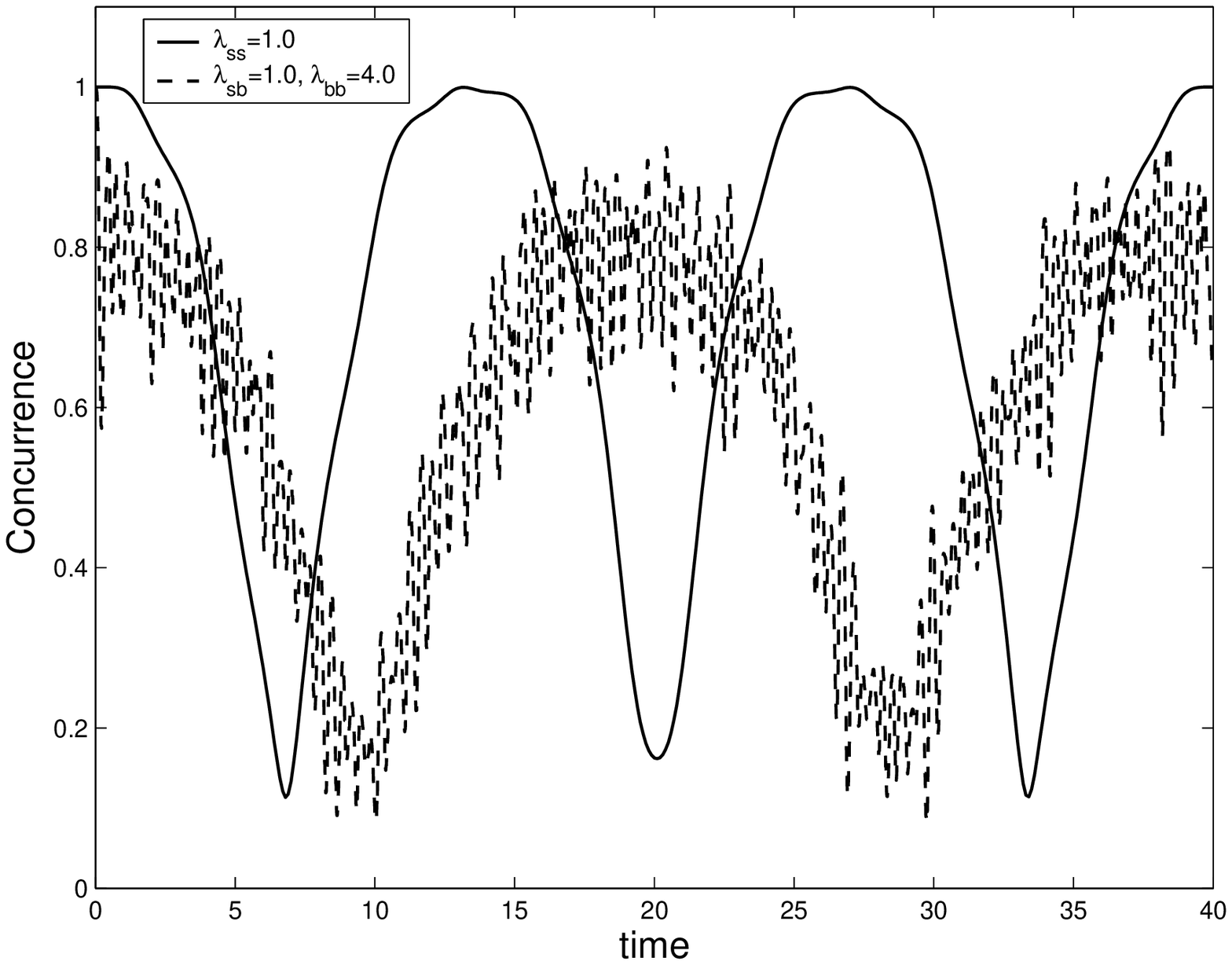}}
\subfigure[$\lambda_{bb}=24.0$]{ \label{fig2:N8C:24}
\includegraphics[width=3in]{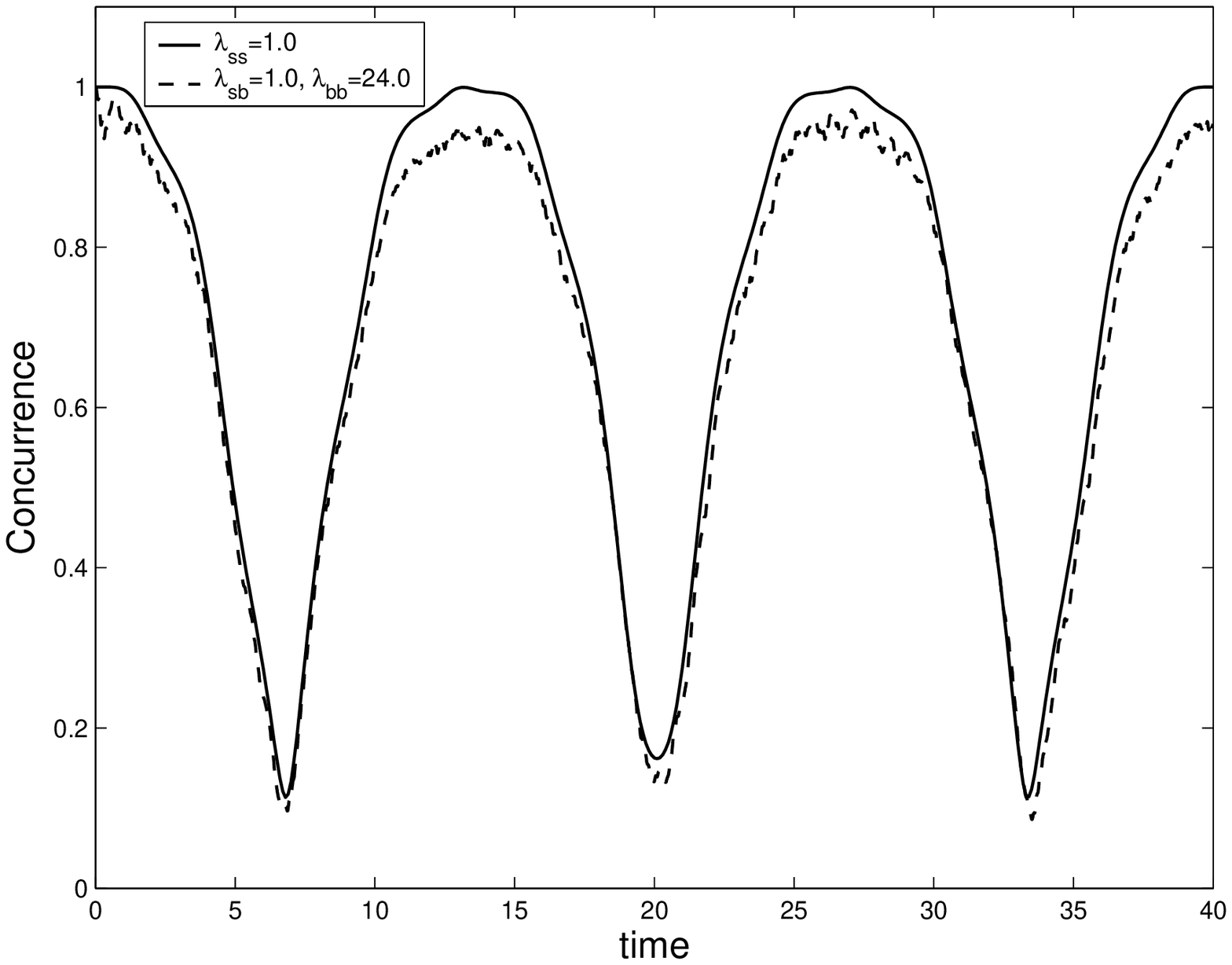}}
\caption{The evolution of concurrence of the open subsystem. Where
there are $8$ spins in the bath and the initial state of the
subsystem is $1/\sqrt{2}(|01\rangle+|10\rangle)$ and
$\lambda_{ss}=\lambda_{sb}=1.0$} \label{fig2:N8C}
\end{figure}

Figures \ref{fig2:N6C}, \ref{fig:N6C} and \ref{fig3:N6C} are plots
of the evolution of concurrence of the three Bell states of the open
subsystem. The number of bath spins is $m=6$ and two coupling
strengths of bath spins, $\lambda_{bb}=6.0$ and $\lambda_{bb}=10.0$,
are used in the calculation. By comparing the results of
$\lambda_{bb}=6.0$ with that of $\lambda_{bb}=10.0$, there are
visible improvements of the concurrence by increasing the coupling
strength of the bath spins. However, it is found that the influence
of the same coupling strengths of bath spins has less effect on the
concurrence than on the polarization correlations. From the figure
we see that even for the strength $\lambda_{bb}=10.0$, the
concurrences are still somewhat different from the isolated cases.
It is also noted that the influence is different for different
initial stats, it can be seen from figure \ref{fig2:N6C} that for
the case of Bell state
$|\psi_S(0)\rangle^2=1/\sqrt{2}(|01\rangle+|10\rangle)$, much better
suppression of decoherence is observed than the other two states.
Thus we provided the results for a larger bath with the initial
state $|\psi_S(0)\rangle^2$ in figure \ref{fig2:N8C}, where there
are $m=8$ bath spins. We found as $\lambda_{bb}=24$, the evolution
of concurrence approaches the dynamics of subsystem evolving in
isolation.

\FloatBarrier
\subsection{Discussion}

All the results of comparison in the above suggest that if we did
not consider the intra-bath coupling, or we cannot control the
intra-bath coupling, the decoherence occurring in our subsystem due
to the bath is very severe. Although the numbers of bath spins in
our simulation is not very big, but these bath spins can be regarded
as the nearest-neighbors to our open subsystem, the interaction
between them $\lambda_{sb}$ is much larger than the coupling exists
between subsystem spin and other degree of freedom in the real-world
environment. So the bath we considered in this study could form a
safeguard device of the open subsystem, which intra-coupling could
be adjusted to a high level to counteract the dissipation by itself
and the real-world environment. \\

To understand the physics behind the decoherence suppression effects
due to the strong coupling among bath spins, we investigated the
states of the bath when the coupling strength changed. The bath
spins coupled antiferromagneticly with each other in the $x$
direction, which cause the bath spins in a kind of frustrated state.
The other terms in the Hamiltonian favors a aligned ordered state.
The final state is the competition between the different terms as
well as the thermal fluctuations and turns out to be very complex.
However, we believe that the decoherence suppression effects are
somehow related to the state ordering of the bath spins.\\

At temperatures not high enough, the properties of the bath are
determined by the few lowest energy levels, so we will concentrate
on the lowest energy levels and try to figure out the
characteristics of the states. In our representation where the $z$
component of the spin is diagonal, it is hard to see the ordering
properties of the states. Since the coupling is in the $x$
direction, we thus transform our states to the $x$ component
diagonal representation for clarity. This is simply achieved by the
following recipe to each spin state,
\begin{equation}
   |\varphi^i\rangle_x=U^{-1}|\varphi^i\rangle_z.
\end{equation}
Where  $U$ is a $2\times 2$ matrix defined as
\begin{equation}\label{}
U=\frac{1}{\sqrt{2}}\left(\begin{array}{cc}
      1 & 1 \\
      1 & -1
    \end{array}\right).
\end{equation}

Now we consider a system with $6$ bath spins and two spins in the
subsystem, the bath states can be expanded with the product state of
the form $|i_1i_2i_3i_4i_5i_6\rangle_x
=|i_1\rangle_x|i_2\rangle_x|i_3\rangle_x|i_4\rangle_x|i_5\rangle_x|i_6\rangle_x$,
here $i_\alpha$ equals $0$ or $1$, the subscript $x$ means the
$x$-component diagonal representation, and $\alpha=1, \cdots, 6$ are
the index of the $6$ bath spins. There are $2^6=64$ product states,
and for each product state, the subsystem can be in four states
$|00\rangle_x$, $|10\rangle_x$,$|01\rangle_x$ and $|11\rangle_x$.
Since the Hamiltonian $H_B$ is invariant under permutations of the
bath spins, so that the states with the same number of ``up'' spins
in the $x$ direction has the same expectation values of energy, thus
we can group the states with the same number of ``up'' spins
together. And using the number of ``up'' spins $n$ to represent such
states and denote it as $||n\rangle$.

\begin{figure}[htbp]
  \centering
  \includegraphics[scale=0.6]{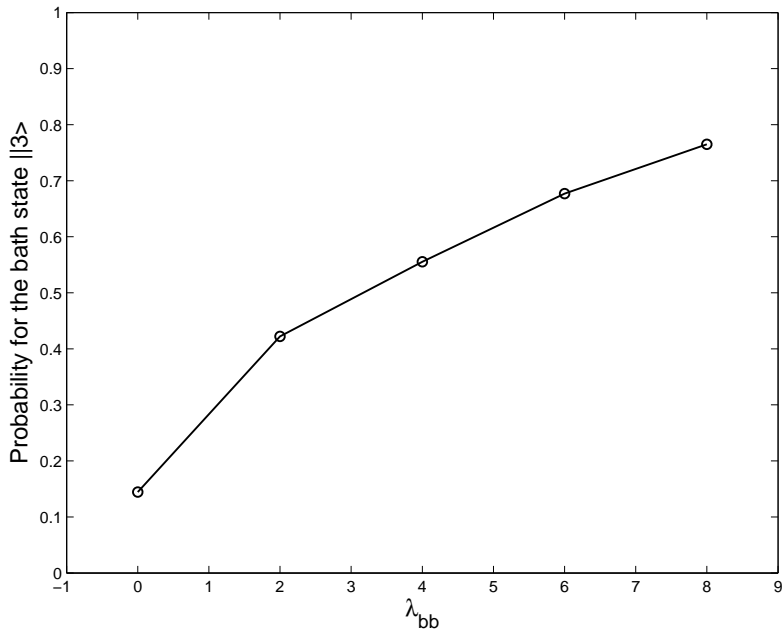}\\
  \caption{The probability that the bath state has half of
the spins in the ``up'' state and half in the ``down'' state as
function of the coupling strength $\lambda_{bb}$, the total number
of bath states is $m=6$.  The initial subsystem state is
$1/\sqrt{2}(|01\rangle+|10\rangle)$ and temperature $T=0.1$.}
\label{fig:prob}
\end{figure}

The probability that each kind of the product states appear in the
bath state can be determined from the full density matrix $\rho(t)$,
i.e.
\begin{equation}\label{prob}
P(n)=\Tr_S \langle n|| \rho(t) ||n\rangle =\Tr_s\sum
\omega_m\left|\langle n|| \Psi_m(t)\rangle \right|^2.
\end{equation}
The $\Tr_s$ means to trace out the subsystem degrees of freedom. In
the case of 6 bath spins, we calculated the probabilities for each
state $||n\rangle$ and found that the state $||3\rangle$, which has
$3$ spins ``up'' and $3$ spins ``down'' in the $x$ direction, has
the majority probability, and the probability increases with the
coupling strength. Figure \ref{fig:prob} is the plot of the
probability $P(3)$. We see that for small coupling strength, the
probability is about $0.1$, and as the coupling strength increases,
the probability increases monotonically and reaches $0.8$ at
$\lambda_{bb}=10$, where the big suppression effects was obtained.
Based on this observation we conclude that as the coupling strength
increases, the bath spins are self organized to a resonant
antiferromagnetic ordered state in $x$ direction, one half of the
spins are in the ``up'' state and the other half are in the ``down''
state. It is this ordering that brings the subsystem to the more
coherent state. This mechanism may play an important role in the
controlling of the subsystem coherence of quantum device, despite
the inevitable influences of thermal noise upon the quantum device,
we may also couple the quantum device to a system as described in
this paper. By changing the coupling strength of the added bath
spins, one may reduce the decoherence to a level for practical
applications.

\section{Conclusion}\label{conclusion}

In this paper, we extended the one-center-spin-spin-bath model
\cite{TWmodel, Milburn} to a 2-center-spin-spin-bath model, which
could be thought as an analog to two correlated qubits in quantum
computer. By calculating the polarization correlation and
concurrence of the subsystem, we found that the spin-bath can play a
revival role in the evolution of polarization correlation and
entanglement between two subsystem spins. In the process of
calculation, we combine the techniques of Ref. \cite{TWmodel} and
Ref. \cite{Jing} to reduce the computer resources greatly. The
physics of this suppression was found to be the effect of the
antiferromagnetic ordering of the bath spins in $x$ direction. We
suggest that the results may be of use in the controlling of
decoherence of quantum devices. \\

This work is supported by the National Nature Science Foundation of
China under grant \#10334020 and \#90103035.

\end{document}